\PassOptionsToPackage{sort&compress}{natbib}
\documentclass[AMA,STIX1COL]{WileyNJD-v2}

\usepackage{lineno,hyperref}
\usepackage{siunitx}
\usepackage{booktabs}
\usepackage{colortbl}
\usepackage{tikz}   
\usepackage{float}  
\usepackage{placeins}
\usepackage{graphicx}
\usepackage[caption=false]{subfig}
\usepackage{float}
\usepackage{multirow}
\usepackage{amsmath,mathtools}
\usepackage{amsfonts}
\usepackage{soul}
\usepackage{comment}
\usepackage{enumitem}
\usepackage{mathrsfs}

\usepackage{todonotes}
\usepackage{nomencl}
\usepackage{tcolorbox}
\makenomenclature
\newtcolorbox[auto counter, number within=section]{nomenclaturebox}[2][]{colframe=black!10!white, colback=black!5!white, coltitle=black, fonttitle=\bfseries, title={}}

\usepackage{algorithm}
\usepackage[algo2e,ruled,vlined]{algorithm2e}
\usepackage{setspace}

\usepackage{cases}

\hypersetup{pdfauthor={Nojus Plungė}}

\newtcolorbox{highlighted}{colback=yellow,coltext=red,breakable}

\newcommand{\FIG}[1]{Fig.~\ref{#1}}

\newcommand{\EQ}[1]{Eq.~\eqref{#1}}

\newcommand{\TAB}[1]{Table~\ref{#1}}

\newlength\myindent
\modulolinenumbers[5]

\newcommand\BibTeX{{\rmfamily B\kern-.05em \textsc{i\kern-.025em b}\kern-.08em
T\kern-.1667em\lower.7ex\hbox{E}\kern-.125emX}}

\articletype{Research Article}%

\received{<day> <Month>, <year>}
\revised{<day> <Month>, <year>}
\accepted{<day> <Month>, <year>}

\begin{document}

\title{A deep energy method formulation for phase-field modelling of ductile fracture}

\author[address1]{Nojus Plungė}
\author[address1]{Peter Brommer}
\author[address2]{Rachel S. Edwards}
\author[address1]{Emmanouil G. Kakouris*}


\address[address1]{\orgdiv{School of Engineering}, \orgname{University of Warwick}, \orgaddress{\country{UK}}}

\address[address2]{\orgdiv{Department of Physics}, \orgname{University of Warwick}, \orgaddress{\country{UK}}}

\corres{*Emmanouil G. Kakouris, School of Engineering, University of Warwick, Coventry, CV4 7AL, UK. \email{Emmanouil.Kakouris@warwick.ac.uk}}


\abstract[Abstract]{This work extends the Deep Energy Method (DEM) to phase-field modelling of ductile fracture in elasto-plastic solids. Its key contribution is an incremental DEM-based formulation that couples history-dependent \(J_2\) plasticity and plasticity-dependent fracture toughness with separate energy-optimised displacement and phase-field representations. The formulation combines the second-order Ambrosio-Tortorelli (AT2) phase-field fracture model with small-strain \(J_2\) plasticity and linear isotropic hardening. A volumetric–deviatoric split defines the elastic crack-driving energy, while accumulated equivalent plastic strain governs the degradation of local fracture toughness. Plastic internal variables are updated locally through radial return. The reduced energy is minimised alternately with respect to the displacement and phase fields, represented by a ReZero residual neural network (NN) and spatially localised Gaussian radial basis functions (RBFs) with trainable centres, respectively. This split addresses the diffuse phase-field profiles observed in preliminary calculations with a shared global network. Alternating optimisation is combined with transfer learning to reuse trainable parameters and optimiser states across loading increments, while the converged constitutive history is carried forward separately at quadrature points. The numerical results demonstrate close agreement with finite-element reference solutions in predicting force-displacement responses, energy evolution, plastic deformation and crack propagation. The framework captures the main features of ductile fracture, although accuracy becomes more sensitive to loading increments near abrupt fracture events.}


\maketitle


\section{Introduction} \label{sec:intro}

Ductile fracture is a critical failure mechanism in metallic materials.\cite{Benzerga2016,Noell2023} It arises from the interaction between plastic deformation, damage accumulation and crack growth. At the macroscopic scale, ductile failure is accompanied by progressive losses of stiffness, strength and ductility before rupture.\cite{Besson2010} At the microscale, damage may develop through void nucleation, growth and coalescence, as well as through micro-shear-band formation and micro-cracking.\cite{Noell2023,Noell2018} The resulting response depends strongly on the stress state, material hardening, geometry and loading path.\cite{Noell2023,Bao2004} Reliable numerical models are therefore essential for assessing structural integrity, predicting crack initiation and estimating the residual load-carrying capacity of engineering components.

Phase-field modelling provides a well-established variational framework for fracture. Cracks are represented by a continuous auxiliary field rather than explicitly tracked discontinuities.\cite{Ambati2015b,Borden2016,Zhang2023} This regularised description allows crack nucleation, propagation, branching and coalescence to be captured on a fixed computational mesh, without remeshing or separate crack-path criteria. The internal length scale controls the width of the diffuse crack representation, although adequate spatial resolution remains necessary near strongly localised regions. For elasto-plastic solids, the phase-field formulation can be coupled with plastic flow and the associated internal variables. Plastic deformation, material degradation and crack growth can then be described within the same continuum framework.\cite{Miehe2016,Yin2020} This provides a systematic basis for representing the transition from distributed plastic deformation to macroscopic fracture.

Ductile phase-field models have progressed from early plasticity-damage couplings to formulations tailored to specific failure mechanisms. Ambati et al.\cite{Ambati2015b} proposed a small-strain elasto-plastic model in which damage evolution is linked to accumulated plastic strain. This formulation captures the transition from plastic deformation to crack initiation and subsequent growth. Miehe et al.\cite{Miehe2016} developed a finite-strain variational theory that couples gradient plasticity and gradient damage through independent length scales. Borden et al.\cite{Borden2016} subsequently introduced finite-deformation balance laws, plastic degradation and stress-triaxiality effects to improve the post-initiation response. Yin and Kaliske\cite{Yin2020} followed a different approach, in which accumulated plastic strain reduces the local fracture toughness while the crack-driving force remains elastic. Han et al.\cite{Han2022} combined a plastic contribution to the crack-driving force with degrading fracture toughness, thereby linking stiffness degradation and the loss of fracture resistance. More recent developments have addressed dynamic fracture, coupled spall and adiabatic shear banding, and thermomechanically coupled finite-strain elasto-plasticity.\cite{Zhang2023,Svolos2025} These studies demonstrate the versatility of ductile phase-field modelling. Their numerical solution, however, remains predominantly based on conventional discretisation techniques.

Ductile phase-field fracture is computationally demanding because the displacement and phase fields are coupled to history-dependent plastic variables. Fine spatial resolution is often required near evolving cracks and plastic zones. The problem must also be solved incrementally, with constitutive variables updated throughout the loading history. These requirements motivate the development of alternative computational approaches based on artificial intelligence and machine learning. Physics-informed neural networks (PINNs) approximate the unknown fields using NNs trained through residuals of the governing equations, initial conditions and boundary conditions.\cite{Raissi2019} Variational neural methods, including the Deep Ritz Method and the DEM, instead employ NNs as trial functions and determine their parameters by minimising an energy functional.\cite{E2018,Samaniego2020,NguyenThanh2020} Energy-based methods are particularly compatible with phase-field fracture because the governing equations follow from variational principles.

Neural approaches to fracture have developed from general physics-informed solvers towards formulations designed specifically for phase-field models. Early studies introduced variational energy-based PINNs for brittle fracture and used transfer learning to reduce the optimisation cost across successive loading increments.\cite{Goswami2020a} Adaptive refinement was subsequently incorporated to improve the resolution of evolving crack paths.\cite{Goswami2020b} Ghaffari Motlagh et al.\cite{GhaffariMotlagh2023} compared several physics-informed and energy-based formulations, with particular attention to boundary-condition enforcement and identification of the phase-field length scale. Manav et al.\cite{Manav2024} demonstrated that the Deep Ritz Method can reproduce brittle crack nucleation, propagation, kinking, branching and coalescence. Their work also highlighted the optimisation difficulties associated with non-convex fracture energies. Further developments have addressed anisotropic and heterogeneous materials, transfer learning, and hybrid finite-element--neural-network strategies.\cite{Plunge2026,Zhang2026,Pantidis2026a,Pantidis2026b} Wang et al.\cite{Wang2026} proposed an extended DEM that places discrete and continuous fracture descriptions within a common neural framework. In its continuous phase-field setting, separate approximations are employed for the displacement and phase fields, and RBFs are used to improve the representation of localised cracks. That formulation, however, considers elastic fracture and does not incorporate the constitutive evolution associated with plastic deformation. More recently, Zhang et al.\cite{ZhangThermo2026} extended the DEM to thermo-mechanical brittle fracture, using separate NNs for displacement and temperature and an explicitly represented sharp crack.

Energy-based neural methods have also been applied independently to elasto-plasticity. He et al.\cite{He2023} extended the DEM to classical \(J_2\) plasticity and incorporated local constitutive updates through a radial-return algorithm. This established a practical treatment of plastic history dependence within an energy-based neural solver, but phase-field fracture was not included. Aldakheel et al.\cite{Aldakheel2025} investigated physics-based machine-learning models for brittle and ductile fracture in elastic-viscoplastic materials. Their constitutive surrogates were trained using finite-element-generated homogeneous one-dimensional responses and were subsequently embedded within boundary-value simulations. Despite these advances, a direct and label-free energy-based neural solver that combines spatial crack evolution, local elasto-plastic history and plasticity-dependent fracture resistance remains underdeveloped.

To address this gap, the present study develops a DEM-based variational formulation for ductile phase-field fracture in isotropic elasto-plastic solids. The constitutive framework follows the small-strain model of Yin and Kaliske.\cite{Yin2020} In this model, accumulated plastic strain progressively reduces the local fracture toughness, whereas the crack-driving force is derived from the elastic strain energy. The formulation therefore distinguishes between the energetic force that drives fracture and the plasticity-induced reduction of material resistance. This setting permits the influence of plastic history on crack evolution to be examined without introducing additional thermo-mechanical fields.

The contribution of this work lies in integrating the ductile phase-field model and the DEM-based solution strategy within one incremental computational procedure. Three features are central to the formulation. Firstly, the displacement and phase fields are represented by trainable approximations, while the plastic strain and hardening variables are advanced locally using a radial-return algorithm. The path-dependent constitutive variables are therefore retained at the integration points and are not introduced as additional trainable fields. Secondly, distinct approximations are adopted for the two global fields. A ReZero residual neural network represents the displacement field, whereas spatially localised Gaussian RBFs with trainable centres resolve the localised phase-field profile. Thirdly, alternating optimisation is combined with transfer learning to reuse trainable parameters and optimiser states across loading increments, while the converged constitutive history is carried forward separately at the quadrature points. The resulting procedure couples \(J_2\) plasticity, plasticity-dependent fracture resistance and irreversible crack evolution without requiring labelled training data.

Numerical benchmarks are used to evaluate force-displacement responses, energy evolution, accumulated equivalent plastic strain and crack propagation against finite-element reference solutions. The examples also examine the sensitivity of the response to loading-increment size, particularly near abrupt fracture events. The results generally agree closely with the reference solutions, with larger differences concentrated near abrupt rupture. The present work is restricted to small strains, linear isotropic hardening and isothermal conditions. More general ductile phase-field formulations, including finite-strain plasticity, thermomechanical coupling and thermo-viscoplastic behaviour,\cite{Svolos2025} require additional fields and constitutive couplings. Their incorporation into the proposed neural framework is left for future investigation.

The remainder of this manuscript is organised as follows. Section~\ref{sec:phase_field_ductile} introduces the adopted ductile phase-field formulation and elasto-plastic constitutive model. Section~\ref{sec:ext_deep_ritz_method} presents the DEM implementation and incremental solution procedure. Section~\ref{sec:num_results} reports the numerical examples and comparisons with the reference solutions. Finally, Section~\ref{sec:conclusions} summarises the main findings and outlines directions for future research.

\section{Phase-field formulation for ductile fracture} \label{sec:phase_field_ductile}

This section defines the ductile phase-field model adopted in this work. The formulation follows the small-strain framework of Yin and Kaliske,\cite{Yin2020} in which accumulated equivalent plastic strain reduces the local fracture toughness. The crack-driving force is derived from the elastic strain energy, while the plastic internal variables evolve through associative $J_{2}$ plasticity with linear isotropic hardening.

\subsection{Variational form} \label{sec:energy_functional}

Consider a deformable body occupying a domain $\Omega\subset\mathbb{R}^{d}$, where $d$ denotes the spatial dimension and $\mathbf{x}\in\Omega$ denotes the position vector. Its boundary is decomposed as $\partial\Omega=\partial\Omega_{\bar{\mathbf{u}}}\cup\partial\Omega_{\bar{\mathbf{h}}}$, with $\partial\Omega_{\bar{\mathbf{u}}}\cap\partial\Omega_{\bar{\mathbf{h}}}=\varnothing$. The displacement satisfies $\mathbf{u}=\bar{\mathbf{u}}$ on $\partial\Omega_{\bar{\mathbf{u}}}$, while the prescribed traction is $\bar{\mathbf{h}}$ on $\partial\Omega_{\bar{\mathbf{h}}}$. A body force $\mathbf{f}^{\mathrm{b}}$ may act in $\Omega$.

The sharp crack set $\Gamma$ is regularised by a scalar phase field $c:\Omega\rightarrow[0,1]$. The value $c=1$ denotes intact material and $c=0$ denotes fully fractured material. The formulation is written in general form, while the numerical examples considered in this work are restricted to two-dimensional plane-strain problems. The $J_{2}$ invariants and tensor contractions retain their standard three-dimensional definitions.

Under quasi-static loading, the incremental total potential energy is written as
\begin{equation}
\label{eqn:ductile_pot_eng}
\Pi(\mathbf{u},c,\boldsymbol{\varepsilon}^{\mathrm{p}},\alpha,\zeta)
=\mathcal{W}^{\mathrm{int}}(\mathbf{u},c,\boldsymbol{\varepsilon}^{\mathrm{p}},\alpha,\zeta)
-\mathcal{W}^{\mathrm{ext}}(\mathbf{u})
+\mathcal{P}^{\mathrm{irr}}(c;c_{n-1}),
\end{equation}
where $\boldsymbol{\varepsilon}^{\mathrm{p}}$ is the plastic strain tensor, $\alpha$ is the isotropic-hardening variable and $\zeta$ is the accumulated equivalent plastic strain. The previously converged phase field, $c_{n-1}$, enters the irreversibility penalty $\mathcal{P}^{\mathrm{irr}}$ as defined in Section~\ref{sec:irreversibility} below.

The external work is
\begin{equation}
\label{eqn:ductile_work_ext}
\mathcal{W}^{\mathrm{ext}}(\mathbf{u})
=\int_{\partial\Omega_{\bar{\mathbf{h}}}}\bar{\mathbf{h}}\cdot\mathbf{u}\,\mathrm{d}S
+\int_{\Omega}\mathbf{f}^{\mathrm{b}}\cdot\mathbf{u} \,\mathrm{d}\Omega.
\end{equation}

The internal contribution is decomposed as
\begin{equation}
\label{eqn:ductile_work_int}
\mathcal{W}^{\mathrm{int}}=\Psi^{\mathrm{e}}+\Psi^{\mathrm{p}}+\Psi^{\mathrm{f}}.
\end{equation}

The elastic contribution contains crack-driving and non-driving parts:
\begin{equation}
\label{eqn:ductile_elast_eng}
\Psi^{\mathrm{e}}
=\int_{\Omega}
\left[
g(c)\psi^{\mathrm{e}+}(\boldsymbol{\varepsilon}^{\mathrm{e}})
+\psi^{\mathrm{e}-}(\boldsymbol{\varepsilon}^{\mathrm{e}})
\right]\,\mathrm{d}\Omega.
\end{equation}
The elastic strain follows the additive small-strain decomposition
\begin{equation}
\label{eqn:elastic_strain_split}
\boldsymbol{\varepsilon}^{\mathrm{e}}
=\boldsymbol{\varepsilon}-\boldsymbol{\varepsilon}^{\mathrm{p}},
\qquad
\boldsymbol{\varepsilon}
=\frac{1}{2}\left(\nabla\mathbf{u}+(\nabla\mathbf{u})^{\mathsf T}\right).
\end{equation}
Only the crack-driving contribution is degraded. With the present convention for $c$, the quadratic degradation function is
\begin{equation}
\label{eqn:ductile_degrad_func}
g(c)=c^{2}.
\end{equation}

The plastic contribution is
\begin{equation}
\label{eqn:ductile_plastic_energy}
\Psi^{\mathrm{p}}=\int_{\Omega}\psi^{\mathrm{p}}(\alpha) \,\mathrm{d}\Omega,
\end{equation}
where $\psi^{\mathrm{p}}$ is defined in Section~\ref{sec:plastic_energy_hardening}.

\begin{figure}[htbp]
	\centering
	\begin{tabular}{cc}
		\subfloat[\label{fig:deform_bdy:a}]{
			\includegraphics[scale=1.0]{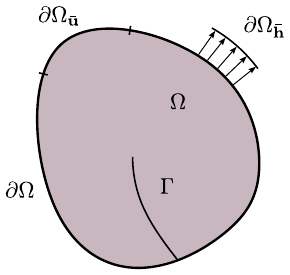}}\hspace{2cm} &
            \subfloat[\label{fig:deform_bdy:b}]{
			\includegraphics[scale=1.00]{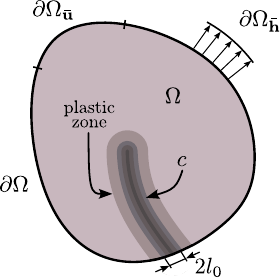}}
	\end{tabular}
	\caption[]{\subref{fig:deform_bdy:a} Solid body subjected to prescribed boundary conditions and capable of undergoing plastic deformation and developing a crack path $\Gamma$, and \subref{fig:deform_bdy:b} its phase-field approximation.}
    \label{fig:deform_bdy}
\end{figure}

The crack-surface contribution follows the AT2 regularisation:
\begin{equation}
\label{eqn:ductile_frac_eng}
\begin{aligned}
\Psi^{\mathrm{f}} = \int_{\Gamma} \mathcal{G}_{c}(\zeta) \,\mathrm{d}\Gamma \approx \int_{\Omega} \mathcal{G}_{c}(\zeta) \mathcal{Z}_{c}(c,\nabla c) \,\mathrm{d}\Omega ,
\end{aligned}
\end{equation}
with
\begin{equation}
\label{eqn:ductile_crack_density}
\begin{aligned}
\mathcal{Z}_{c} = \frac{(1-c)^{2}}{2l_{0}} + \frac{l_{0}}{2} \left| \nabla c \right|^{2},
\end{aligned}
\end{equation}
where $l_{0} \in \mathbb{R}^{+}$ is the phase-field length-scale parameter. The dependence of $\mathcal{G}_{c}$ on $\zeta$ introduces the toughness-degradation mechanism used to represent ductile fracture. A schematic representation of the phase-field approximation for ductile fracture is presented in \FIG{fig:deform_bdy}.

\subsection{Elastic energy split} \label{sec:elastic_energy_split}

The crack-driving and non-driving elastic energy densities in \EQ{eqn:ductile_elast_eng} are defined using the volumetric-deviatoric split adopted by Yin and Kaliske.\cite{Yin2020} This split is consistent with classical $J_{2}$ plasticity and excludes the compressive volumetric contribution from the crack-driving energy.

The first invariant and deviatoric part of the elastic strain are
\begin{equation}
\label{eqn:first_invariant_elastic_strain}
I_{1}=\operatorname{tr}(\boldsymbol{\varepsilon}^{\mathrm{e}}),
\qquad
\boldsymbol{\varepsilon}^{\mathrm{e}}_{\mathrm{dev}}
=\boldsymbol{\varepsilon}^{\mathrm{e}}-\frac{1}{3}I_{1}\mathbf{I},
\end{equation}
where $\mathbf{I}$ is the second-order identity tensor. The two energy densities are
\begin{equation}
\label{eqn:elastic_energy_split}
\begin{aligned}
\psi^{\mathrm{e}+} \left( \boldsymbol{\varepsilon}^{\mathrm{e}} \right) &= \frac{1}{2} \kappa \left\langle I_{1} \right\rangle_{+}^{2} + \mu \left( \boldsymbol{\varepsilon}^{\mathrm{e}}_{\mathrm{dev}} : \boldsymbol{\varepsilon}^{\mathrm{e}}_{\mathrm{dev}} \right),
\qquad
\psi^{\mathrm{e}-} \left( \boldsymbol{\varepsilon}^{\mathrm{e}} \right) &= \frac{1}{2} \kappa \left\langle I_{1} \right\rangle_{-}^{2},
\end{aligned}
\end{equation}
where $\kappa$ and $\mu$ are the bulk and shear moduli. The Macaulay brackets are defined by
\begin{equation}
\label{eqn:macaulay_brackets}
\langle x\rangle_{+}=\frac{1}{2}(x+\lvert x\rvert),
\qquad
\langle x\rangle_{-}=\frac{1}{2}(x-\lvert x\rvert).
\end{equation}

The corresponding stress response is
\begin{equation}
\label{eqn:stress_response}
\boldsymbol{\sigma}
=\kappa\left[g(c)\langle I_{1}\rangle_{+}
+\langle I_{1}\rangle_{-}\right]\mathbf{I}
+2\mu g(c)\boldsymbol{\varepsilon}^{\mathrm{e}}_{\mathrm{dev}}.
\end{equation}
Consequently, the deviatoric stress entering the plasticity model is
\begin{equation}
\label{eqn:deviatoric_stress}
\boldsymbol{\sigma}_{\mathrm{dev}}
=2\mu g(c)\boldsymbol{\varepsilon}^{\mathrm{e}}_{\mathrm{dev}}.
\end{equation}
Thus, the tensile volumetric contribution and the full deviatoric contribution enter the crack-driving energy, whereas the compressive volumetric contribution remains undegraded.

\subsection{Plastic energy and hardening} \label{sec:plastic_energy_hardening}

The plastic response is described by associative $J_{2}$ plasticity with linear isotropic hardening. The plastic energy density is
\begin{equation}
\label{eqn:plastic_energy_density}
\psi^{\mathrm{p}}(\alpha) =\sigma_{\mathrm{y}}\alpha+\frac{1}{2}H\alpha^{2},
\end{equation}
where $\sigma_{\mathrm{y}}$ is the initial yield stress and $H\geq0$ is the hardening modulus. The thermodynamic force conjugate to $\alpha$ is
\begin{equation}
\label{eqn:plastic_thermodynamic_force}
t_{\alpha} =\frac{\partial\psi^{\mathrm{p}}}{\partial\alpha}=\sigma_{\mathrm{y}}+H\alpha.
\end{equation}

The yield function is
\begin{equation}
\label{eqn:j2_yield_function}
\mathcal{F}(\boldsymbol{\sigma},\alpha)
=\lVert\boldsymbol{\sigma}_{\mathrm{dev}}\rVert
-\sqrt{\frac{2}{3}}\textbf{\,}t_{\alpha}\leq0,
\end{equation}
where $\lVert\cdot\rVert$ denotes the Frobenius norm. For plastic loading, the associative flow rule and hardening evolution are
\begin{equation}
\label{eqn:plastic_flow_rule}
\dot{\boldsymbol{\varepsilon}}^{\mathrm{p}}=\dot{\chi}\mathbf{N},
\qquad
\mathbf{N}=\frac{\boldsymbol{\sigma}_{\mathrm{dev}}}{\lVert\boldsymbol{\sigma}_{\mathrm{dev}}\rVert},
\qquad
\dot{\alpha}=\sqrt{\frac{2}{3}}\textbf{\,}\dot{\chi},
\end{equation}
where $\dot{\chi}$ is the plastic multiplier. Plastic incompressibility follows from
$\operatorname{tr}(\dot{\boldsymbol{\varepsilon}}^{\mathrm{p}})=0$. The admissible evolution satisfies the Kuhn-Tucker conditions
\begin{equation}
\label{eqn:kuhn_tucker_conditions}
\dot{\chi}\geq0,
\qquad
\mathcal{F}\leq0,
\qquad
\mathcal{F}\dot{\chi}=0.
\end{equation}

The accumulated equivalent plastic strain controlling the fracture toughness is
\begin{equation}
\label{eqn:equivalent_plastic_strain}
\zeta(t)=\sqrt{\frac{2}{3}}\int_{0}^{t}\sqrt{\dot{\boldsymbol{\varepsilon}}^{\mathrm{p}}:\dot{\boldsymbol{\varepsilon}}^{\mathrm{p}}}\,\mathrm{d}\tau.
\end{equation}
Since $\mathbf{N}:\mathbf{N}=1$, \EQ{eqn:plastic_flow_rule} gives $\dot{\zeta}=\sqrt{2/3}\textbf{\,}\dot{\chi}$. Therefore, $\alpha$ and $\zeta$ evolve identically when their initial values coincide. They are retained as separate symbols to distinguish their roles in isotropic hardening and fracture-toughness degradation.

\subsection{Fracture-toughness degradation} \label{sec:toughness_degradation}

Following Yin and Kaliske,\cite{Yin2020} ductile effects are introduced by allowing the accumulated equivalent plastic strain to reduce the local fracture toughness:
\begin{equation}
\label{eqn:fracture_toughness_degradation}
\mathcal{G}_{c}(\zeta)=f(\zeta)\mathcal{G}_{c}^{0},
\end{equation}
where $\mathcal{G}_{c}^{0}$ is the undegraded fracture toughness. It represents the critical energy release rate of the corresponding brittle response before the plasticity-induced reduction becomes active. The degradation function satisfies
\begin{equation}
\label{eqn:toughness_degradation_conditions}
0<f(\zeta)\leq1,
\qquad
f(\zeta)=1\quad\text{for }\zeta\leq\zeta^{\mathrm{cr}},
\qquad
\frac{\mathrm{d}f}{\mathrm{d}\zeta}\leq0
\quad\text{where defined for }\zeta>\zeta^{\mathrm{cr}}.
\end{equation}
The parameter $\zeta^{\mathrm{cr}}$ denotes the accumulated equivalent plastic strain threshold at which toughness degradation begins.

The function used in this work is
\begin{equation}
\label{eqn:toughness_degradation_function}
\begin{aligned}
f(\zeta) =
\begin{cases}
1, 
& \zeta < \zeta^{\mathrm{cr}}, \\[2mm]
\dfrac{1-b}{\xi^{2}}
\left(\zeta-\zeta^{\mathrm{cr}}-\xi
\right)^{2} + b,
& \zeta^{\mathrm{cr}} \leq \zeta < \xi+\zeta^{\mathrm{cr}}, \\[3mm]
b,
& \zeta \geq \xi+\zeta^{\mathrm{cr}},
\end{cases}
\end{aligned}
\end{equation}
where $\xi>0$ controls the strain interval over which the toughness decreases and $0<b\leq1$ is its residual fraction. A strictly positive value of $b$ prevents the local fracture toughness from vanishing.

Consistent with the assumption adopted by Yin and Kaliske,\cite{Yin2020} $f(\zeta)$ is treated as a known history-dependent coefficient during each variational subproblem. Its derivative with respect to the current plastic variables is therefore not included when differentiating the subproblem. This is an algorithmic assumption, rather than a general identity for \EQ{eqn:toughness_degradation_function}. Because $f(\zeta)$ remains inside the fracture-energy integral in \EQ{eqn:ductile_frac_eng}, its spatial variation is retained. In an equivalent strong form, this variation generates a contribution involving $\nabla f(\zeta)\cdot\nabla c$.

\subsection{Radial-return algorithm} \label{sec:radial_return}

The plastic internal variables are updated locally using the radial-return algorithm of associative $J_{2}$ plasticity.\cite{HanReddy2013,Yin2020} Consider the increment from $t_{n}$ to $t_{n+1}$. The converged history $(\boldsymbol{\varepsilon}^{\mathrm{p}}_{n},\alpha_{n},\zeta_{n})$ is known. For a candidate strain $\boldsymbol{\varepsilon}_{n+1}$ and a fixed phase-field value $c_{n+1}$, the trial elastic deviatoric strain and trial deviatoric stress are
\begin{equation}
\label{eqn:trial_state}
\boldsymbol{\varepsilon}^{\mathrm{e,tr}}_{\mathrm{dev},n+1}=\boldsymbol{\varepsilon}_{\mathrm{dev},n+1}-\boldsymbol{\varepsilon}^{\mathrm{p}}_{n},
\qquad
\boldsymbol{\sigma}^{\mathrm{tr}}_{\mathrm{dev},n+1}=2\mu g(c_{n+1})\boldsymbol{\varepsilon}^{\mathrm{e,tr}}_{\mathrm{dev},n+1}.
\end{equation}

The corresponding trial yield function is
\begin{equation}
\label{eqn:trial_yield_function}
\mathcal{F}^{\mathrm{tr}}_{n+1}=\left\lVert\boldsymbol{\sigma}^{\mathrm{tr}}_{\mathrm{dev},n+1}\right\rVert-\sqrt{\frac{2}{3}}\left(\sigma_{\mathrm{y}}+H\alpha_{n}\right).
\end{equation}

If $\mathcal{F}^{\mathrm{tr}}_{n+1}\leq0$, the step is elastic and
\begin{equation}
\label{eqn:elastic_trial_update}
\boldsymbol{\varepsilon}^{\mathrm{p}}_{n+1}=\boldsymbol{\varepsilon}^{\mathrm{p}}_{n},
\qquad
\alpha_{n+1}=\alpha_{n},
\qquad
\zeta_{n+1}=\zeta_{n},
\qquad
\boldsymbol{\sigma}_{\mathrm{dev},n+1}=\boldsymbol{\sigma}^{\mathrm{tr}}_{\mathrm{dev},n+1}.
\end{equation}

If $\mathcal{F}^{\mathrm{tr}}_{n+1}>0$, the trial state is returned to the yield surface along
\begin{equation}
\label{eqn:plastic_flow_direction}
\mathbf{N}_{n+1}
=\frac{\boldsymbol{\sigma}^{\mathrm{tr}}_{\mathrm{dev},n+1}} {\left\lVert\boldsymbol{\sigma}^{\mathrm{tr}}_{\mathrm{dev},n+1}\right\rVert}.
\end{equation}
For linear isotropic hardening, the plastic-multiplier increment has the closed form
\begin{equation}
\label{eqn:plastic_multiplier_increment}
\Delta\chi=\frac{\mathcal{F}^{\mathrm{tr}}_{n+1}}{2\mu g(c_{n+1})+\dfrac{2}{3}H}.
\end{equation}
The corrected variables are then
\begin{equation}
\label{eqn:plastic_corrector_update}
\begin{aligned}
\boldsymbol{\varepsilon}^{\mathrm{p}}_{n+1}
&=\boldsymbol{\varepsilon}^{\mathrm{p}}_{n}+\Delta\chi\mathbf{N}_{n+1},\\
\alpha_{n+1}
&=\alpha_{n}+\sqrt{\frac{2}{3}}\Delta\chi,\\
\zeta_{n+1}
&=\zeta_{n}+\sqrt{\frac{2}{3}}\Delta\chi,\\
\boldsymbol{\sigma}_{\mathrm{dev},n+1}
&=\boldsymbol{\sigma}^{\mathrm{tr}}_{\mathrm{dev},n+1}
-2\mu g(c_{n+1})\Delta\chi\mathbf{N}_{n+1}.
\end{aligned}
\end{equation}
The plastic correction enforces $\mathcal{F}_{n+1}=0$ up to numerical precision. The complete stress is recovered as
\begin{equation}
\label{eqn:updated_stress_response}
\boldsymbol{\sigma}_{n+1}
=\kappa\left[g(c_{n+1})\langle I_{1,n+1}\rangle_{+}+\langle I_{1,n+1}\rangle_{-}\right]\mathbf{I}+\boldsymbol{\sigma}_{\mathrm{dev},n+1}.
\end{equation}

\begin{figure}[htbp!]
\centering
\includegraphics[width=0.8\textwidth]{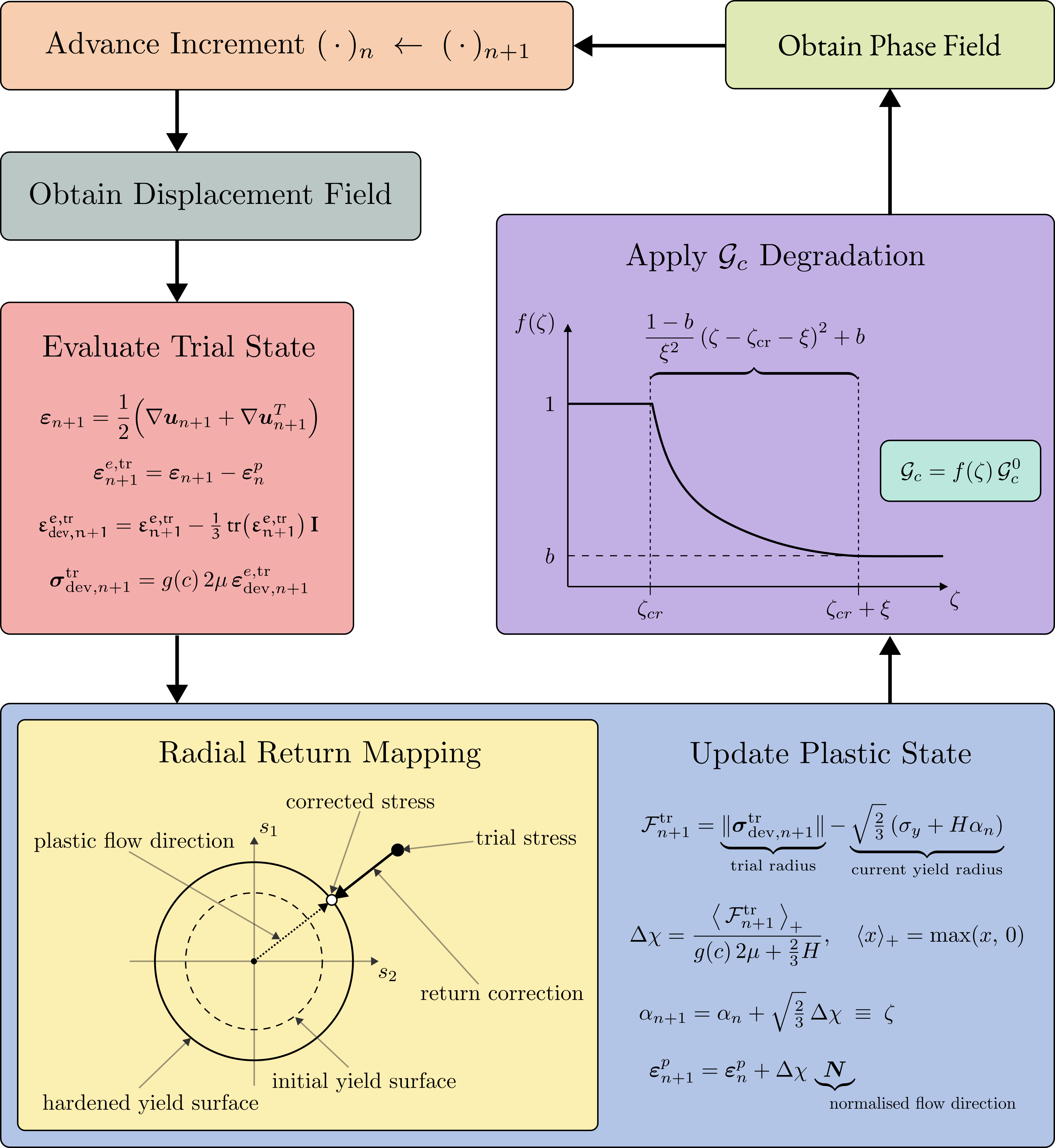}
\caption[]{Radial-return mapping for the degraded $J_{2}$ constitutive response. The trial state is accepted when $\mathcal{F}^{\mathrm{tr}}\leq0$ and otherwise projected onto the current yield surface by the plastic corrector.}
\label{fig:radial_return}
\end{figure}

After the return mapping, $f(\zeta_{n+1})$ determines the local degraded fracture toughness used in the phase-field subproblem. During the global iterations at a given loading increment, every constitutive trial update starts from the same converged history at $t_n$. The internal variables at $t_{n+1}$ are committed only after the displacement and phase fields have converged. This procedure preserves the constitutive history without introducing $\boldsymbol{\varepsilon}^{\mathrm{p}}$, $\alpha$ or $\zeta$ as trainable fields. A schematic representation of the return algorithm is presented in \FIG{fig:radial_return}.

\subsection{Irreversibility} \label{sec:irreversibility}

Crack healing is prevented between successive loading increments. Since $c=1$ denotes intact material and $c=0$ denotes fully fractured material, irreversibility requires
\begin{equation}
\label{eqn:irreversibility_condition}
c_{n}\leq c_{n-1}
\qquad\text{in }\Omega.
\end{equation}

Following the penalty treatment used in variational phase-field formulations,\cite{Gerasimov2019,Manav2024,Plunge2026} and accounting for the present convention for $c$, the constraint is enforced by adding
\begin{equation}
\label{eqn:irreversibility_penalty}
\mathcal{P}^{\mathrm{irr}}
=\int_{\Omega}\frac{1}{2}\kappa_{\mathrm{irr}}
\left\langle c_{n-1}-c_{n}\right\rangle_{-}^{2}
\,\mathrm{d}\Omega.
\end{equation}
The penalty is active only when $c_{n}>c_{n-1}$, which represents an inadmissible recovery of material integrity.

For the AT2 crack density in \EQ{eqn:ductile_crack_density}, the penalty parameter is chosen using the AT2 scaling of Gerasimov and De Lorenzis,\cite{Gerasimov2019} with the undegraded fracture toughness used as the reference scale:
\begin{equation}
\label{eqn:irreversibility_penalty_parameter}
\kappa_{\mathrm{irr}}=\frac{\mathcal{G}_{c}^{0}}{l_{0}}\left(\frac{1}{\mathrm{tol}_{\mathrm{irr}}^{2}}-1\right),
\end{equation}
where $0<\mathrm{tol}_{\mathrm{irr}}\leq1$ is the prescribed irreversibility tolerance, set to 0.02 in this work. Using $\mathcal{G}_{c}^{0}$ provides a fixed global penalty scale and prevents the irreversibility penalty from decreasing as the local fracture toughness degrades.

\subsection{Non-dimensionalisation scheme} \label{sec:non_dims}

A non-dimensionalisation scheme is introduced in this work for the governing quantities prior to neural optimisation to improve numerical conditioning, building on the scaling approaches of Manav et al.\cite{Manav2024} and Plungė et al.\cite{Plunge2026} Let $L$ be a characteristic dimension of the body and let $E_{\mathrm{ref}}=E$ be the Young's modulus of the isotropic solid, where
\begin{equation}
E=\frac{9\kappa\mu}{3\kappa+\mu}.
\end{equation}

The scaling is based on the undegraded fracture toughness $\mathcal{G}_{c}^{0}$ and the AT2 length scale $l_{0}$, providing fixed reference quantities throughout the loading history. The reference strain, stress, displacement and energy scales are
\begin{equation}
\label{eqn:reference_scales}
\varepsilon_{\mathrm{ref}}
=\sqrt{\frac{\mathcal{G}_{c}^{0}}
{E_{\mathrm{ref}}l_{0}}},
\qquad
\sigma_{\mathrm{ref}}
=E_{\mathrm{ref}}\varepsilon_{\mathrm{ref}}
=\sqrt{\frac{E_{\mathrm{ref}}\mathcal{G}_{c}^{0}}{l_{0}}},
\qquad
u_{\mathrm{ref}}=L\varepsilon_{\mathrm{ref}},
\qquad
\Pi_{\mathrm{ref}}=\frac{L^{d}\mathcal{G}_{c}^{0}}{l_{0}}.
\end{equation}

The dimensionless spatial, kinematic and history variables are defined by
\begin{equation}
\label{eqn:dimensionless_primary_variables}
\begin{gathered}
\widetilde{\mathbf{x}}=\frac{\mathbf{x}}{L},
\qquad
\widetilde{l}_{0}=\frac{l_{0}}{L},
\qquad
\widetilde{\mathbf{u}}=\frac{\mathbf{u}}{u_{\mathrm{ref}}},
\qquad
\widetilde{\boldsymbol{\varepsilon}}
=\frac{\boldsymbol{\varepsilon}}{\varepsilon_{\mathrm{ref}}},
\qquad
\widetilde{\boldsymbol{\varepsilon}}^{\mathrm{p}}
=\frac{\boldsymbol{\varepsilon}^{\mathrm{p}}}{\varepsilon_{\mathrm{ref}}},
\\[2mm]
\widetilde{\alpha}
=\frac{\alpha}{\varepsilon_{\mathrm{ref}}},
\qquad
\widetilde{\zeta}
=\frac{\zeta}{\varepsilon_{\mathrm{ref}}},
\qquad
\widetilde{\zeta}^{\mathrm{cr}}
=\frac{\zeta^{\mathrm{cr}}}{\varepsilon_{\mathrm{ref}}},
\qquad
\widetilde{\xi}
=\frac{\xi}{\varepsilon_{\mathrm{ref}}}.
\end{gathered}
\end{equation}

The differential operators and integration measures transform according to
\begin{equation}
\widetilde{\nabla}=L\nabla,
\qquad
\mathrm{d}\Omega=L^{d}\,\mathrm{d}\widetilde{\Omega},
\qquad
\mathrm{d}S=L^{d-1}\,\mathrm{d}\widetilde{S}.
\end{equation}

The phase field $c$ and the residual toughness fraction $b$ are already dimensionless. The dimensionless material parameters are
\begin{equation}
\label{eqn:dimensionless_material_parameters}
\begin{gathered}
\widetilde{\kappa}=\frac{\kappa}{E_{\mathrm{ref}}},
\qquad
\widetilde{\mu}=\frac{\mu}{E_{\mathrm{ref}}},
\qquad
\widetilde{H}=\frac{H}{E_{\mathrm{ref}}},
\qquad
\widetilde{\sigma}_{\mathrm{y}}
=\frac{\sigma_{\mathrm{y}}}{\sigma_{\mathrm{ref}}},
\qquad
\widetilde{\boldsymbol{\sigma}}
=\frac{\boldsymbol{\sigma}}{\sigma_{\mathrm{ref}}}.
\end{gathered}
\end{equation}
For isotropic elasticity,
\begin{equation}
\widetilde{\kappa}
=\frac{1}{3(1-2\nu)},
\qquad
\widetilde{\mu}
=\frac{1}{2(1+\nu)},
\end{equation}
where $\nu$ is Poisson's ratio.

The arguments of the toughness-degradation law are scaled consistently. Since $\zeta=\varepsilon_{\mathrm{ref}}\widetilde{\zeta}$, $\zeta^{\mathrm{cr}}=\varepsilon_{\mathrm{ref}}\widetilde{\zeta}^{\mathrm{cr}}$ and $\xi=\varepsilon_{\mathrm{ref}}\widetilde{\xi}$,
\begin{equation}
\label{eqn:dimensionless_toughness}
\widetilde{\mathcal{G}}_{c}
=\frac{\mathcal{G}_{c}}{\mathcal{G}_{c}^{0}}
=f\!\left(
\widetilde{\zeta};
\widetilde{\zeta}^{\mathrm{cr}},
\widetilde{\xi},b
\right).
\end{equation}
The dimensionless degradation function therefore retains the same piecewise form as \EQ{eqn:toughness_degradation_function}.

The penalty parameter becomes
\begin{equation}
\label{eqn:dimensionless_irreversibility_parameter}
\widetilde{\kappa}_{\mathrm{irr}}
=\frac{l_{0}\kappa_{\mathrm{irr}}}{\mathcal{G}_{c}^{0}}
=\frac{1}{\mathrm{tol}_{\mathrm{irr}}^{2}}-1.
\end{equation}

With $\widetilde{\Pi}=\Pi/\Pi_{\mathrm{ref}}$, the dimensionless internal and irreversibility contributions are
\begin{equation}
\label{eqn:dimensionless_energy}
\begin{aligned}
\widetilde{\mathcal{W}}^{\mathrm{int}}
&=
\int_{\widetilde{\Omega}}
\Bigg[
g(c)\widetilde{\psi}^{\mathrm{e}+}
+\widetilde{\psi}^{\mathrm{e}-}
+\widetilde{\sigma}_{\mathrm{y}}\widetilde{\alpha}
+\frac{1}{2}\widetilde{H}\widetilde{\alpha}^{2} +f(\widetilde{\zeta})
\left(
\frac{(1-c)^{2}}{2}
+\frac{\widetilde{l}_{0}^{2}}{2}
\left\lvert\widetilde{\nabla}c\right\rvert^{2}
\right)
\Bigg]
\,\mathrm{d}\widetilde{\Omega},
\\[2mm]
\widetilde{\mathcal{P}}^{\mathrm{irr}}
&=
\int_{\widetilde{\Omega}}
\frac{1}{2}\widetilde{\kappa}_{\mathrm{irr}}
\left\langle c_{n-1}-c_{n}\right\rangle_{-}^{2}
\,\mathrm{d}\widetilde{\Omega},
\end{aligned}
\end{equation}
where
\begin{equation}
\widetilde{\psi}^{\mathrm{e}\pm}
=\frac{l_{0}}{\mathcal{G}_{c}^{0}}\psi^{\mathrm{e}\pm}.
\end{equation}

The dimensionless traction and body force are
\begin{equation}
\widetilde{\mathbf{h}}
=\frac{\bar{\mathbf{h}}}{\sigma_{\mathrm{ref}}},
\qquad
\widetilde{\mathbf{f}}^{\mathrm{b}}
=\frac{L\mathbf{f}^{\mathrm{b}}}{\sigma_{\mathrm{ref}}}.
\end{equation}

Finally, defining
\begin{equation}
\widetilde{\mathcal{F}}
=\frac{\mathcal{F}}{\sigma_{\mathrm{ref}}},
\qquad
\Delta\widetilde{\chi}
=\frac{\Delta\chi}{\varepsilon_{\mathrm{ref}}},
\end{equation}
allows the yield function and radial-return equations to retain their original algebraic form. For notational convenience, the tilde notation is omitted in the subsequent governing and computational equations. Dimensional material parameters are nevertheless reported in physical units in the numerical examples and are non-dimensionalised before optimisation.

\section{DEM-based formulation for phase-field ductile fracture} \label{sec:ext_deep_ritz_method}

\subsection{Deep energy method} \label{sec:deep_ritz_method}

The DEM is a variational physics-informed approach in which trainable functions are used as trial approximations of the solution fields and their parameters are determined through minimisation of an energy-based objective.\cite{Samaniego2020,NguyenThanh2020} It follows the general structure of the Deep Ritz Method, combining a trainable approximation space, numerical integration of the variational functional and optimisation of the resulting parameterised problem.\cite{E2018} Unlike standard residual-based PINNs, the governing equations are not imposed through pointwise strong-form residuals. Instead, the solution fields are determined through minimisation of the variational functional.\cite{Manav2024,Plunge2026}

In the present problem, the global trainable fields are the {displacement and phase fields},
\begin{equation}
\label{eqn:dem_trial_fields}
\mathbf{u}(\mathbf{x}) \approx \mathbf{u}_{\boldsymbol{\theta}_{u}}(\mathbf{x}),
\qquad
c(\mathbf{x})
\approx
c_{\boldsymbol{\theta}_{c}}(\mathbf{x}),
\end{equation}
where $\boldsymbol{\theta}_{u}$ and $\boldsymbol{\theta}_{c}$ collect the trainable parameters of the corresponding representations. The two fields use separate trainable representations, as described in Section~\ref{sec:trainable_architecture}.

The elastoplastic internal variables are not introduced as additional trainable fields. Instead, for a given trial state they are evaluated locally at the quadrature points using the radial-return algorithm of Section~\ref{sec:radial_return}. Denoting the local constitutive state by
\begin{equation}
\label{eqn:constitutive_map}
\mathbf{q}_{n}^{\star} = \left( \boldsymbol{\varepsilon}_{n}^{\mathrm{p}\star}, \alpha_{n}^{\star}, \zeta_{n}^{\star} \right),
\end{equation}
the local constitutive update may be written symbolically as
\begin{equation}
\label{eqn:local_constitutive_map}
\mathbf{q}_{n}^{\star} = \mathcal{R} \left( \boldsymbol{\varepsilon} \left( \mathbf{u}_{\boldsymbol{\theta}_{u}} \right), c_{\boldsymbol{\theta}_{c}}; \mathbf{q}_{n-1} \right),
\end{equation}
where $\mathcal{R}$ denotes the radial-return operator and $\mathbf{q}_{n-1}$ is the constitutive history committed at the preceding loading increment. Every trial constitutive evaluation within increment $n$ starts from this same committed history. The updated variables are committed only after convergence of the global fields.

The trainable fields are evaluated at the nodes of the background mesh and interpolated to the quadrature points, where the strain, phase-field gradient and constitutive response are calculated. The resulting energy contributions define the training objective introduced in Section~\ref{sec:loss_function}. Numerical integration is discussed separately in Section~\ref{sec:numeric_quadrature_rule}.

\begin{figure}[htbp!]
\centering
\begin{tabular}{>{\centering\arraybackslash}m{0.55\textwidth}
>{\centering\arraybackslash}m{0.4\textwidth}}
\subfloat[\label{fig:training_block}]{
\includegraphics[width=0.54\textwidth]{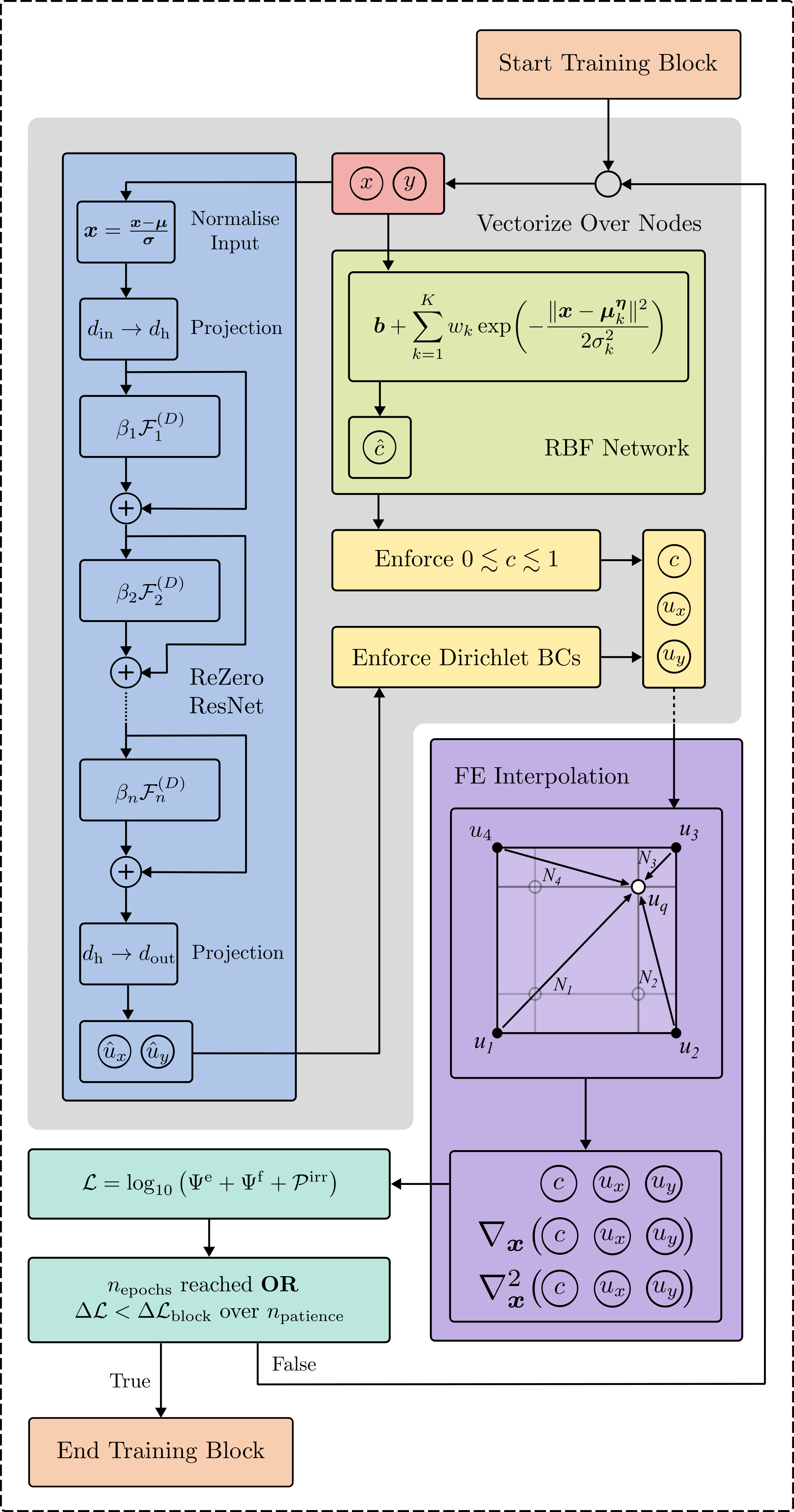}} &
\subfloat[\label{fig:training_scheme}]{
\includegraphics[width=0.39\textwidth]{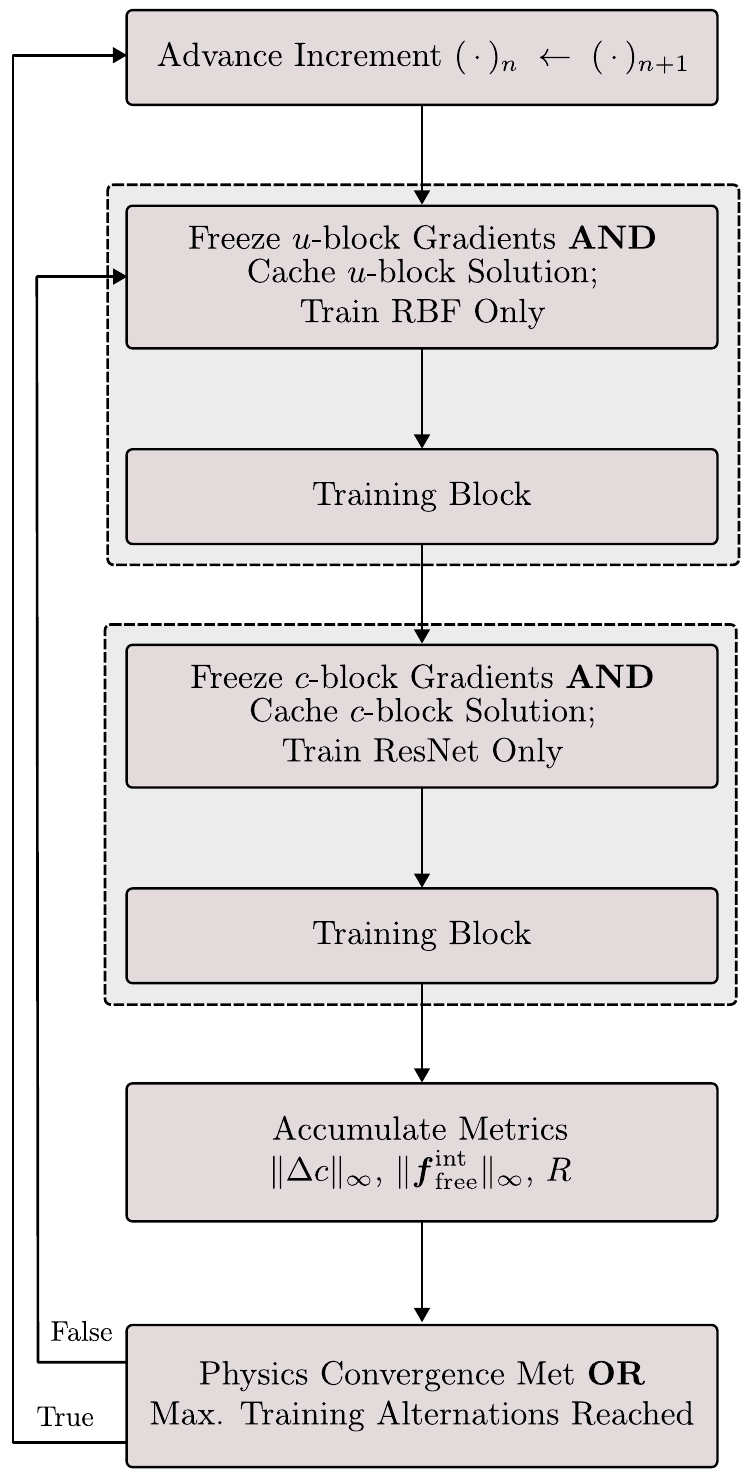}}
\end{tabular}
\caption[]{Overview of the proposed incremental DEM. (a) Evaluation of a training block using the trainable field representations, admissibility and Dirichlet mappings, finite-element interpolation, the energy-based loss and the block-level stopping criterion. (b) Alternating optimisation of the phase and displacement fields, followed by physics-based convergence checks and commitment of the converged constitutive history.}
\label{fig:method_diagram}
\end{figure}

Figure~\ref{fig:method_diagram} summarises the incremental procedure. The trainable fields are updated through alternating optimisation, while the constitutive variables remain local quantities governed by the material model.

\subsection{Training loss} \label{sec:loss_function}

The training objective is constructed from the elastic, fracture and irreversibility contributions introduced in Section~\ref{sec:phase_field_ductile}. For the displacement-controlled problems considered in this work, loading is imposed through prescribed displacements and no prescribed tractions or body forces are applied. Consequently, the explicit external-work term $\mathcal{W}^{\mathrm{ext}}$ defined in \EQ{eqn:ductile_work_ext} vanishes. 

The reduced training energy is therefore
\begin{equation}
\label{eqn:dem_energy_surrogate}
\mathcal{E}_{n} \left( \boldsymbol{\theta}_{u}, \boldsymbol{\theta}_{c} \right) = \Psi^{\mathrm{e}}_{n} + \Psi^{\mathrm{f}}_{n} + \mathcal{P}^{\mathrm{irr}}_{n},
\end{equation}
where the elastic contribution is evaluated using the local constitutive state returned by \EQ{eqn:local_constitutive_map}.

The plastic contribution $\Psi^{\mathrm{p}}$ is not included explicitly in the global training objective. This reflects the local treatment of the plastic internal variables rather than their omission from the constitutive model. During optimisation of the displacement field, the radial-return algorithm evaluates the plastic state and corresponding stress for each candidate displacement field. The resulting constitutive state is treated as fixed during differentiation, so gradients are not propagated through the return-mapping update. The resulting first variation is therefore obtained from the elastic contribution evaluated at the return-mapped constitutive state.

During optimisation of the phase field, the most recently available trial constitutive state is held fixed, according to the adopted staggered constitutive treatment. Consequently, $\Psi^{\mathrm{p}}$ is constant with respect to the active phase-field parameters. The toughness-degradation factor $f(\zeta_{n}^{\star})$ is treated in the same manner. Its value is retained locally within the fracture-energy integral, but gradients are not propagated through the constitutive history, consistently with Section~\ref{sec:toughness_degradation}. Thus, plastic deformation influences the phase-field problem through the spatially varying degraded toughness without introducing the plastic variables into the global trainable parameter space.

With the non-dimensionalisation introduced in Section~\ref{sec:non_dims}, $\mathcal{E}_{n}$ is dimensionless. For $\mathcal{E}_{n}>0$, the scalar loss used for optimisation is
\begin{equation}
\label{eq:loss_functional}
\mathcal{L}_{n} = \log_{10} \left[ \mathcal{E}_{n} \left( \boldsymbol{\theta}_{u}, \boldsymbol{\theta}_{c} \right) \right].
\end{equation}

The use of a logarithmic energy objective follows the scaling strategy used in previous energy-based neural fracture formulations.\cite{Manav2024,Plunge2026} It reduces the dynamic range of the objective and facilitates the use of consistent optimisation and stopping criteria across loading increments.

Differentiating the loss gives
\begin{equation}
\label{eqn:log_energy_gradient}
\nabla_{\boldsymbol{\theta}} \mathcal{L}_{n}
=
\frac{1}{\ln(10)\mathcal{E}_{n}}
\nabla_{\boldsymbol{\theta}} \mathcal{E}_{n},
\end{equation}
where $\boldsymbol{\theta}$ denotes the active trainable parameter block. The logarithmic transformation therefore multiplies the energy gradient by a positive scalar. It preserves the stationary points and the component-wise signs of the gradient while rescaling its magnitude.

\subsection{Numerical quadrature}
\label{sec:numeric_quadrature_rule}

The energy functional is evaluated over a background spatial discretisation comprising four-node quadrilateral elements.\cite{deBorst2012} The trainable displacement and phase fields are first evaluated at the element nodes, while their values and spatial gradients at the quadrature points are obtained using the corresponding bilinear shape functions. This follows the numerical-gradient strategy of Manav et al.,\cite{Manav2024} who reported spurious high-frequency oscillations in the learned fields when their gradients were computed using automatic differentiation. A related interpolation-based differentiation strategy using higher-order B-spline basis functions was adopted in Ref.~\citenum{Plunge2026}. The energy contributions are then integrated element-wise using Gauss quadrature. At each quadrature point, the interpolated strain and phase-field value are supplied to the local constitutive update, and the resulting constitutive state is used to evaluate the elastic response and the plasticity-dependent fracture toughness. The interpolated phase-field gradient enters directly into the fracture-energy contribution.

\subsection{Trainable field representations} \label{sec:trainable_architecture}

The displacement and phase fields exhibit markedly different spatial behaviour. The displacement field is relatively smooth away from the fracture zone, although steep gradients may develop near crack localisation. The phase field develops a localised transition whose characteristic width is controlled by the phase-field length scale $l_{0}$. Preliminary calculations in the present study using a single shared global neural-network representation for both fields did not reproduce this localisation reliably. In particular, the displacement response could converge while the phase field remained excessively diffuse. Separate trainable representations are therefore adopted for $\mathbf{u}$ and $c$. This choice also supports the alternating optimisation strategy introduced in Section~\ref{sec:optimisation_strategy}.

\subsubsection{Displacement} \label{sec:displacement_architecture}

The displacement field is represented by a ReZero-based residual NN adapted from the architecture previously developed for brittle anisotropic fracture.\cite{Plunge2026} Residual networks learn corrections to an identity mapping through shortcut connections.\cite{He2015} ReZero modifies this structure by introducing a trainable scalar coefficient on each residual branch and initialising it to zero.\cite{Bachlechner2020} Each residual block therefore acts initially as an identity mapping and progressively develops non-trivial transformations during optimisation.

Following the non-dimensionalisation introduced in Section~\ref{sec:non_dims}, the spatial coordinates supplied to the displacement network are additionally standardised component-wise. This removes differences in coordinate scale associated with the spatial extent of the domain in each direction. For coordinate component $i=1,\ldots,d$, the network input is defined as
\begin{equation}
\label{eqn:coordinate_standardisation}
x^{\mathrm{in}}_{i}
=
s_{\mathrm{x}}
\frac{x_i-\mu_i}{\sigma_i},
\qquad
i=1,\ldots,d,
\end{equation}
where $\mu_i$ and $\sigma_i$ are the mean and standard deviation, respectively, of the nodal coordinates in direction $i$. The scalar $s_{\mathrm{x}}$ is fixed at $s_{\mathrm{x}}=1$.

The standardised coordinate vector is first projected onto a hidden representation $\mathbf{S}_{0}\in\mathbb{R}^{v}$. For residual block $I\in\{1,\ldots,N_{\mathrm{b}}\}$,
\begin{equation}
\label{eq:rezero}
\begin{aligned}
\mathbf{S}_{I}
&=
\mathbf{S}_{I-1} + \beta_{I} \mathcal{F}_{I}^{(D)} \left( \mathbf{S}_{I-1} \right),
\\
\mathbf{z}_{I,0}
&=
\mathbf{S}_{I-1},
\\
\mathbf{z}_{I,j}
&=
\phi \left[ \mathbf{A}_{I,j} \left( \mathbf{z}_{I,j-1} \right) \right],
\qquad
j=1,\ldots,D-1,
\\
\mathcal{F}_{I}^{(D)} \left( \mathbf{S}_{I-1} \right)
&=
\mathbf{A}_{I,D} \left( \mathbf{z}_{I,D-1} \right),
\\
\mathbf{A}_{I,j}(\mathbf{y})
&=
\mathbf{W}_{I,j}\mathbf{y} + \mathbf{b}_{I,j},
\end{aligned}
\end{equation}
where $v$ is the hidden width, $D$ is the depth of the residual branch and $\beta_I$ is the ReZero scaling parameter. The coefficients $\beta_I$ are initialised to zero. An output projection maps the final hidden representation to the two displacement components.

For the numerical examples considered in this work, the architecture parameters are set to
\begin{equation}
D=3,
\qquad
v=256,
\qquad
N_{\mathrm{b}}=4.
\end{equation}
These values were selected empirically through preliminary numerical experiments. Smaller architectures exhibited greater variability across random initialisations, whereas larger architectures did not provide a clear improvement in the obtained solutions and incurred greater computational cost. The adopted configuration therefore provides a practical balance between robustness across initialisations and computational efficiency.

\paragraph{Activation function}

A scaled hyperbolic-tangent activation is employed,
\begin{equation}
\label{eqn:adaptive_tanh}
\phi(\mathbf{z})
=
\tanh(m\mathbf{z}),
\end{equation}
where the operation is applied component-wise and $m$ is a shared trainable parameter. The coefficient $m$ controls the slope of the activation and is initialised as $m=1$. The same form of adaptive activation was adopted previously for variational brittle-fracture modelling.\cite{Plunge2026}

\paragraph{Weight regularisation}

Weight regularisation was required in previous neural brittle-fracture formulations to control excessive displacement gradients and associated strain magnitudes near localised fracture zones.\cite{Manav2024,Plunge2026} No comparable instability was observed in the present numerical experiments when the displacement representation was coupled with the local elastoplastic return mapping. No additional parameter-norm regularisation is therefore applied to the displacement network. This choice is specific to the present formulation and numerical examples and is not assumed to hold generally for ductile-fracture problems.

\paragraph{Boundary-condition enforcement}

Following the strong-imposition strategy used in variational neural fracture formulations,\cite{Manav2024,Plunge2026} Dirichlet boundary conditions are imposed exactly through smooth boundary-window functions. For displacement component $i=1,\ldots,d$, the admissible field is constructed as
\begin{equation}
\label{eqn:displacement_constraint_map}
u_{\boldsymbol{\theta}_{u},i}(\mathbf{x})
=
\left[1-w^{\mathrm{F}}_{i}(\mathbf{x})\right]
\left[1-w^{\mathrm{L}}(\mathbf{x})\right]
\widehat{u}_{\boldsymbol{\theta}_{u},i}(\mathbf{x})
+
w^{\mathrm{L}}(\mathbf{x})\overline{u}_{i},
\end{equation}
where $\widehat{u}_{\boldsymbol{\theta}_{u},i}$ denotes the unconstrained network output. The functions $w^{\mathrm{F}}_{i}$ and $w^{\mathrm{L}}$ equal unity on the corresponding fixed and loaded boundaries and decay smoothly away from them. Their supports are chosen not to overlap. \EQ{eqn:displacement_constraint_map} therefore imposes $u_i=0$ on the fixed boundary and $u_i=\overline{u}_i$ on the loaded boundary without introducing additional penalty terms into the loss.

\subsubsection{Phase field}
\label{sec:phase_field_architecture} 
A different representation is adopted for the phase field. In the present ductile model, accumulated equivalent plastic strain can reduce the fracture toughness over a region wider than the final localised crack. Preliminary calculations using the global ReZero representation for $c$ tended to produce diffuse phase-field profiles within these regions of reduced toughness rather than the required localised fracture zone. Wang et al.\cite{Wang2026} similarly employ separate representations for the displacement and phase fields and recommend a radial basis function (RBF) network for the phase field in the continuous formulation. Motivated by these observations, the present work represents the phase field using spatially localised Gaussian RBFs.\cite{Broomhead1988}

The unconstrained phase-field representation is
\begin{equation}
\label{eq:rbf}
\widehat{c}_{\boldsymbol{\theta}_{c}}(\mathbf{x}) = b_{c} + \sum_{k=1}^{K} w^{c}_{k} \exp \left[ -\frac{ \left\lVert \mathbf{x}-\boldsymbol{\mu}^{\eta}_{k} \right\rVert^{2} }{ 2\sigma_{k}^{2} } \right],
\end{equation}
where $w^{c}_{k}$ are trainable coefficients, $b_c$ is a trainable bias (set to $-3.99$) and $\boldsymbol{\mu}^{\eta}_{k}$ denotes the centre of the $k$th basis function. One initial centre is assigned to each node of the background mesh, such that $K=N_{\mathrm{node}}$.

The width of each Gaussian is fixed according to the local nodal spacing,
\begin{equation}
\label{eqn:rbf_width}
\sigma_{k} = s_{\mathrm{RBF}}h_{k},
\end{equation}
where $h_k$ is the mean distance from the initial centre $\boldsymbol{\mu}^{0}_{k}$ to its $N_{\mathrm{nb}}=12$ nearest neighbouring centres. The dimensionless coefficient $s_{\mathrm{RBF}}$ controls the overlap between neighbouring basis functions. It is treated as a fixed hyperparameter, with $s_{\mathrm{RBF}}=1$ used throughout this work. This definition allows the Gaussian widths to reflect local variations in the resolution of the background mesh.

The centres are also allowed to move within a bounded neighbourhood of their initial positions. For spatial component
$i=1,\ldots,d$,
\begin{equation}
\label{eqn:bounded_rbf_centres}
\mu^{\eta}_{k,i} = \mu^{0}_{k,i} + \rho_{k} \tanh \left( \frac{\eta_{k,i}}{\rho_{k}} \right),
\qquad
\rho_{k} = \frac{h_k}{2},
\end{equation}
where $\eta_{k,i}$ is trainable and is initialised to zero. This parameterisation allows each centre to adjust to the evolving phase-field profile while limiting its movement in each coordinate direction to $\rho_k$ from its initial position. The Gaussian widths are kept fixed during optimisation to reduce additional parameter coupling and numerical sensitivity.

\paragraph{Phase-field constraint mapping and initialisation}

The unconstrained RBF field is mapped through the piecewise linear soft-bound function adopted previously in variational brittle-fracture modelling:\cite{Plunge2026}
\begin{equation}
\label{eqn:phase_field_constraint_map}
f_c(z) =
\begin{cases}
\beta_c(z+\ell_s),
& z<-\ell_s, \\[1mm]
\dfrac{z}{2\ell_s} +\dfrac{1}{2},
& \lvert z\rvert\leq\ell_s, \\[2mm]
1+\beta_c(z-\ell_s),
& z>\ell_s,
\end{cases}
\qquad
c_{\boldsymbol{\theta}_{c}}(\mathbf{x}) = f_c\!\left(
\widehat{c}_{\boldsymbol{\theta}_{c}}(\mathbf{x}) \right).
\end{equation}
The parameter $\ell_s$ defines the half-width of the central linear interval and is unrelated to the material length scale $l_0$. The parameter $\beta_c$ controls the slope of the two outer branches. The values
$\ell_s=4.0$ and
$\beta_c=10^{-3}$ are used throughout this work, based on the empirical observations made in previous works.\cite{Manav2024, Plunge2026} The mapping is therefore a soft constraint rather than an exact projection onto $[0,1]$. The small but non-zero slope outside the central interval avoids the zero gradients that would arise from hard clipping.

The RBF coefficients $w_k^c$ and centre parameters $\eta_{k,i}$ are initialised to zero. The bias $b_c$ is selected close to the upper end of, but within, the central interval so that the initial field represents intact material,
$c\approx1$, while retaining a non-zero optimisation gradient. No additional nonlinear activation function or parameter-norm regularisation is applied to the RBF representation.

\subsection{Optimisation strategy}\label{sec:optimisation_strategy}

The solution is advanced incrementally under prescribed displacement loading. 
Transfer learning is utilised by each block for computational efficiency. Each block has its own loss-based stopping criterion, while convergence of the coupled increment is assessed after each complete phase-field--displacement cycle using physics-based convergence criteria.

\subsubsection{Transfer learning}\label{sec:transfer_learning}

The converged trainable parameters from increment $n-1$ are used to initialise the displacement and phase-field representations at increment $n$. The optimiser state associated with each field, including the parameter-wise step sizes, is also transferred between increments. The committed plastic history is carried forward separately at the quadrature points. This warm start exploits the similarity between neighbouring loading states and reduces the optimisation required at successive increments.\cite{Manav2024,Plunge2026}

At the first loading increment, the ReZero coefficients of the displacement network are initialised to zero, while the RBF representation is initialised to the intact phase-field state described in Section~\ref{sec:phase_field_architecture}.

\subsubsection{Alternating training scheme}
\label{sec:training_training}

The displacement and phase-field representations are optimised alternately. Both blocks minimise the same loss defined in \EQ{eq:loss_functional}; no separate energy functional is introduced for either block. Instead, only the parameters of the active representation are varied:
\begin{equation}
\label{eqn:block_losses}
\mathcal{L}_{n}^{c} \left( \boldsymbol{\theta}_{c} \right) = \mathcal{L}_{n} \left( \boldsymbol{\theta}_{u}^{\mathrm{fixed}}, \boldsymbol{\theta}_{c} \right),
\qquad
\mathcal{L}_{n}^{u}
\left( \boldsymbol{\theta}_{u} \right) = \mathcal{L}_{n} \left( \boldsymbol{\theta}_{u}, \boldsymbol{\theta}_{c}^{\mathrm{fixed}} \right).
\end{equation}

Each coupled cycle begins with the phase-field block. The displacement representation and the current trial constitutive state are held fixed, while $\boldsymbol{\theta}_{c}$ is optimised. The fracture-toughness degradation factor $f(\zeta_{n}^{\star})$ is therefore also held fixed during this subproblem, consistent with Section~\ref{sec:toughness_degradation}.

The phase-field parameters are then frozen and the displacement block is optimised. For each candidate displacement state, the plastic variables are evaluated through the radial-return algorithm. Every return-mapping evaluation starts from the history committed at increment $n-1$. The resulting trial plastic state is not committed during the optimisation iterations.

After both blocks have been completed, the convergence criteria in Section~\ref{sec:physics_convergence} are evaluated. If they are not satisfied, another phase-field--displacement cycle is performed. Once convergence is reached, the final plastic state is committed as the constitutive history for increment $n+1$.

For computational efficiency, when one representation is frozen, its nodal values are cached and reused until its parameters are updated again. This avoids unnecessary repeated evaluations without modifying the optimisation problem.

\subsubsection{Optimiser}\label{sec:optimiser}

Both parameter blocks are optimised using resilient backpropagation (RPROP).\cite{riedmiller1993direct} RPROP assigns an individual step size to each trainable parameter and determines the update direction from the sign of the corresponding loss gradient. If the gradient retains the same sign between successive iterations, the step size is increased. A change in sign reduces the step size.

RPROP has shown favourable performance in previous variational neural fracture studies.\cite{Manav2024,Plunge2026} In particular, Manav et al.\cite{Manav2024} found RPROP more effective than the alternative optimisation algorithms considered in their study for resolving localised phase-field fracture.

The parameters used in the present work are
\begin{equation}
\label{eqn:rprop_parameters}
\Delta_{0}=10^{-3},
\qquad
\eta^{+}=1.2,
\qquad
\eta^{-}=0.5,
\qquad
\Delta_{\min}=10^{-9},
\qquad
\Delta_{\max}=50,
\end{equation}
where $\Delta_{0}$ is the initial parameter-wise step size, $\eta^{+}$ and $\eta^{-}$ are the increase and decrease factors, and $\Delta_{\min}$ and $\Delta_{\max}$ define the allowable step-size range. Automatic differentiation is applied to the discretely integrated loss to obtain its gradient with respect to the active trainable parameters. This is distinct from the spatial differentiation of the displacement and phase fields, which is performed using the interpolation procedure described in Section~\ref{sec:numeric_quadrature_rule}.

\subsubsection{Block-level stopping}
\label{sec:block_stopping}

Optimisation of an individual displacement or phase-field block is terminated when the relative change in the loss remains below $\mathrm{tol}_{\mathcal{L}} = 10^{-6}$ over a patience window of $N_{\mathrm{pat}}=50$ consecutive epochs. A prescribed maximum number of epochs, $N_{\mathrm{epochs}}$, provides an additional termination condition.


\subsubsection{Physics-based convergence} \label{sec:physics_convergence}

Loss stagnation within the individual optimisation blocks does not by itself establish convergence of the coupled problem. Convergence of load increment $n$ is therefore assessed after each complete phase-field--displacement cycle using the phase-field evolution and two displacement-based indicators.

\paragraph{Phase field}

Convergence monitoring between successive staggered iterations is commonly employed in variational phase-field fracture formulations. In the present scheme, stabilisation of the phase field is assessed from its change between successive coupled cycles using the maximum norm:
\begin{equation}
\label{eqn:phase_convergence_indicator}
\left\lVert c^{(j)} - c^{(j-1)} \right\rVert_{\infty} < \mathrm{tol}_{c},
\qquad
\mathrm{tol}_{c} = 1.5\times10^{-3},
\end{equation}
where $c^{(j)}$ denotes the vector of nodal phase-field values at the end of coupled cycle $j$. To avoid terminating on a temporary plateau, this condition must be satisfied for $N_{\mathrm{plateau},c}=3$ consecutive cycles.

\paragraph{Displacement field}

The displacement convergence indicators are constructed from the internal forces associated with the stresses returned by the radial-return algorithm. The assembled internal-force vector is
\begin{equation}
\label{eqn:assembled_internal_force}
\mathbf{f}^{\mathrm{int}} = \sum_{e=1}^{N_{\mathrm{el}}}\sum_{g=1}^{N_{\mathrm{gp}}} w_{g} \left| \det\mathbf{J}_{eg} \right| \mathbf{B}_{eg}^{\mathsf{T}} \boldsymbol{\sigma}_{eg},
\end{equation}
where $\mathbf{B}_{eg}$ is the strain--displacement matrix at quadrature point $g$ of element $e$.\cite{deBorst2012}

Since the numerical examples are quasi-static and displacement controlled, no external nodal forces act on the unconstrained degrees of freedom. From the admissible displacement mapping in \EQ{eqn:displacement_constraint_map}, variations of the unconstrained network output are weighted by the factor
\begin{equation}
\label{eqn:boundary_window_weight}
\omega_{ai} = \left(1-w_{a}^{\mathrm{L}}\right) \left(1-w_{ai}^{\mathrm{F}}\right),
\end{equation}
where $a$ denotes the node index and $i$ the displacement component. The corresponding weighted equilibrium residual is therefore
\begin{equation}
\label{eqn:free_internal_force}
\mathbf{f}^{\mathrm{int}}_{\mathrm{free}} = \mathbf{f}^{\mathrm{int}} \odot \boldsymbol{\omega},
\end{equation}
where $\odot$ denotes the Hadamard product.

The first displacement criterion is
\begin{equation}
\label{eqn:force_residual_criterion}
\left\lVert \mathbf{f}^{\mathrm{int}}_{\mathrm{free}} \right\rVert_{\infty} < f^{\mathrm{int}}_{\mathrm{tol}},
\qquad
f^{\mathrm{int}}_{\mathrm{tol}} = 10^{-5},
\end{equation}
where the force is non-dimensional according to Section~\ref{sec:non_dims}.

A second indicator monitors stabilisation of the generalised reaction force associated with the prescribed displacement. For a prescribed displacement magnitude $U_p$ acting in the unit direction $\widehat{\mathbf{e}}$, \EQ{eqn:displacement_constraint_map} gives $\partial u_{ai}/\partial U_p=w_a^{\mathrm{L}}\widehat{e}_i$. The reaction work-conjugate to $U_p$ is therefore
\begin{equation}
\label{eqn:reaction_force}
R=\sum_{i} \widehat{e}_{i} \sum_{a=1}^{N_{\mathrm{node}}} w_{a}^{\mathrm{L}} f_{ai}^{\mathrm{int}}.
\end{equation}
This reaction measure is also used to construct the force-displacement response reported in Section~\ref{sec:num_results}.

Over the most recent $N_{\mathrm{plateau},u}$ coupled cycles, the reaction-force spread is defined as
\begin{equation}
\label{eqn:reaction_force_spread}
R_{\mathrm{spread}}^{(j)} = \max_{k\in\mathcal{W}_{j}}R^{(k)} - \min_{k\in\mathcal{W}_{j}}R^{(k)},
\qquad
\mathcal{W}_{j} = \left\{ j-N_{\mathrm{plateau},u}+1,\ldots,j \right\}.
\end{equation}
The displacement response is considered stationary when
\begin{equation}
\label{eqn:reaction_force_criterion}
R_{\mathrm{spread}}^{(j)} < R_{\mathrm{tol}}R_{\mathrm{max}},
\qquad
R_{\mathrm{tol}} = 5\times10^{-4},
\qquad
N_{\mathrm{plateau},u} = 3,
\end{equation}
where $R_{\mathrm{max}}$ is the maximum absolute reaction force reached up to the current stage of the simulation. Using the spread over several cycles reduces sensitivity to small oscillations in the reaction force.

The displacement requirement is satisfied when either
\EQ{eqn:force_residual_criterion} or
\EQ{eqn:reaction_force_criterion} is met. The complete increment is accepted as converged only when the phase-field criterion in
\EQ{eqn:phase_convergence_indicator} and one of the displacement criteria are satisfied.

During the coupled optimisation, the toughness-degradation factor $f(\zeta_{n}^{\star})$ is evaluated from the current trial constitutive state rather than being lagged to the value from increment $n-1$. Its current value therefore enters the forward evaluation of the fracture-energy contribution. However, the dependence of $f$ on $\zeta_{n}^{\star}$ is excluded from automatic differentiation, so gradients are not propagated through the degradation law or the constitutive history. During the phase-field block, the most recently evaluated $f(\zeta_{n}^{\star})$ is held fixed. This treatment is consistent with the assumption $\partial f/\partial\zeta=0$ adopted by Yin and Kaliske,\cite{Yin2020} while retaining the current spatially varying degraded fracture toughness.

A maximum of $N_{\mathrm{max}}=500$ coupled optimisation cycles is imposed to prevent indefinite iteration. If this limit is reached before the physics-based criteria are satisfied, the current state is retained and the simulation proceeds to the next loading increment. Reaching $N_{\mathrm{max}}$ is therefore an algorithmic termination condition and is not interpreted as convergence of the coupled increment.

\section{Numerical results} \label{sec:num_results}

Five ductile fracture specimens are considered to assess the proposed DEM formulation. The DEM predictions are compared with reference finite element method (FEM) solutions in terms of the force-displacement response, energy evolution, accumulated equivalent plastic strain and phase-field distributions. Parameters common to all DEM simulations are summarised in Table~\ref{tab:common_dem_settings}. Specimen-specific geometries, material parameters and loading schedules are reported in the corresponding subsections.

The FEM reference solutions are obtained using a staggered iterative scheme. Convergence of each staggered cycle is assessed using the normalised energy-iteration measure $\gamma$, defined in Eq.~(43) of Ambati et al.\cite{Ambati2015a} and subsequently used in their ductile-fracture formulation.\cite{Ambati2015b} A convergence tolerance of $\gamma \leq \qty{5}{\degree}$ is adopted in the present calculations.

The FEM computations were performed on an Intel Core i9-14900K CPU using the MUMPS sparse direct solver with $20$ threads, whereas the DEM simulations were carried out on NVIDIA A100 $40$ GB PCIe GPUs.

All specimens are analysed in two dimensions under plane-strain conditions. For computational efficiency, the prescribed displacement history is divided into coarse and fine loading stages. Coarser increments are used during the predominantly elastic regime, followed by finer increments during material softening and rupture. The displacement ranges and corresponding increment counts are specified separately for each specimen.

In DEM, each specimen is simulated using eight random seeds to assess sensitivity to random initialisation. The displacement-step-size study in Section~\ref{sec:num_parametric} uses $6$ seeds for each increment count. Where ensemble curves are shown, they represent the mean response, with shaded bands spanning $\pm 1$ standard deviation.

The code and data used to generate the FEM and DEM results are available through the associated Zenodo repository.\cite{plunge2026zenodo}

\begin{table}[htbp!]
    \centering
    \caption{Numerical parameters adopted for the DEM field representations, optimisation and convergence criteria. All numbers are dimensionless.}
    \label{tab:common_dem_settings}
    \begin{tabular}{llll}
        \toprule
        \textbf{Category} & \textbf{Parameter} & \textbf{Notation} & \textbf{Value} \\
        \midrule

        \multirow{6}{*}{Displacement representation}
        & Residual-branch depth       & $D$               & $3$   \\
        & Hidden width                & $v$               & $256$ \\
        & Number of ReZero blocks     & $N_{\mathrm{b}}$  & $4$   \\
        & Initial ReZero scaling      & $\beta_I$         & $0$   \\
        & Coordinate scaling  & $s_x$          & $1$   \\
        & Initial activation scaling  & $m$               & $1$   \\
        \midrule

        \multirow{5}{*}{Phase-field representation}
        & Number of nearest neighbours & $N_{\mathrm{nb}}$  & $12$      \\
        & RBF overlap coefficient      & $s_{\mathrm{RBF}}$ & $1$       \\
        & Soft-bound half-width        & $\ell_s$           & $4.0$     \\
        & Outer-branch slope        & $\beta_c$           & $10^{-3}$     \\
        & Irreversibility tolerance & $\mathrm{tol}_\mathrm{irr}$          & $0.02$ \\
        \midrule

        \multirow{5}{*}{RPROP optimiser}
        & Initial step size        & $\Delta_0$      & $10^{-3}$ \\
        & Step-increase factor     & $\eta^{+}$      & $1.2$     \\
        & Step-decrease factor     & $\eta^{-}$      & $0.5$     \\
        & Minimum step size        & $\Delta_{\min}$ & $10^{-9}$ \\
        & Maximum step size        & $\Delta_{\max}$ & $50$      \\
        \midrule

        \multirow{3}{*}{Block stopping}
        & Relative loss-improvement tolerance & $\mathrm{tol}_{\mathcal{L}}$ & $10^{-6}$ \\
        & Patience window                    & $N_{\mathrm{pat}}$            & $50$      \\
        & Maximum number of epochs        & $N_{\mathrm{epochs}}$ & $250$      \\

        \midrule

        \multirow{6}{*}{Coupled convergence}
        & Phase-field tolerance       & $\mathrm{tol}_{c}$                   & $1.5\times10^{-3}$ \\
        & Phase-field plateau length  & $N_{\mathrm{plateau},c}$             & $3$                \\
        & Internal-force tolerance    & $f_{\mathrm{tol}}^{\mathrm{int}}$    & $10^{-5}$          \\
        & Reaction-force tolerance    & $R_{\mathrm{tol}}$                   & $5\times10^{-4}$          \\
        & Reaction plateau length     & $N_{\mathrm{plateau},u}$             & $3$                \\
        & Maximum coupled cycles      & $N_{\max}$                           & $500$               \\
        \bottomrule
    \end{tabular}
\end{table}

\subsection{Ductile fracture of asymmetrically notched and symmetrically notched specimens} \label{sec:num_exp_1}

\begin{figure}[htbp!]
    \centering
    \setlength{\tabcolsep}{3pt}
    \begin{tabular}{>{\centering\arraybackslash}m{0.33\textwidth}
                    >{\centering\arraybackslash}m{0.33\textwidth}
                    >{\centering\arraybackslash}m{0.33\textwidth}}
        \subfloat[\label{fig:hole_notched:a}]{
            \includegraphics[width=0.324\textwidth]{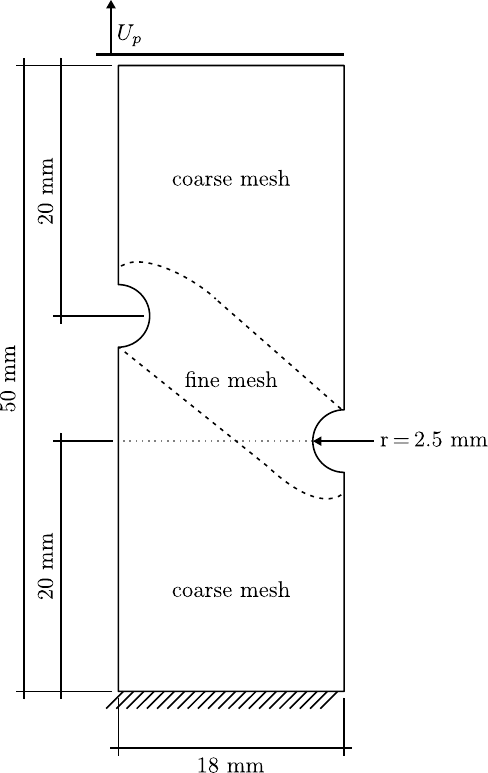}} &
        \subfloat[\label{fig:hole_notched:b}]{
            \includegraphics[width=0.325\textwidth]{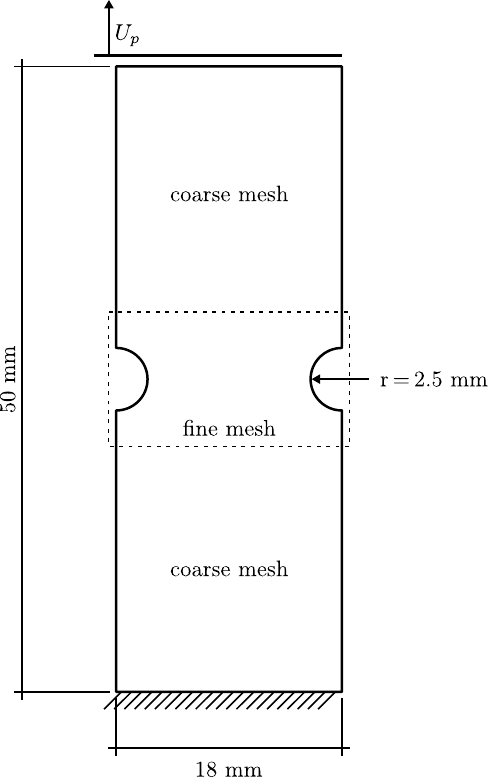}} &
        \subfloat[\label{fig:hole_notched:c}]{
            \includegraphics[width=0.31\textwidth]{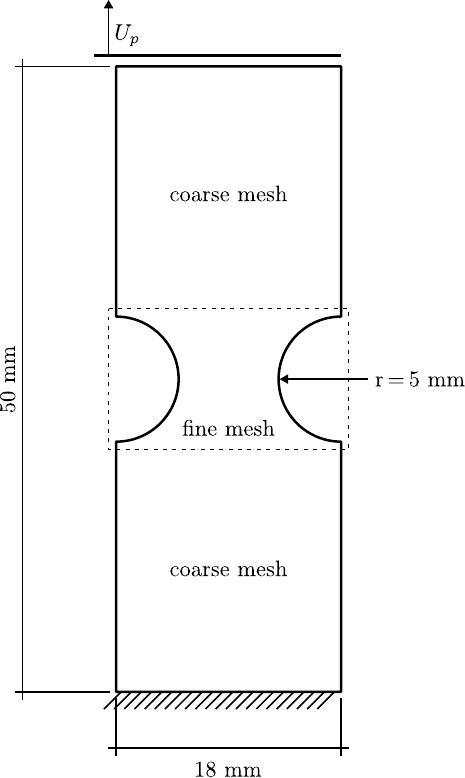}}
    \end{tabular}
    \caption[]{Geometry and boundary conditions for the notched specimens:
    \subref{fig:hole_notched:a} asymmetrically notched specimen,
    \subref{fig:hole_notched:b} symmetrically notched specimen with
    $r=\qty{2.5}{\milli\metre}$, and
    \subref{fig:hole_notched:c} symmetrically notched specimen with
    $r=\qty{5}{\milli\metre}$.}
    \label{fig:asym_sym_notched_geometries}
\end{figure}

Three notched specimens with the same material parameters are considered to investigate the influence of geometry on plastic localisation and fracture evolution. The geometries and boundary conditions are shown in \FIG{fig:asym_sym_notched_geometries}, and the corresponding material parameters are given in \TAB{tab:holenotch_params}. The asymmetrically notched specimen is used to assess crack propagation and linking between offset notches. The two symmetric specimens, with notch radii $r=\qty{2.5}{\milli\metre}$ and $r=\qty{5}{\milli\metre}$, are used to examine the effect of notch radius on the localisation pattern and subsequent rupture response.

\begin{table}[htbp!]
    \centering
    \caption{Material parameters adopted for the asymmetrically and symmetrically notched specimens.}
    \label{tab:holenotch_params}
    \begin{tabular}{llll}
        \toprule
        \textbf{Parameter} & \textbf{Notation} & \textbf{Value} & \textbf{Unit} \\
        \midrule
        Young's modulus                    & $E$                         & \num{72624.3}        & MPa  \\
        Poisson's ratio                    & $\nu$                       & $0.3311$           & --   \\
        Yield stress                       & $\sigma_{\mathrm{y}}$       & $345$              & MPa  \\
        Hardening modulus                  & $H$                         & $250$              & MPa  \\
        Undegraded fracture toughness      & $\mathcal{G}_{c}^{0}$       & $95$               & N/mm \\
        Phase-field length scale           & $l_0$                       & $0.3$              & mm   \\
        Toughness-degradation interval     & $\xi$                       & $0.4$              & --   \\
        Residual toughness fraction        & $b$                         & $1\times10^{-6}$   & --   \\
        Degradation threshold              & $\zeta^{\mathrm{cr}}$       & $0.02$             & --   \\
        \bottomrule
    \end{tabular}
\end{table}

\paragraph{Asymmetrically notched specimen}

The FEM reference used \num{10000} load increments. The first $100$ increments covered loading up to $U_p=\qty{0.9}{\milli\metre}$, while the remaining \num{9900} reached $U_p=\qty{1.65}{\milli\metre}$. The computation required \qty{6.23}{\hour} on the hardware described above. Each DEM run used \num{230} load increments: \num{30} increments up to $U_p=\qty{0.9}{\milli\metre}$ and \num{200} up to $U_p=\qty{1.65}{\milli\metre}$. Across the eight seeds, the mean DEM wall-clock time was \qty{16.5\pm1.1}{\hour}, with a range of \qtyrange{15.5}{18.6}{\hour}. Because the FEM and DEM computations were performed on different hardware, these wall-clock times are reported as computational costs on the stated hardware rather than as a hardware-normalised efficiency comparison. No DEM run reached the \num{500}-cycle limit before rupture. Seed \num{8} reached the limit once, during the rupture increment. The limit therefore did not affect the pre-rupture response.

\begin{figure}[htbp!]
	\centering
    \captionsetup[subfloat]{farskip=0pt,captionskip=1pt}
    \setlength{\tabcolsep}{4pt}
    \begin{tabular}{@{}c@{\hspace{4pt}}c@{}}
        \subfloat[\label{fig:asymmnotch_plots:a}]{
            \includegraphics[width=0.5\textwidth]{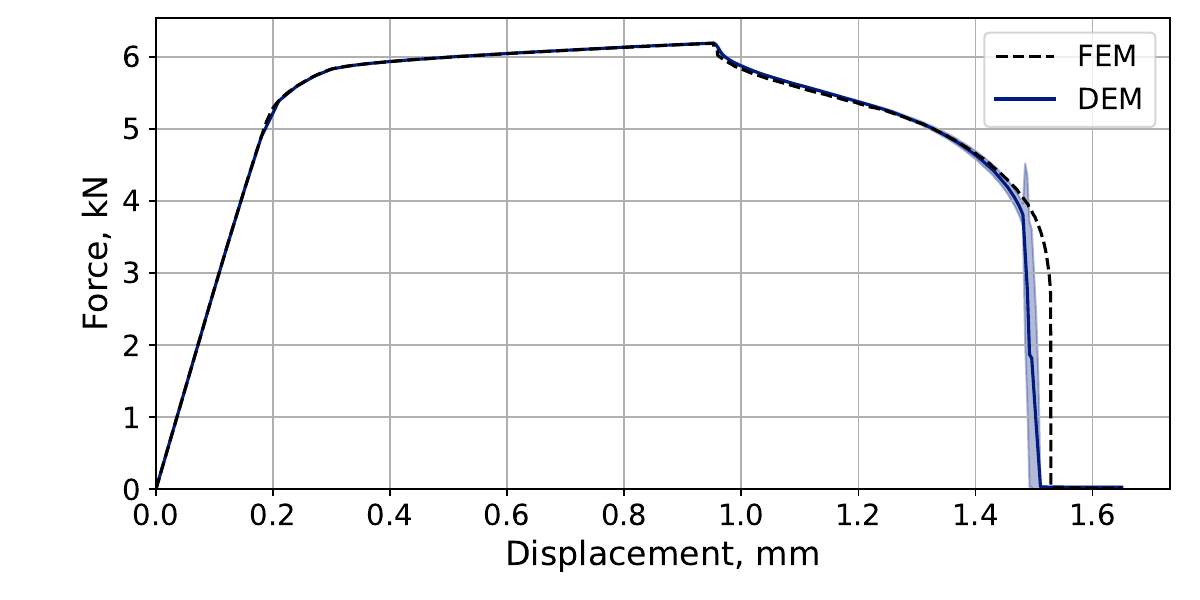}} &
        \subfloat[\label{fig:asymmnotch_plots:b}]{
            \includegraphics[width=0.5\textwidth]{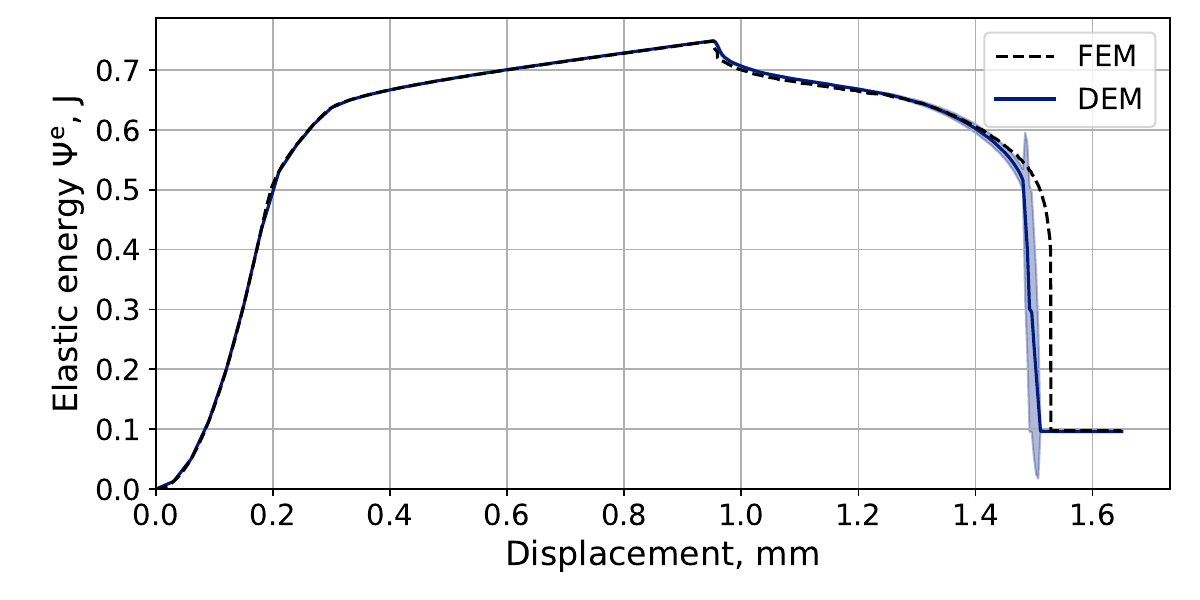}} \\[-2pt]
        \subfloat[\label{fig:asymmnotch_plots:c}]{
            \includegraphics[width=0.5\textwidth]{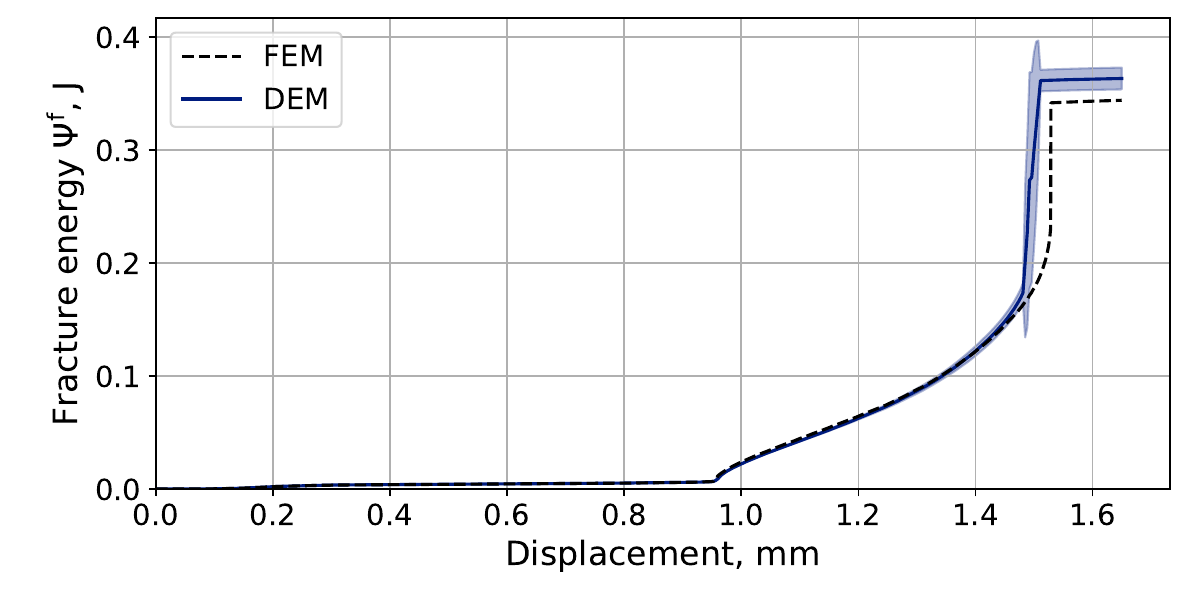}} &
        \subfloat[\label{fig:asymmnotch_plots:d}]{
            \includegraphics[width=0.5\textwidth]{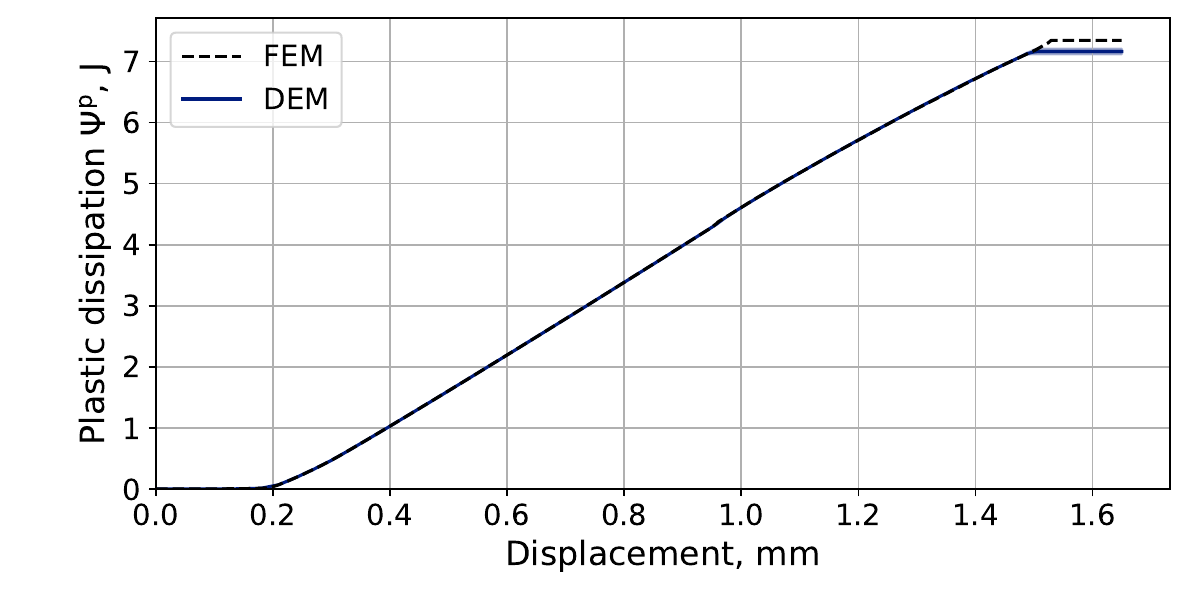}} \\[-2pt]
        \multicolumn{2}{c}{%
        \subfloat[\label{fig:asymmnotch_plots:e}]{
            \includegraphics[width=\textwidth]{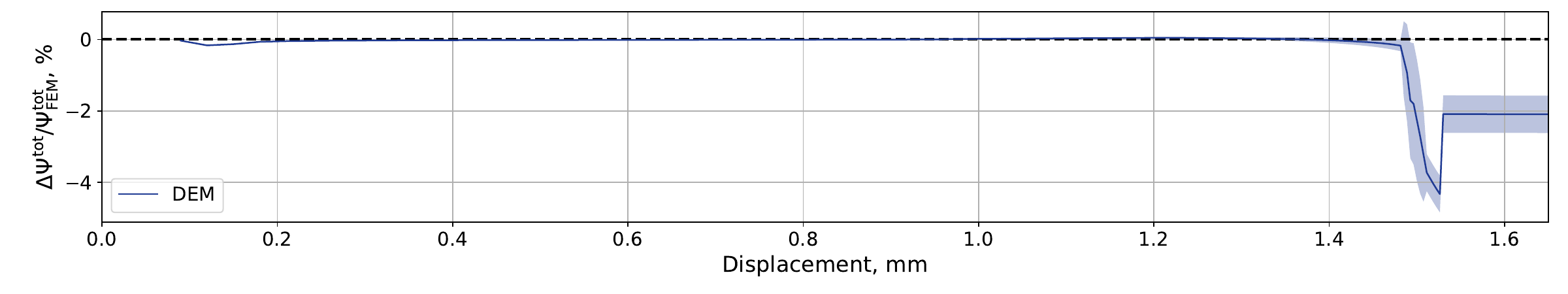}}} \\
	\end{tabular}
    \caption[]{Asymmetrically notched specimen: global response obtained with the proposed DEM and the FEM reference: \subref{fig:asymmnotch_plots:a} force-displacement response; \subref{fig:asymmnotch_plots:b} elastic energy $\Psi^{\mathrm{e}}$; \subref{fig:asymmnotch_plots:c} fracture energy $\Psi^{\mathrm{f}}$; \subref{fig:asymmnotch_plots:d} plastic energy $\Psi^{\mathrm{p}}$; and \subref{fig:asymmnotch_plots:e} relative difference in total energy, $(\Psi^{\mathrm{tot}}_{\mathrm{DEM}}-\Psi^{\mathrm{tot}}_{\mathrm{FEM}}) /\Psi^{\mathrm{tot}}_{\mathrm{FEM}}$, where $\Psi^{\mathrm{tot}}=\Psi^{\mathrm{e}}+\Psi^{\mathrm{f}}+\Psi^{\mathrm{p}}$. All quantities are plotted against the prescribed displacement $U_p$. The DEM curves represent the mean over eight seeds, and the shaded bands span $\pm1$ standard deviation.}
    \label{fig:asymmnotch_plots}
\end{figure}

The global responses are compared in \FIG{fig:asymmnotch_plots}. The DEM ensemble closely reproduces the FEM force-displacement curve throughout the stable loading regime. It captures the peak load and the gradual post-peak reduction with little variation among the seeds. The elastic, fracture, and plastic energy contributions also remain close to the FEM reference over most of the loading history.

Rupture occurs at a slightly smaller prescribed displacement in the DEM simulations. This difference appears only during the final unstable stage and may partly reflect the coarser DEM load discretisation. The plastic history and the associated toughness degradation are updated incrementally. Consequently, the predicted onset of an abrupt terminal event can be sensitive to the number and sequence of load increments. This interpretation is consistent with the load-step sensitivity examined in Section~\ref{sec:num_parametric}. The earlier rupture is accompanied by a slightly lower final plastic contribution and an earlier increase in fracture energy. The magnitude of the total-energy difference remains small throughout stable loading, increases briefly during rupture, and then decreases after complete separation. The discrepancy is therefore concentrated around the terminal instability and does not affect the close pre-rupture agreement.

\begin{figure}[htbp!]
	\centering
    \setlength{\tabcolsep}{3pt}
    \begin{tabular}{>{\centering\arraybackslash}m{0.23\textwidth}
                    >{\centering\arraybackslash}m{0.23\textwidth}
                    >{\centering\arraybackslash}m{0.23\textwidth}
                    >{\centering\arraybackslash}m{0.23\textwidth}}
        \subfloat[\label{fig:asymmnotch_fields:a}]{
            \includegraphics[width=0.23\textwidth]{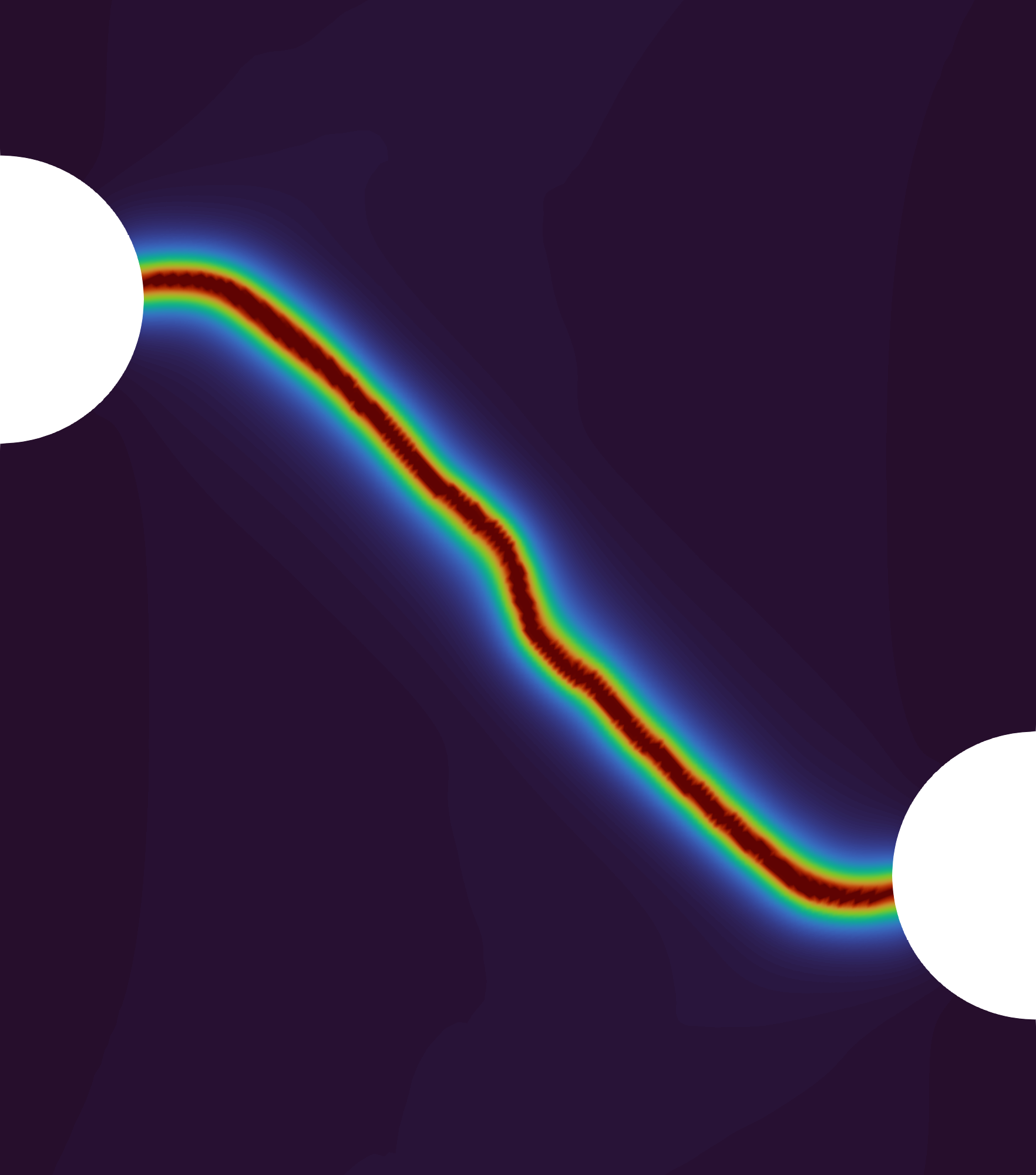}} &
        \subfloat[\label{fig:asymmnotch_fields:b}]{
            \includegraphics[width=0.23\textwidth]{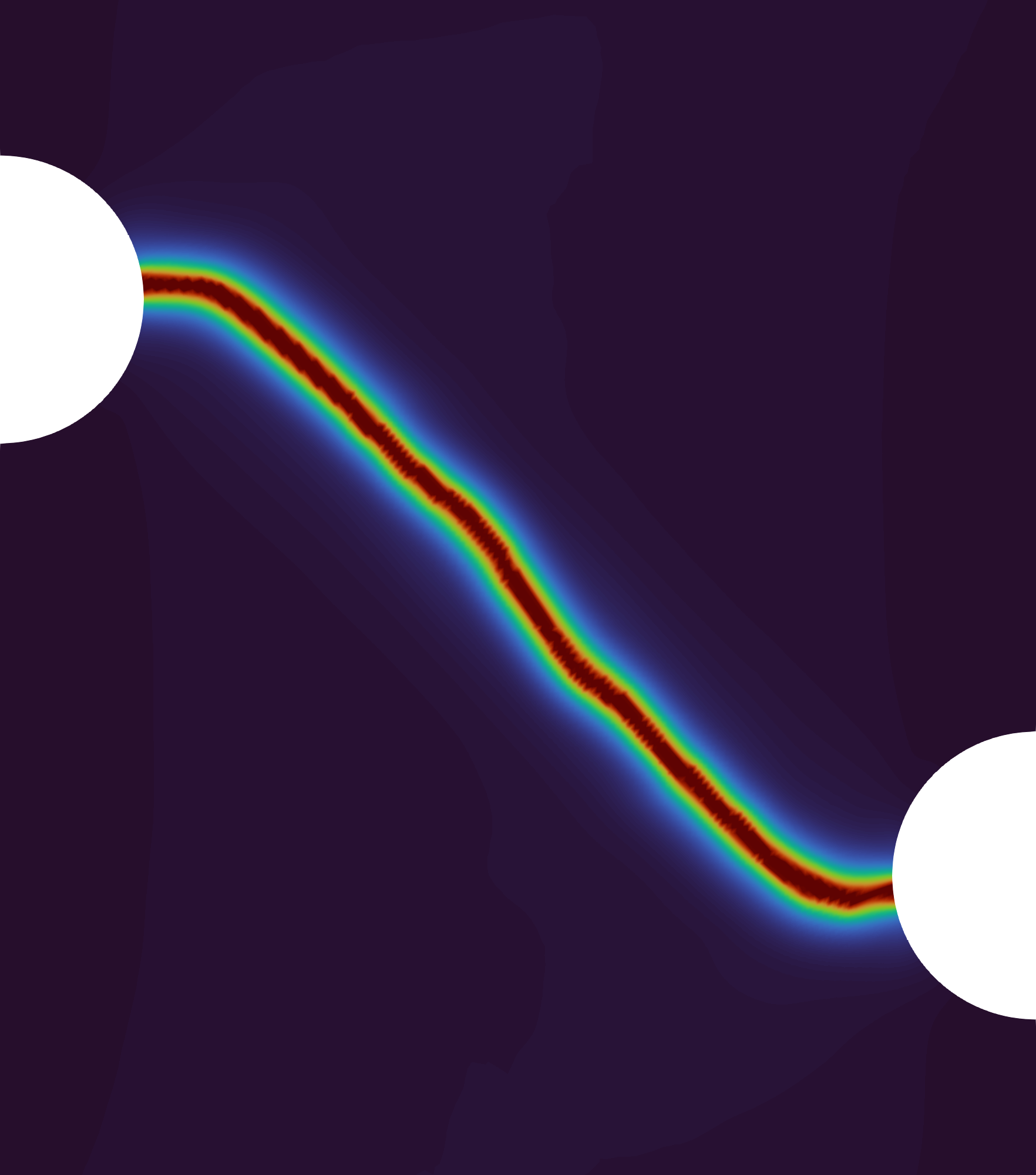}} &
        \subfloat[\label{fig:asymmnotch_fields:c}]{
            \includegraphics[width=0.23\textwidth]{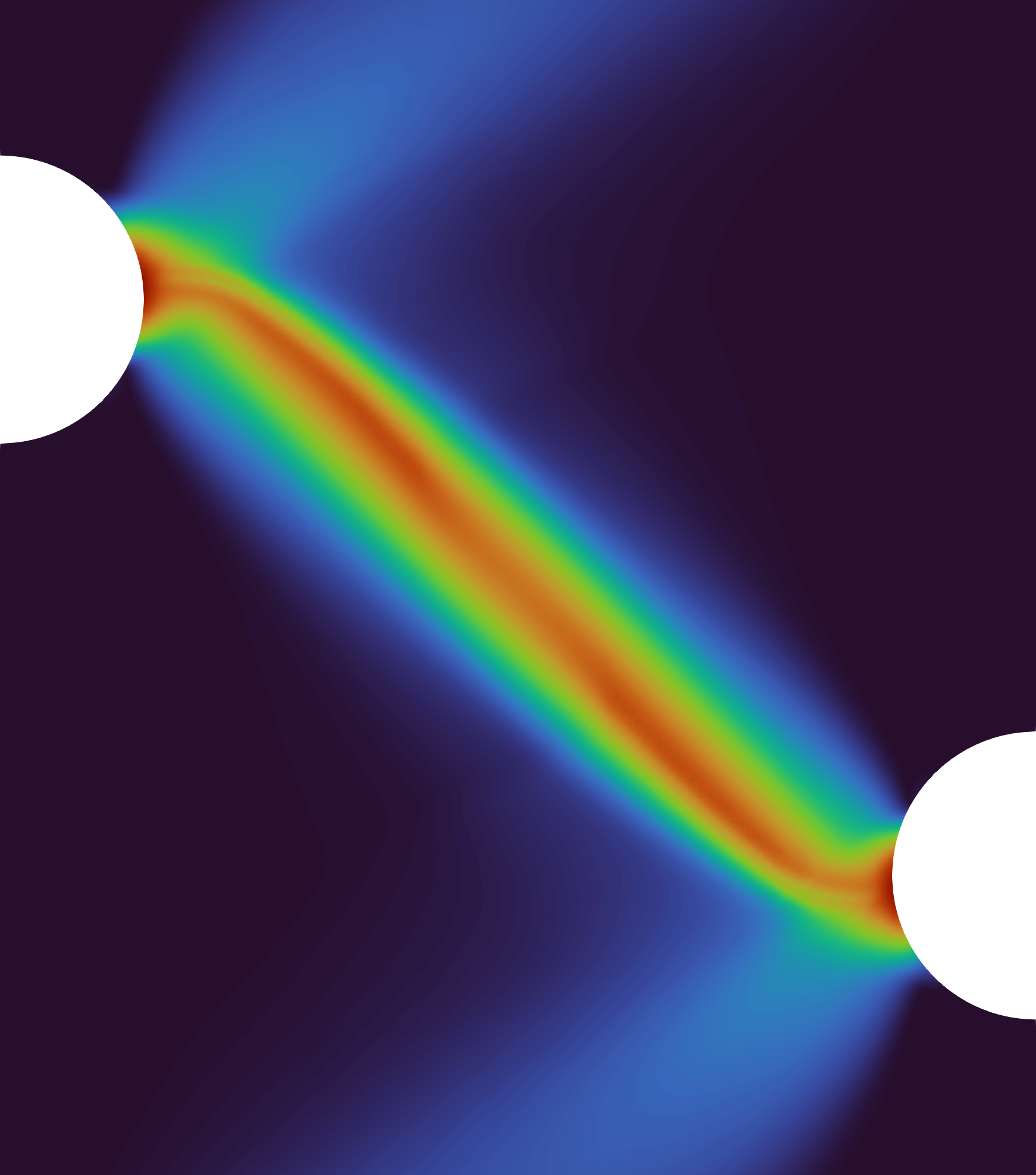}} &
        \subfloat[\label{fig:asymmnotch_fields:d}]{
            \includegraphics[width=0.23\textwidth]{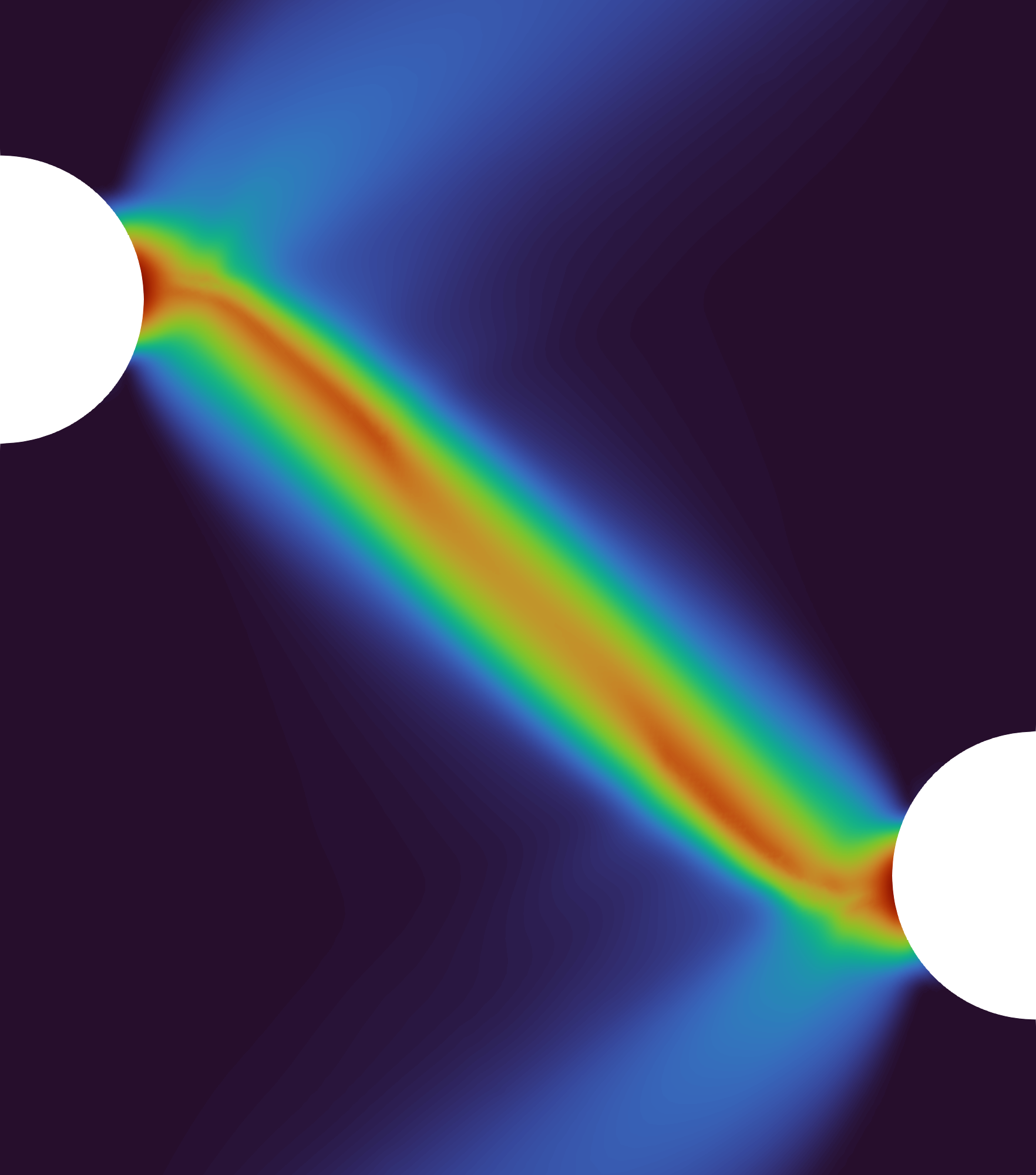}} \\
        \multicolumn{2}{c}{\includegraphics[width=0.25\textwidth]{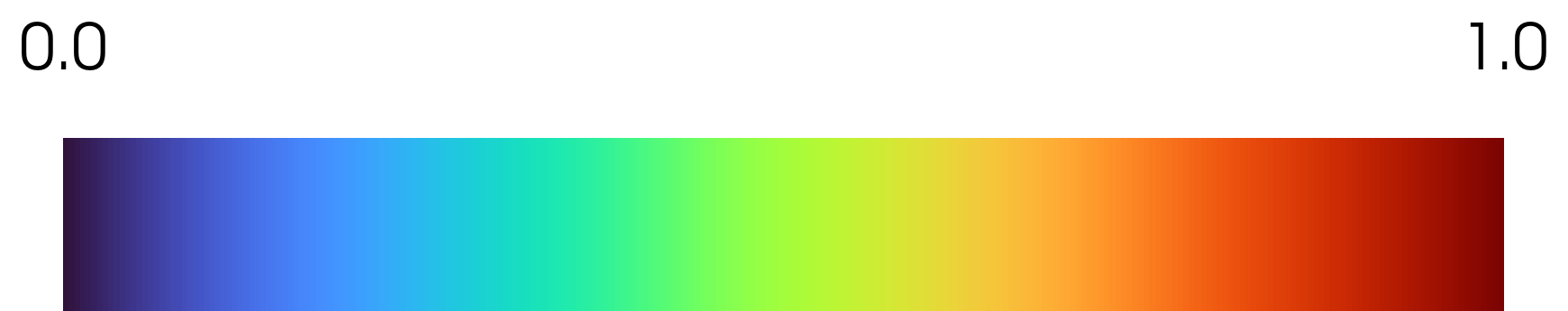}} &
        \multicolumn{2}{c}{\includegraphics[width=0.25\textwidth]{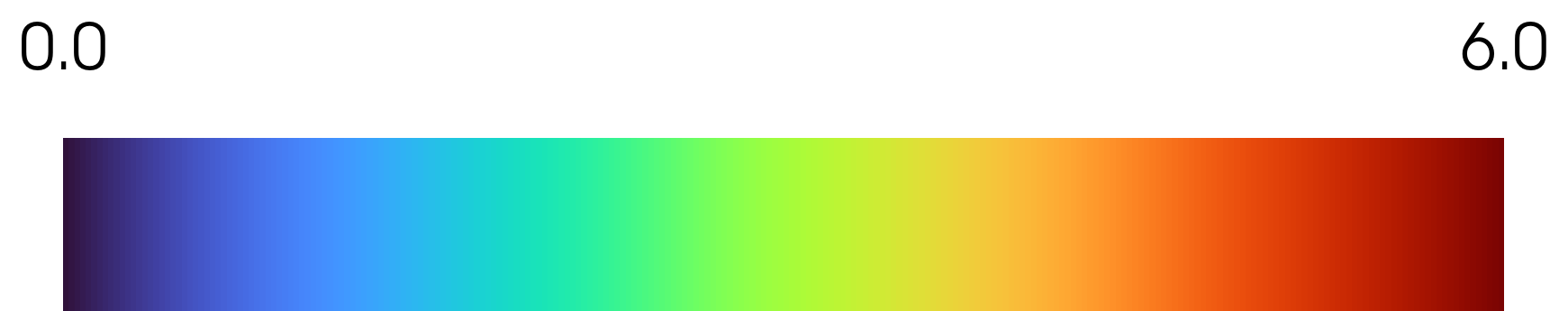}} \\
	\end{tabular}
    \caption[]{Asymmetrically notched specimen. Phase-field and accumulated equivalent plastic strain distributions at the respective rupture states. The phase field $c$ is shown for \subref{fig:asymmnotch_fields:a} FEM and \subref{fig:asymmnotch_fields:b} DEM, while the accumulated equivalent plastic strain $\zeta$ is shown for \subref{fig:asymmnotch_fields:c} FEM and \subref{fig:asymmnotch_fields:d} DEM. The DEM fields correspond to seed \num{8}.}
    \label{fig:asymmnotch_fields}
\end{figure}

The fields at the respective rupture states are shown in \FIG{fig:asymmnotch_fields}. Both methods predict a curved crack joining the two notch roots, with similar overall trajectories. This crack-linking mechanism is consistent with that reported by Ambati et al.\cite{Ambati2015b} The FEM crack has somewhat greater local variation in the central linking region, whereas the DEM path is smoother. The distributions of accumulated equivalent plastic strain $\zeta$ are also similar. The slightly lower DEM accumulation in the central region may be associated with its earlier rupture and coarser load-increment history. These local differences are concentrated in the central linking region and do not change the overall fracture mechanism. The small spread across the \num{8} seeds further indicates that the predicted global response is not strongly affected by network initialisation.

\paragraph{Symmetrically-notched specimen $r=\qty{2.5}{\milli\metre}$}

The FEM reference used \num{10000} load increments. The first \num{100} increments covered the loading range up to $U_p=\qty{0.4}{\milli\metre}$, followed by \num{9900} increments up to $U_p=\qty{3.6}{\milli\metre}$. The computation required \qty{6.08}{\hour}. Each DEM simulation used \num{230} load increments, with \num{30} increments up to $U_p=\qty{0.4}{\milli\metre}$ and \num{200} increments up to $U_p=\qty{3.6}{\milli\metre}$. Across the eight seeds, the mean DEM wall-clock time was \qty{28.0\pm3.7}{\hour}, with a range of \qtyrange{23}{34.3}{\hour}. This was the most computationally demanding of the \num{5} benchmark cases because of the extended softening regime, during which approximately \numrange{15}{30} coupled cycles were required per increment. Seeds \num{2} and \num{7} each reached the \num{500}-cycle limit once, during their respective rupture increments. No seed reached this limit before rupture.

\begin{figure}[htbp!]
	\centering
    \captionsetup[subfloat]{farskip=0pt,captionskip=1pt}
    \setlength{\tabcolsep}{4pt}
    \begin{tabular}{@{}c@{\hspace{4pt}}c@{}}
        \subfloat[\label{fig:holenotch_2_5_plots:a}]{
            \includegraphics[width=0.48\textwidth]{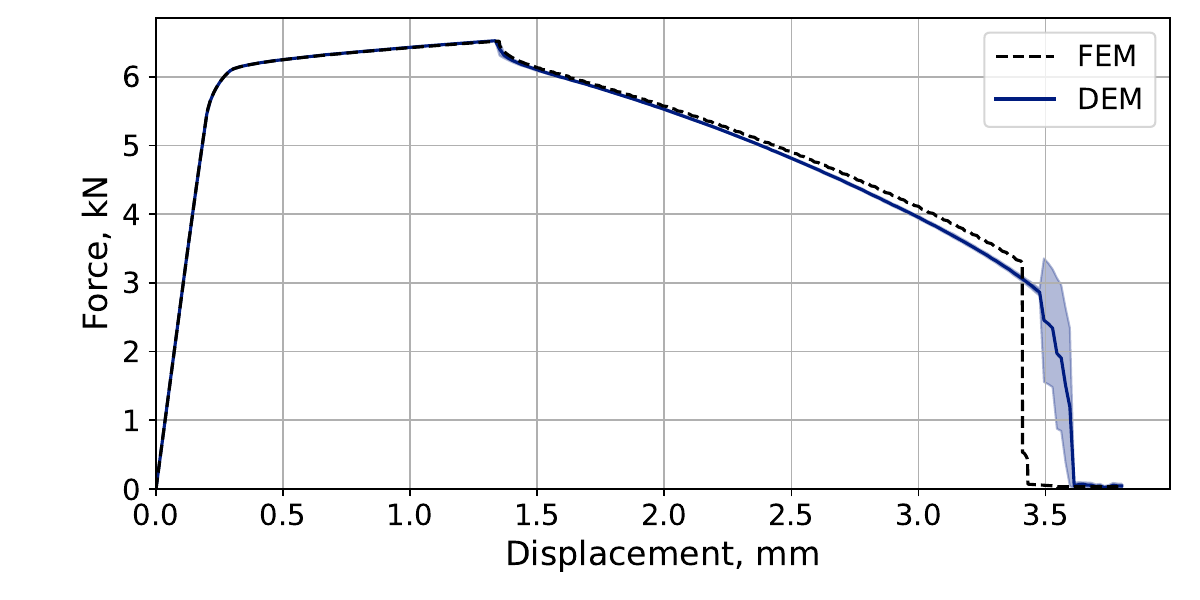}} &
        \subfloat[\label{fig:holenotch_2_5_plots:b}]{
            \includegraphics[width=0.48\textwidth]{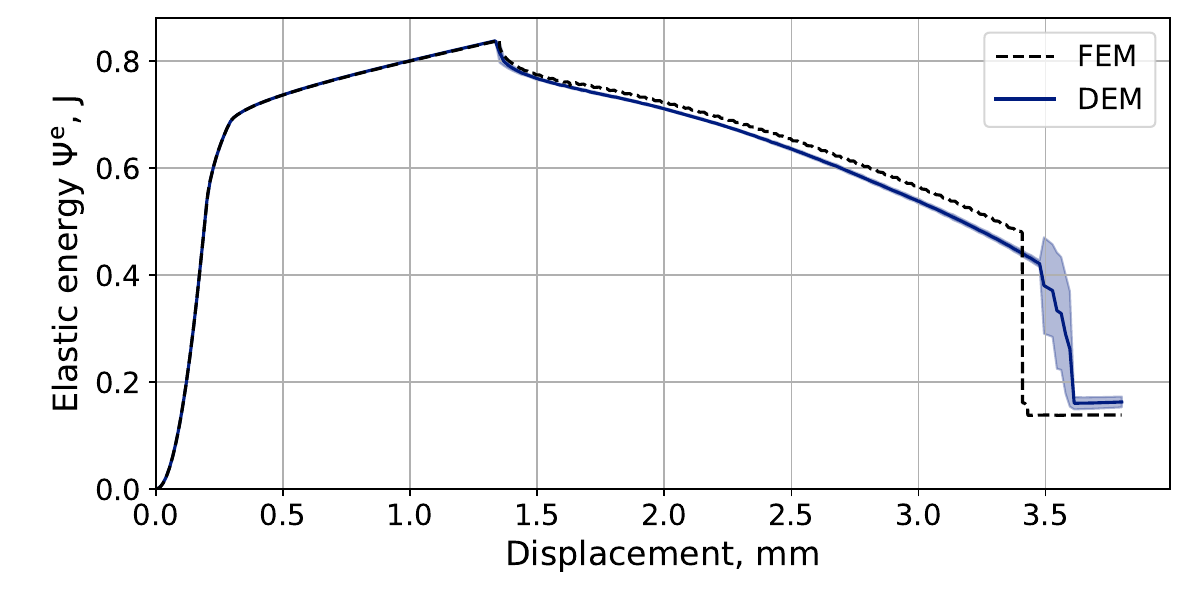}} \\[-2pt]
        \subfloat[\label{fig:holenotch_2_5_plots:c}]{
            \includegraphics[width=0.48\textwidth]{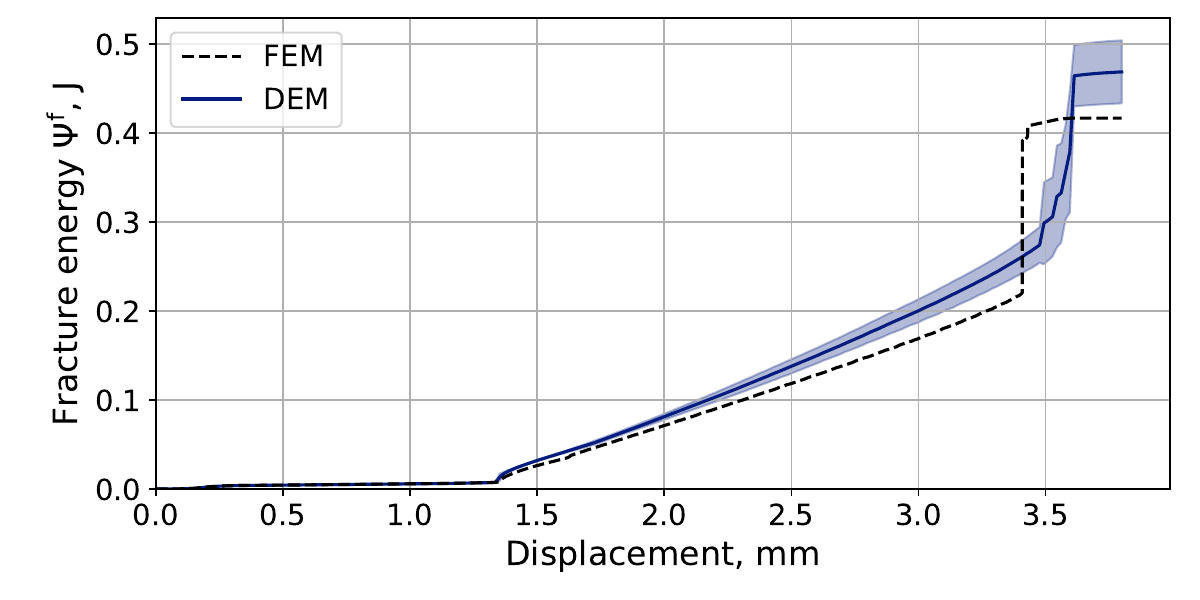}} &
        \subfloat[\label{fig:holenotch_2_5_plots:d}]{
            \includegraphics[width=0.48\textwidth]{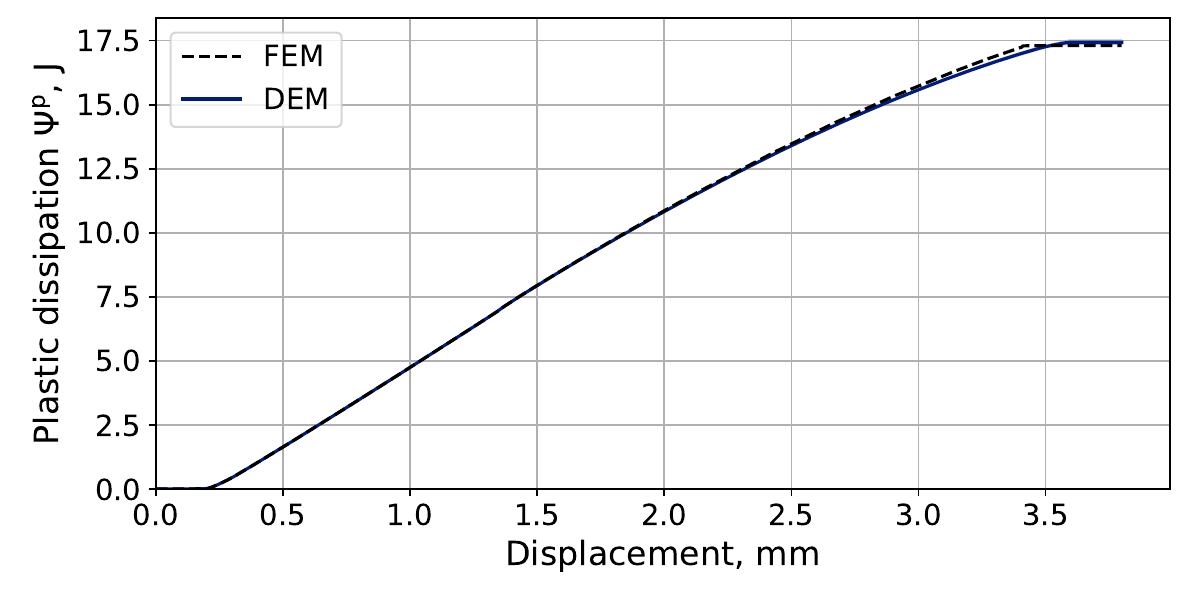}} \\[-2pt]
        \multicolumn{2}{c}{%
        \subfloat[\label{fig:holenotch_2_5_plots:e}]{
            \includegraphics[width=0.96\textwidth]{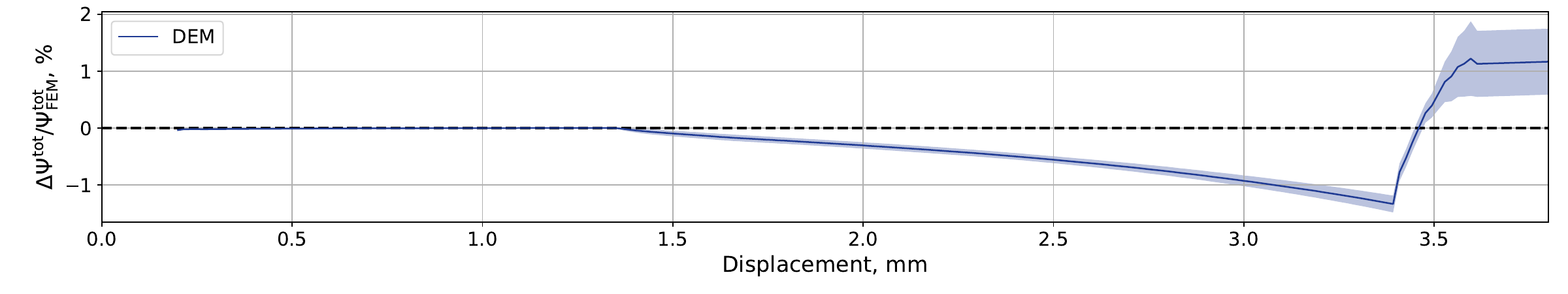}}} \\
	\end{tabular}
    \caption[]{Symmetrically notched $r=\qty{2.5}{\milli\metre}$ specimen. Global response obtained with the proposed DEM and the FEM reference: \subref{fig:holenotch_2_5_plots:a} force-displacement response; \subref{fig:holenotch_2_5_plots:b} elastic energy $\Psi^{\mathrm{e}}$; \subref{fig:holenotch_2_5_plots:c} fracture energy $\Psi^{\mathrm{f}}$; and \subref{fig:holenotch_2_5_plots:d} plastic energy $\Psi^{\mathrm{p}}$; \subref{fig:holenotch_2_5_plots:e} relative difference in total energy, $100(\Psi^{\mathrm{tot}}_{\mathrm{DEM}}-\Psi^{\mathrm{tot}}_{\mathrm{FEM}})/\Psi^{\mathrm{tot}}_{\mathrm{FEM}}$, where $\Psi^{\mathrm{tot}}=\Psi^{\mathrm{e}}+\Psi^{\mathrm{f}}+\Psi^{\mathrm{p}}$. All quantities are plotted against the prescribed displacement $U_p$. Shaded bands span $\pm 1$ standard deviation over the eight DEM seeds.}
    \label{fig:holenotch_2_5_plots}
\end{figure}

The global responses are compared in \FIG{fig:holenotch_2_5_plots}. The DEM reproduces the loading branch and the extended softening response closely, but predicts final rupture at a somewhat larger prescribed displacement than the FEM reference. The plastic-energy evolution remains in close agreement, while the elastic and fracture-energy responses exhibit a corresponding shift as rupture is approached. The relative difference in total energy becomes increasingly negative beyond approximately $U_p=\qty{1.4}{\milli\metre}$ and changes sign after the FEM rupture event, reaching a small positive difference close to the final DEM rupture. The spread across the eight DEM seeds remains narrow over most of the loading history. The delayed rupture is therefore reproducible across the random initialisations considered here, rather than being associated with an isolated optimisation trajectory. Given the incremental and path-dependent constitutive evolution, the substantially coarser DEM loading discretisation may also contribute to the small shift in the onset of the final instability.

\begin{figure}[htbp!]
	\centering
    \setlength{\tabcolsep}{3pt}
    \begin{tabular}{>{\centering\arraybackslash}m{0.23\textwidth}
                    >{\centering\arraybackslash}m{0.23\textwidth}
                    >{\centering\arraybackslash}m{0.23\textwidth}
                    >{\centering\arraybackslash}m{0.23\textwidth}}
        \subfloat[\label{fig:holenotch_2_5_fields:a}]{
            \includegraphics[width=0.23\textwidth]{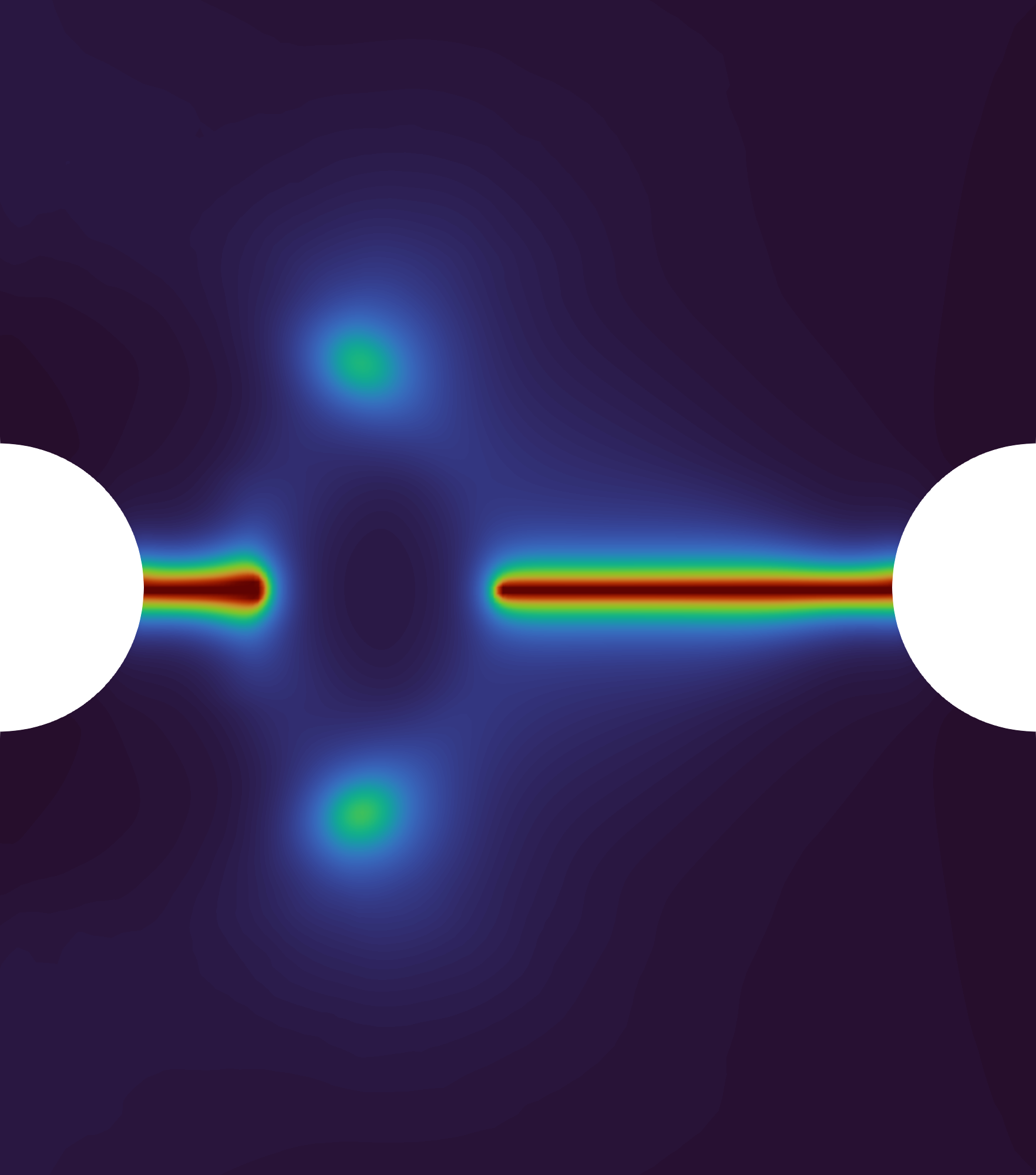}} &
        \subfloat[\label{fig:holenotch_2_5_fields:b}]{
            \includegraphics[width=0.23\textwidth]{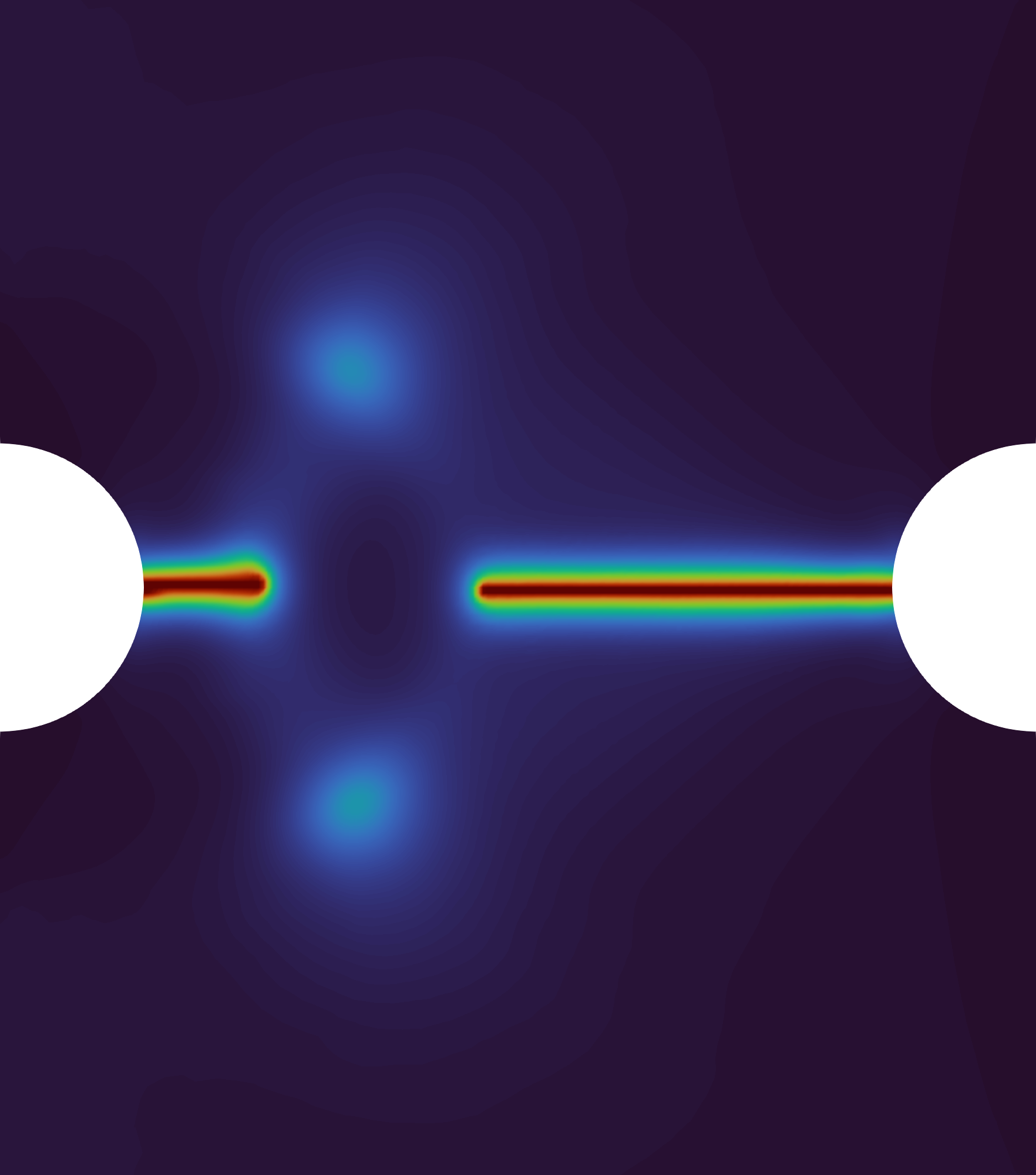}} &
        \subfloat[\label{fig:holenotch_2_5_fields:c}]{
            \includegraphics[width=0.23\textwidth]{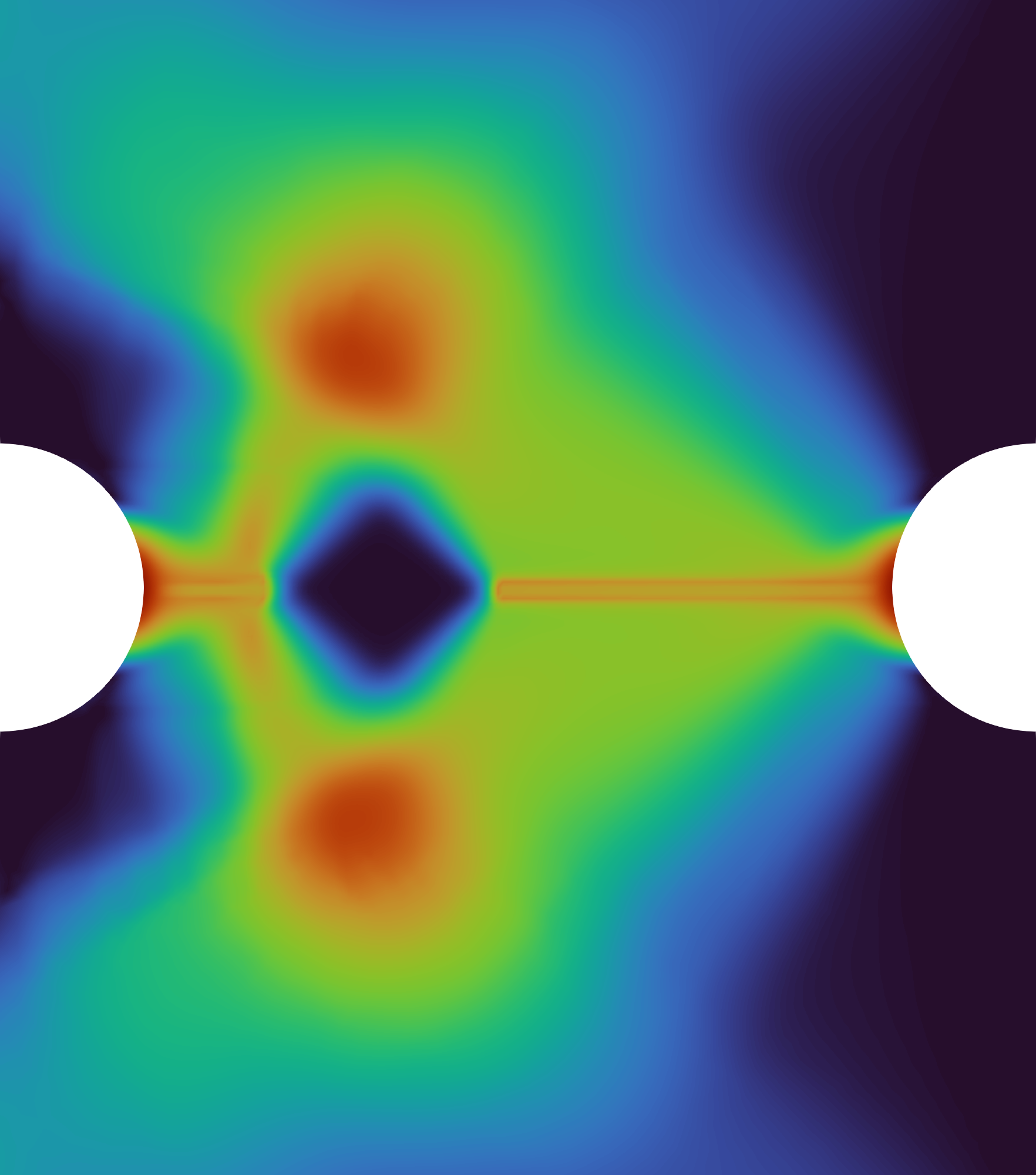}} &
        \subfloat[\label{fig:holenotch_2_5_fields:d}]{
            \includegraphics[width=0.23\textwidth]{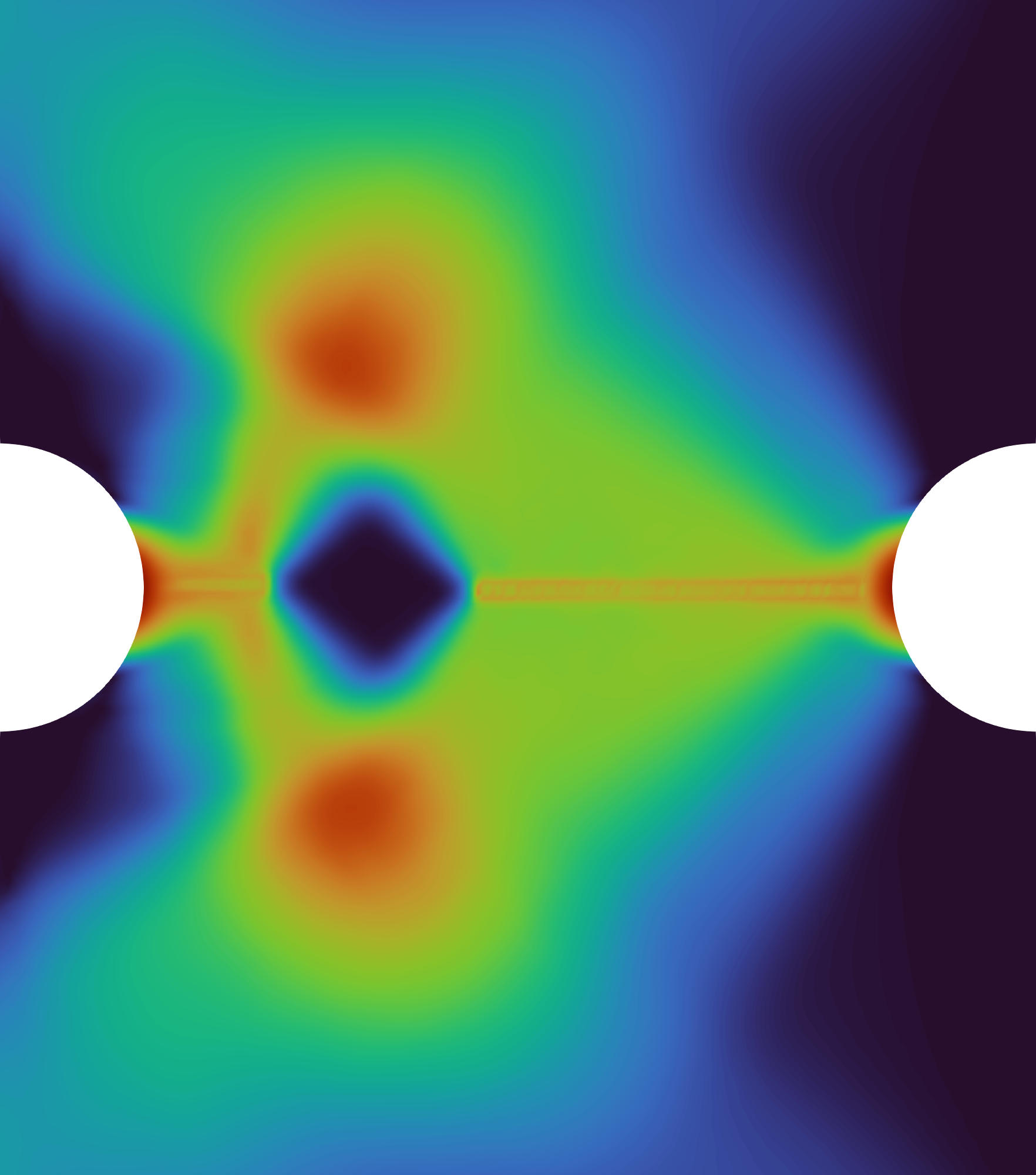}} \\
        \multicolumn{2}{c}{\includegraphics[width=0.25\textwidth]{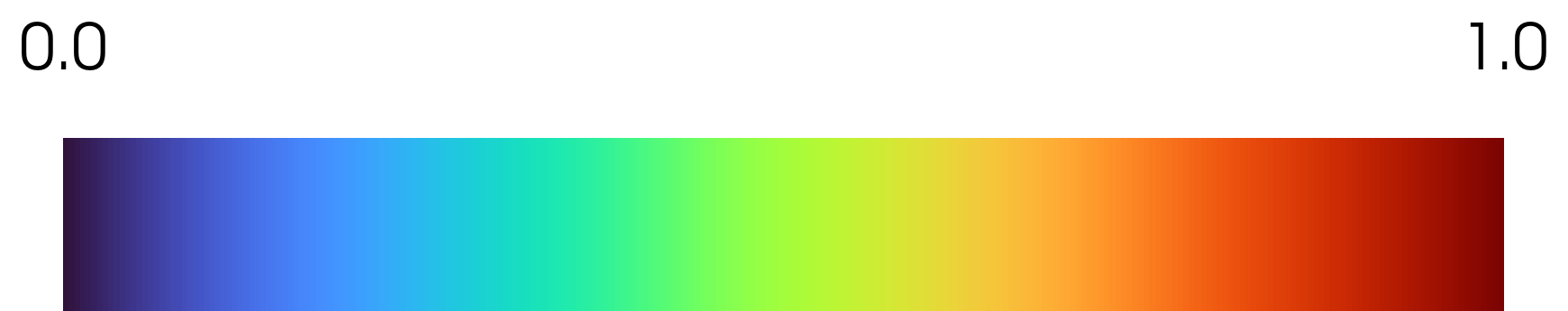}} &
        \multicolumn{2}{c}{\includegraphics[width=0.25\textwidth]{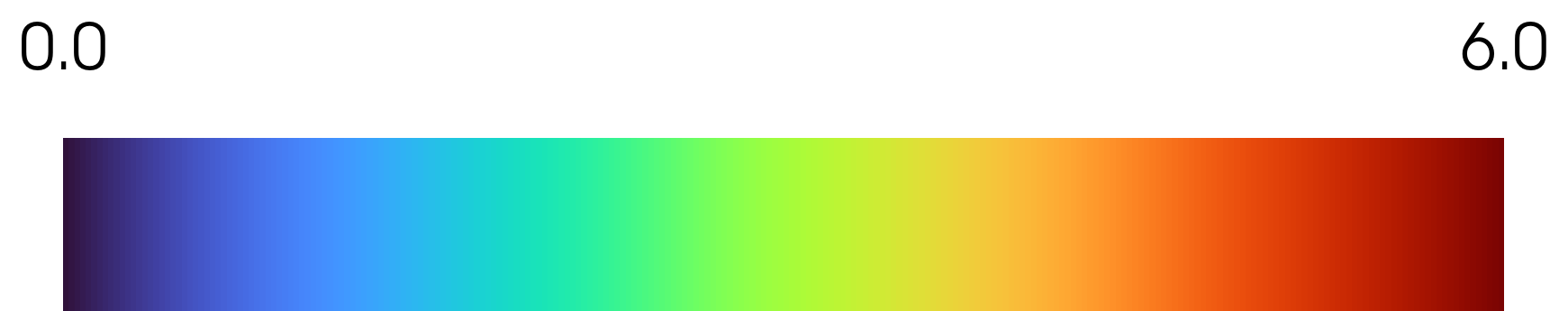}} \\
	\end{tabular}
    \caption[]{Symmetrically notched $r=\qty{2.5}{\milli\metre}$ specimen. Phase-field and accumulated equivalent plastic strain distributions in the ligament region at $U_p=\qty{3.407}{\milli\metre}$. The phase field $c$ is shown for \subref{fig:holenotch_2_5_fields:a} FEM and \subref{fig:holenotch_2_5_fields:b} DEM, while the accumulated equivalent plastic strain $\zeta$ is shown for \subref{fig:holenotch_2_5_fields:c} FEM and \subref{fig:holenotch_2_5_fields:d} DEM. The DEM fields correspond to seed~$1$.}
    \label{fig:holenotch_2_5_fields_a}
\end{figure}

\begin{figure}[htbp!]
	\centering
    \setlength{\tabcolsep}{3pt}
    \begin{tabular}{>{\centering\arraybackslash}m{0.23\textwidth}
                    >{\centering\arraybackslash}m{0.23\textwidth}
                    >{\centering\arraybackslash}m{0.23\textwidth}
                    >{\centering\arraybackslash}m{0.23\textwidth}}
        \subfloat[\label{fig:holenotch_2_5_fields:e}]{
            \includegraphics[width=0.23\textwidth]{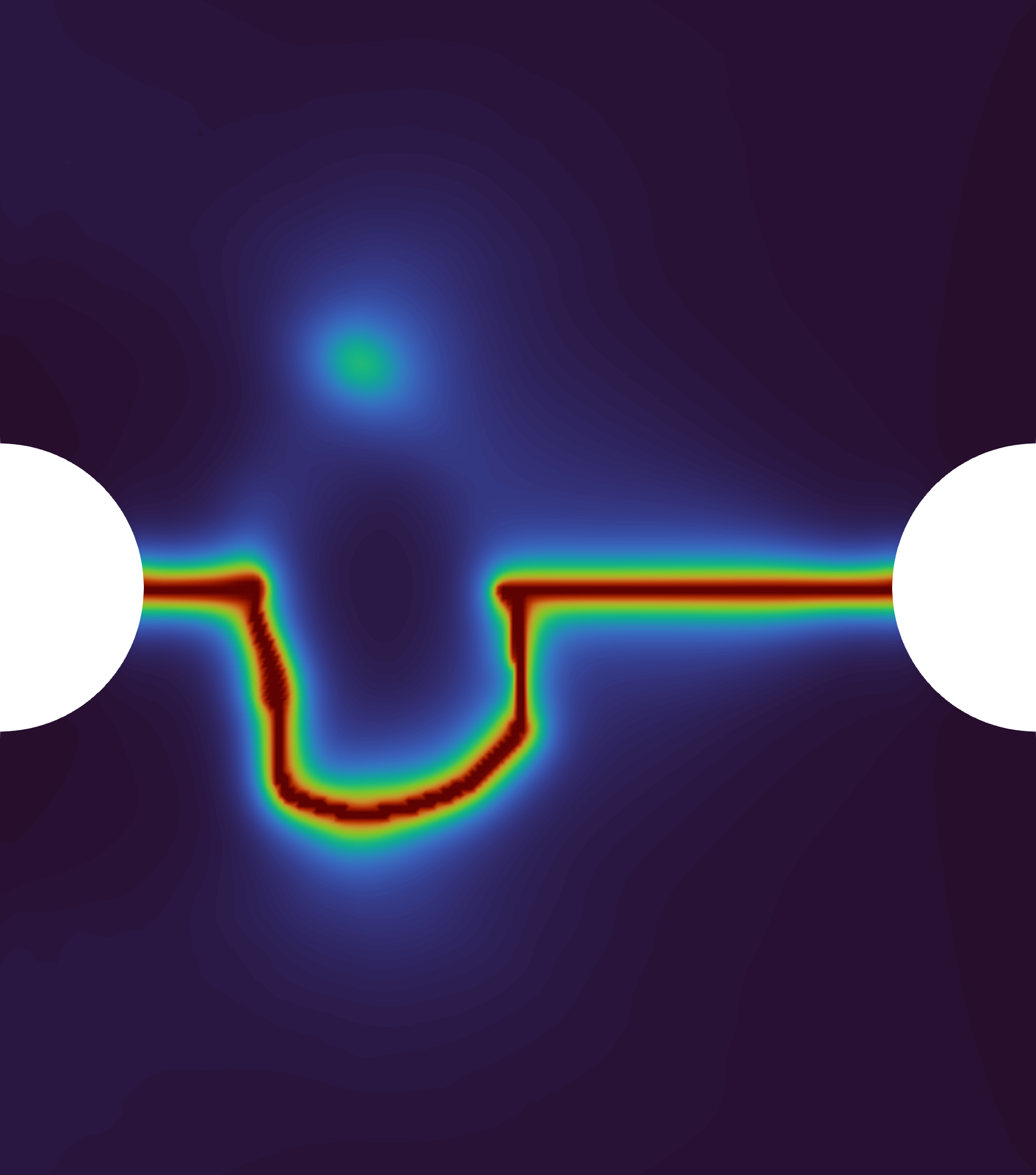}} &
        \subfloat[\label{fig:holenotch_2_5_fields:f}]{
            \includegraphics[width=0.23\textwidth]{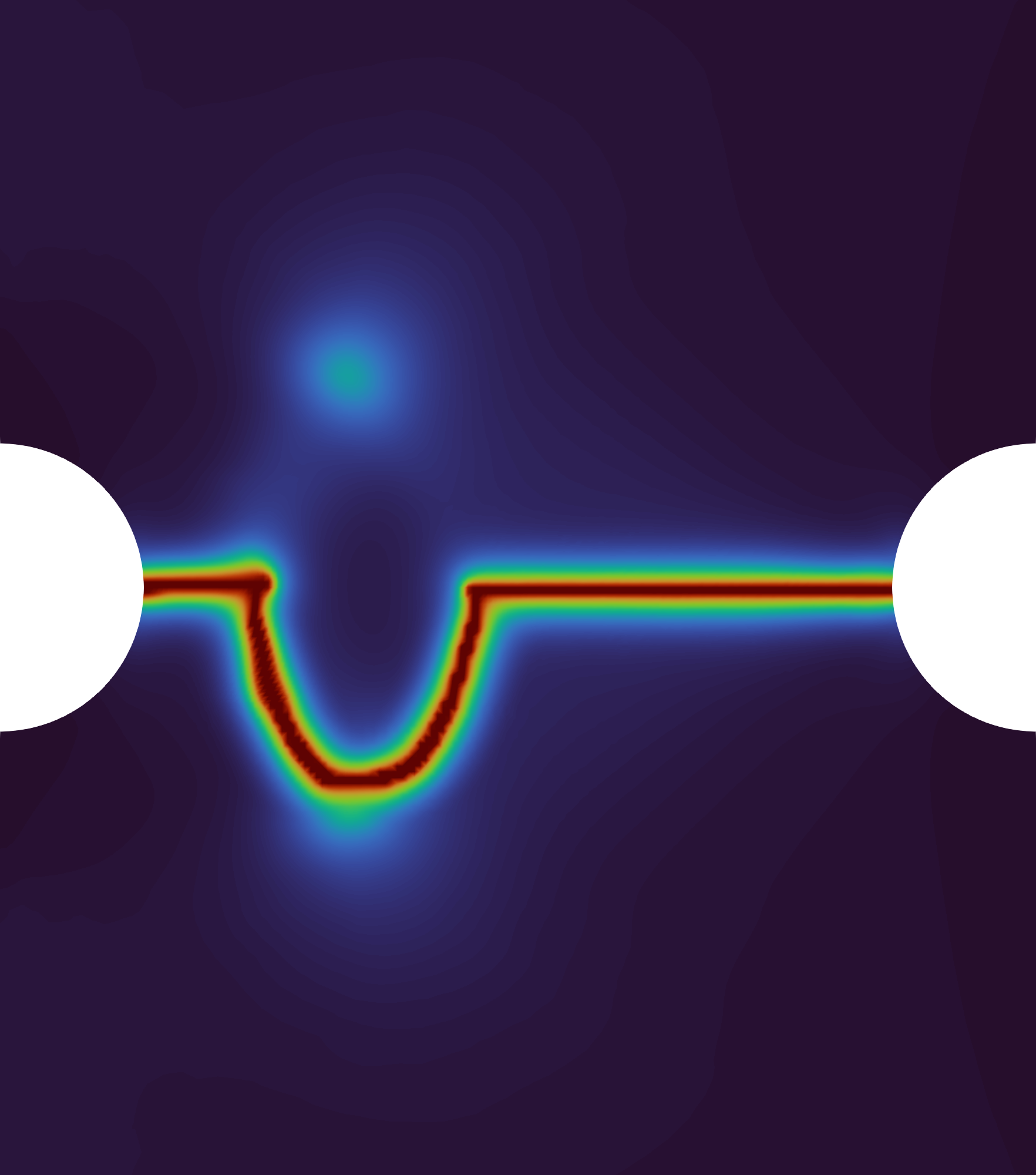}} &
        \subfloat[\label{fig:holenotch_2_5_fields:g}]{
            \includegraphics[width=0.23\textwidth]{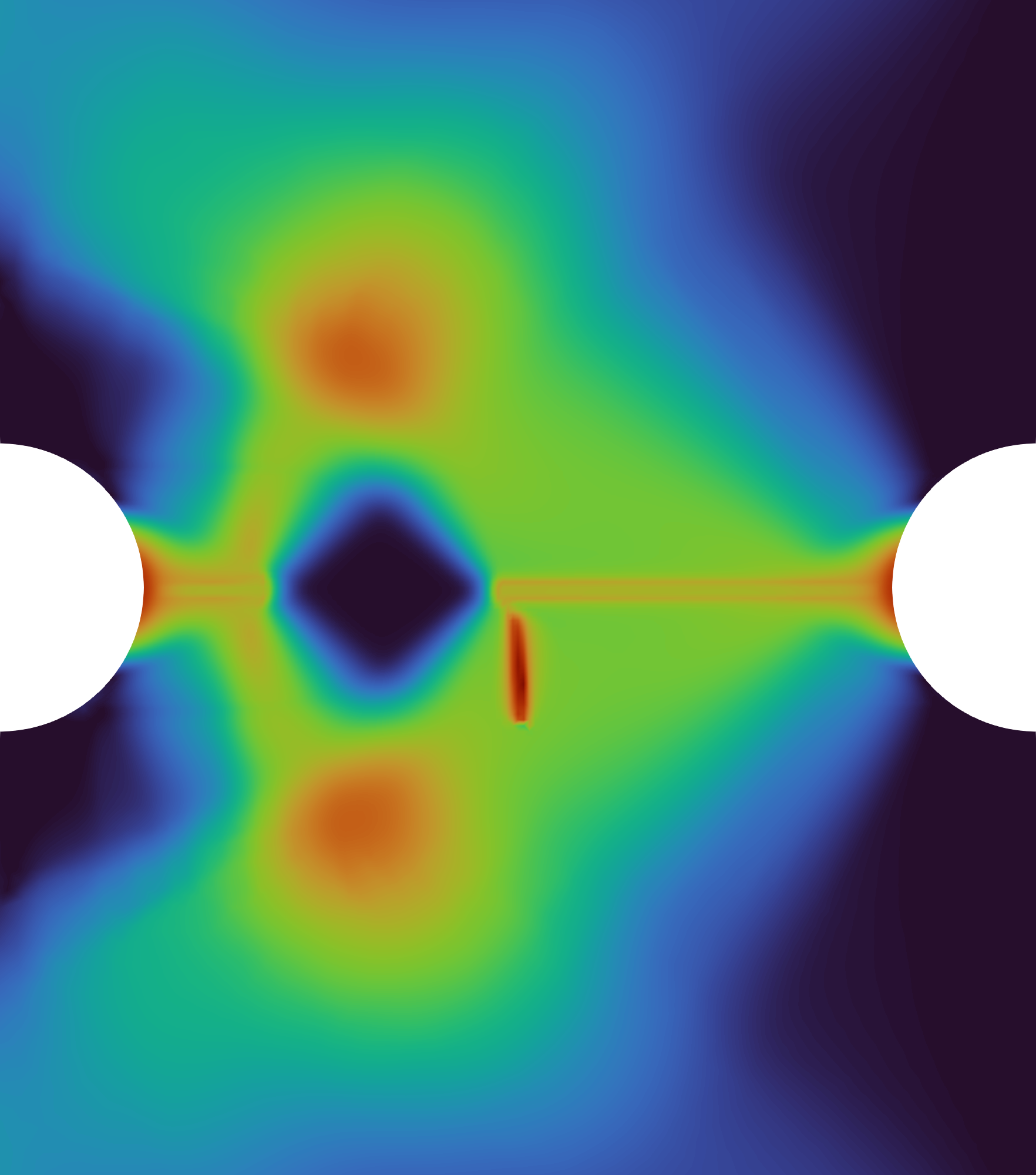}} &
        \subfloat[\label{fig:holenotch_2_5_fields:h}]{
            \includegraphics[width=0.23\textwidth]{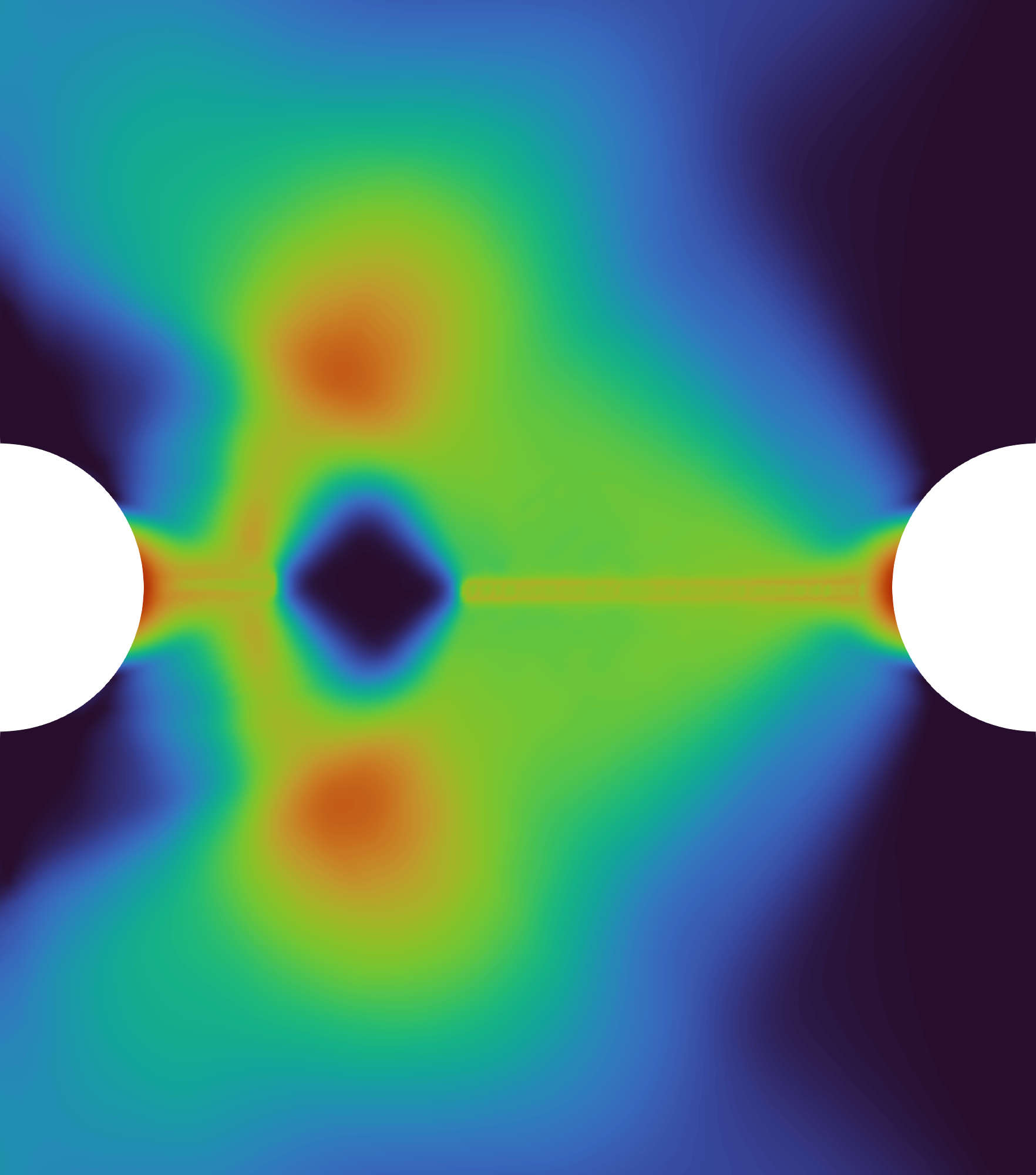}} \\
        \multicolumn{2}{c}{\includegraphics[width=0.25\textwidth]{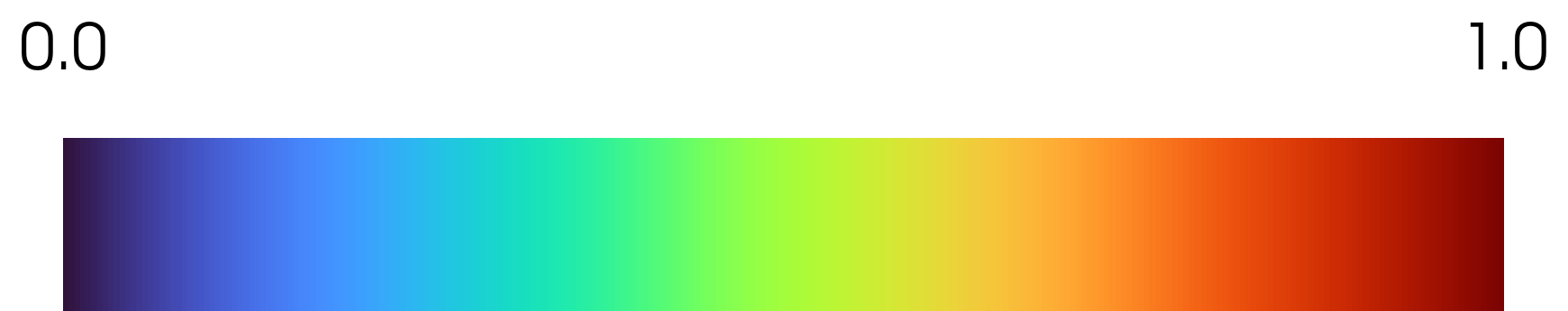}} &
        \multicolumn{2}{c}{\includegraphics[width=0.25\textwidth]{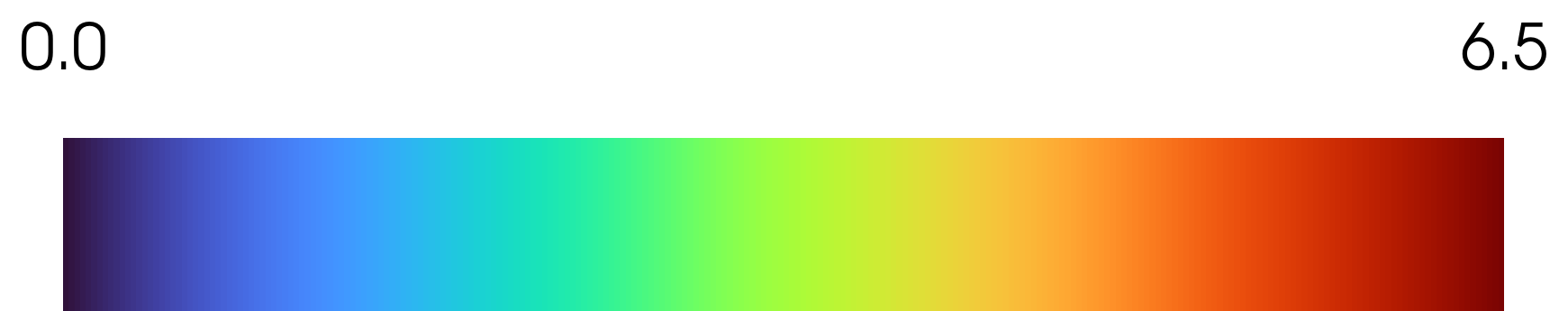}} \\
	\end{tabular}
    \caption[]{Symmetrically notched $r=\qty{2.5}{\milli\metre}$ specimen. Phase-field and accumulated equivalent plastic strain distributions in the ligament region after rupture. The phase field $c$ is shown for
    \subref{fig:holenotch_2_5_fields:e} FEM and
    \subref{fig:holenotch_2_5_fields:f} DEM, while the accumulated equivalent plastic strain $\zeta$ is shown for
    \subref{fig:holenotch_2_5_fields:g} FEM and
    \subref{fig:holenotch_2_5_fields:h} DEM.}
    \label{fig:holenotch_2_5_fields_b}
\end{figure}

The phase-field and accumulated equivalent plastic strain distributions are shown in \FIG{fig:holenotch_2_5_fields_a} at $U_p=\qty{3.407}{\milli\metre}$ and in \FIG{fig:holenotch_2_5_fields_b} after rupture. Before rupture, the two methods predict closely comparable crack initiation from the notch roots and similar plastic-localisation patterns. This behaviour is also consistent with the failure mechanism reported for the smaller-radius notched specimen by Ambati et al.\cite{Ambati2015b}. The main local difference is the narrower phase-field band obtained with the DEM. Since the RBF widths are defined by $\sigma_k=s_{\mathrm{RBF}}h_k$, the locally refined nodal spacing reduces the individual kernel widths for a fixed $s_{\mathrm{RBF}}$ and may contribute to this sharper representation. The same value of $s_{\mathrm{RBF}}$ is retained for all specimens, avoiding specimen-specific tuning. The physical phase-field length scale $l_0$ is unchanged.

After rupture, both methods predict the same overall inward crack-linking mechanism, although the final transition is resolved differently. The FEM solution develops the final connection over two load increments, whereas the DEM completes the corresponding linking event within a single increment. The additional intermediate load state in the FEM solution is accompanied by a localised region of higher accumulated equivalent plastic strain.

\paragraph{Symmetrically-notched specimen $r=\qty{5}{\milli\metre}$}

The FEM reference used \num{2000} load increments. The first \num{100} increments covered the loading range up to $U_p=\qty{1.6}{\milli\metre}$, followed by \num{1900} increments up to $U_p=\qty{1.8}{\milli\metre}$. The computation required \qty{0.98}{\hour}. Each DEM simulation used \num{130} load increments, with \num{30} increments up to $U_p=\qty{1.6}{\milli\metre}$ and a further \num{100} increments up to $U_p=\qty{1.8}{\milli\metre}$.

Five of the eight DEM runs completed in \qty{2.3\pm0.1}{\hour}. Seeds \num{6}, \num{7} and \num{8} required \qtylist{13.4;8.2;13.9}{\hour}, respectively. For these runs, the reaction force became very small after complete separation and increasingly sensitive to numerical fluctuations. As a result, the reaction-force stabilisation criterion was no longer satisfied within the prescribed cycle limit, and the affected post-rupture increments reached the maximum of $500$ coupled cycles. No seed reached this limit before rupture. When all eight runs are included, the mean DEM wall-clock time is \qty{5.9\pm5.2}{\hour}. The large variation in runtime is therefore associated with the post-rupture termination criterion rather than with the prediction of the loading response or onset of rupture.

\begin{figure}[htbp!]
	\centering
    \captionsetup[subfloat]{farskip=0pt,captionskip=1pt}
    \setlength{\tabcolsep}{4pt}
    \begin{tabular}{@{}c@{\hspace{4pt}}c@{}}
        \subfloat[\label{fig:holenotch_5_plots:a}]{
            \includegraphics[width=0.48\textwidth]{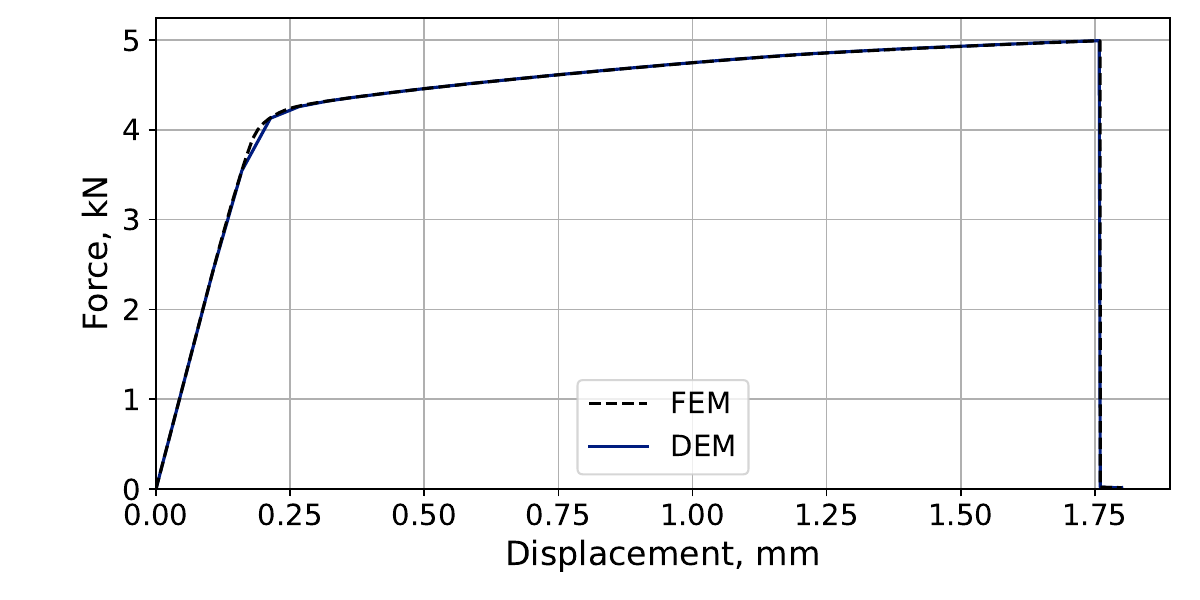}} &
        \subfloat[\label{fig:holenotch_5_plots:b}]{
            \includegraphics[width=0.48\textwidth]{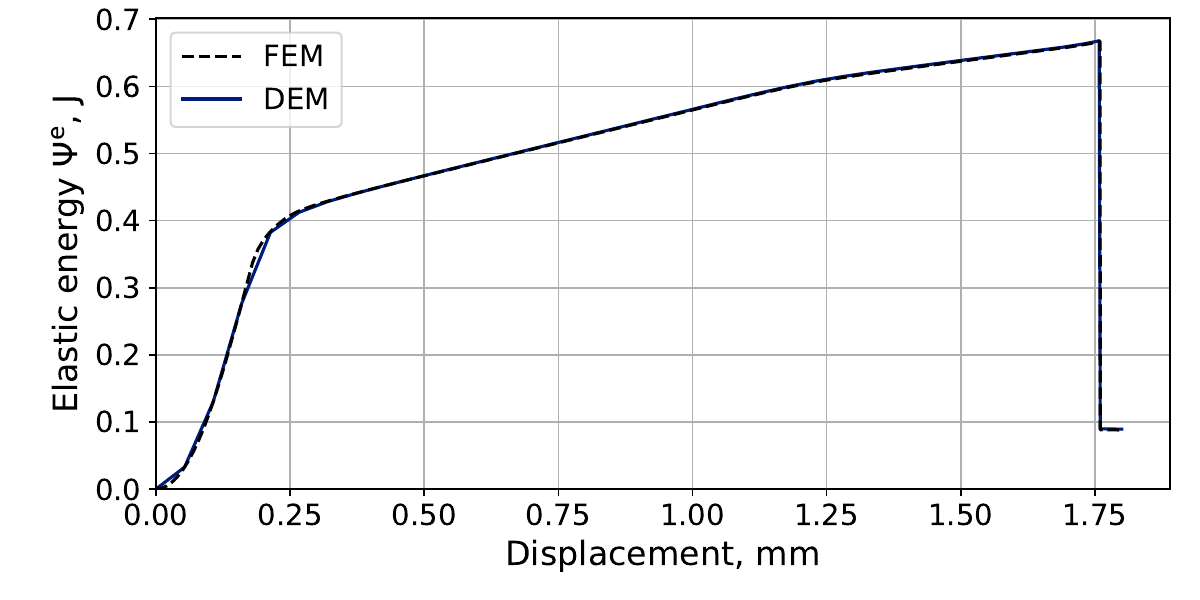}} \\[-2pt]
        \subfloat[\label{fig:holenotch_5_plots:c}]{
            \includegraphics[width=0.48\textwidth]{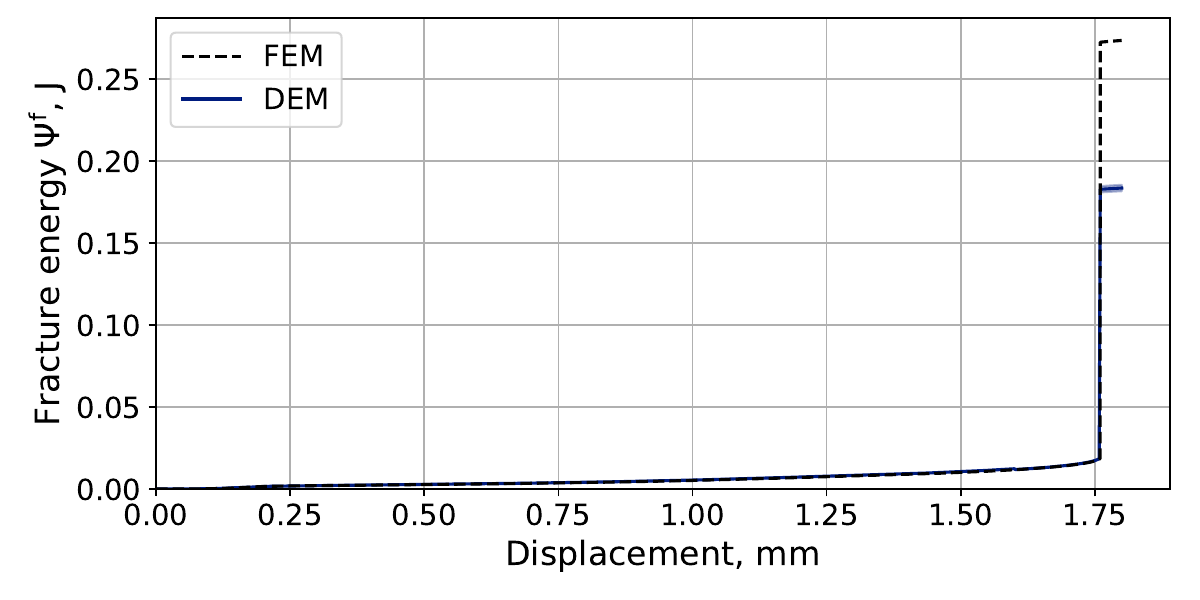}} &
        \subfloat[\label{fig:holenotch_5_plots:d}]{
            \includegraphics[width=0.48\textwidth]{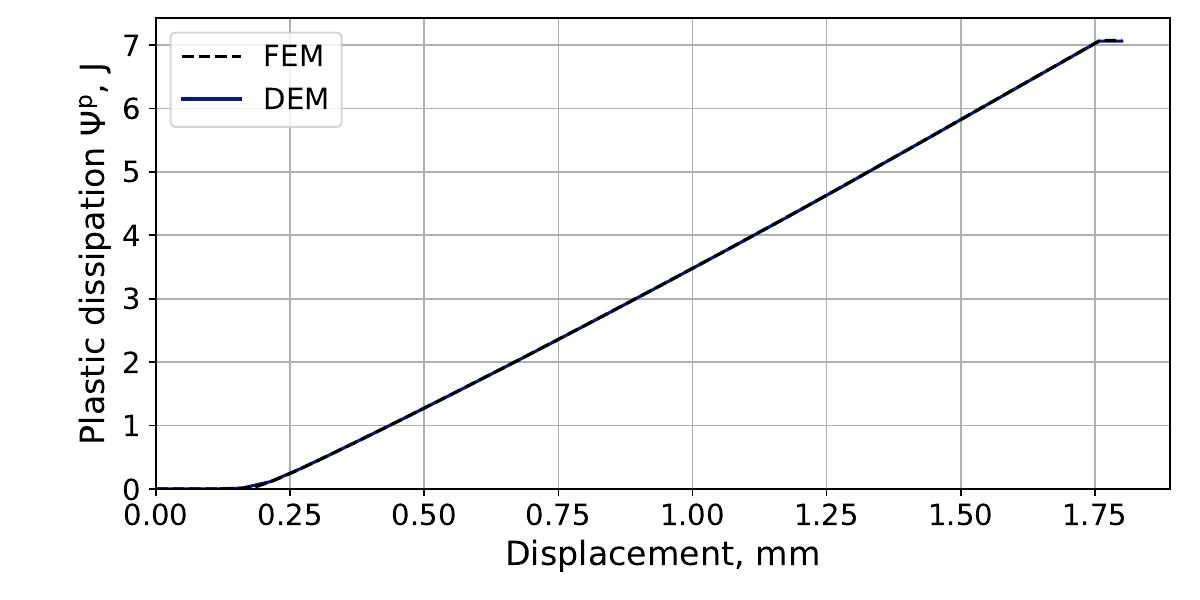}} \\[-2pt]
        \multicolumn{2}{c}{%
        \subfloat[\label{fig:holenotch_5_plots:e}]{
            \includegraphics[width=0.96\textwidth]{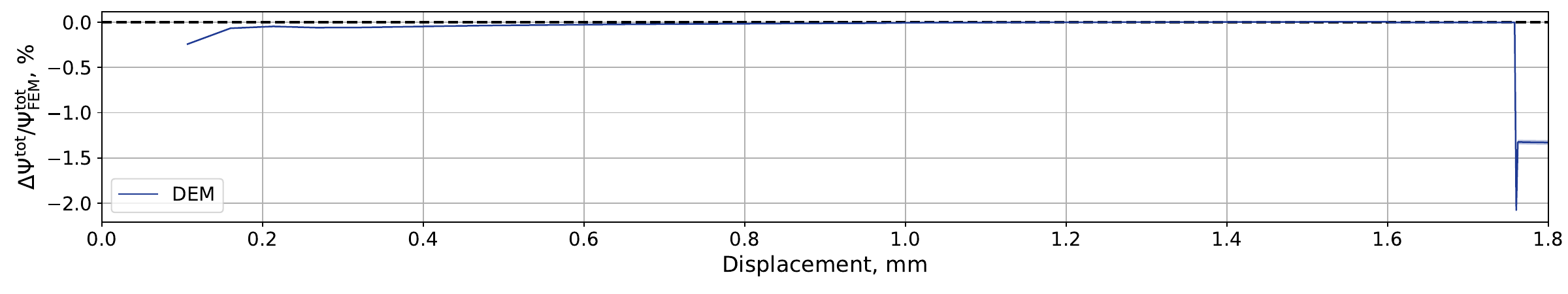}}} \\
	\end{tabular}
    \caption[]{Symmetrically notched $r=\qty{5}{\milli\metre}$ specimen. Global response obtained with the proposed DEM and the FEM reference: \subref{fig:holenotch_5_plots:a} force--displacement response; \subref{fig:holenotch_5_plots:b} elastic energy $\Psi^{\mathrm{e}}$; \subref{fig:holenotch_5_plots:c} fracture energy $\Psi^{\mathrm{f}}$; and \subref{fig:holenotch_5_plots:d} plastic energy $\Psi^{\mathrm{p}}$; \subref{fig:holenotch_5_plots:e} relative difference in total energy, $100(\Psi^{\mathrm{tot}}_{\mathrm{DEM}}-\Psi^{\mathrm{tot}}_{\mathrm{FEM}})/\Psi^{\mathrm{tot}}_{\mathrm{FEM}}$, where $\Psi^{\mathrm{tot}}=\Psi^{\mathrm{e}}+\Psi^{\mathrm{f}}+\Psi^{\mathrm{p}}$. All quantities are plotted against the prescribed displacement $U_p$. Shaded bands span $\pm1$ standard deviation over the eight DEM seeds.}
    \label{fig:holenotch_5_plots}
\end{figure}

The global responses are compared in \FIG{fig:holenotch_5_plots}. The DEM and FEM solutions remain closely aligned throughout loading and softening, apart from a small initial difference associated with the different load discretisations. The main discrepancy develops during the terminal fracture event and is concentrated in the fracture-energy contribution. The elastic and plastic energies remain closely matched, whereas the two methods select different final crack-linking branches. The resulting DEM state has a total energy approximately \qty{1.3}{\percent} below the FEM reference.

All eight DEM seeds select the same terminal fracture branch. The difference is therefore reproducible with respect to random initialisation and is not associated with an isolated optimisation trajectory. We interpret this as the selection of an alternative fracture branch during the abrupt final instability, rather than as evidence that either discrete solution represents a unique global energy minimum. Before this instability, the two approaches remain in close agreement.

\begin{figure}[htbp!]
	\centering
    \setlength{\tabcolsep}{3pt}
    \begin{tabular}{>{\centering\arraybackslash}m{0.23\textwidth}
                    >{\centering\arraybackslash}m{0.23\textwidth}
                    >{\centering\arraybackslash}m{0.23\textwidth}
                    >{\centering\arraybackslash}m{0.23\textwidth}}
        \subfloat[\label{fig:holenotch_5_fields:a}]{
            \includegraphics[width=0.23\textwidth]{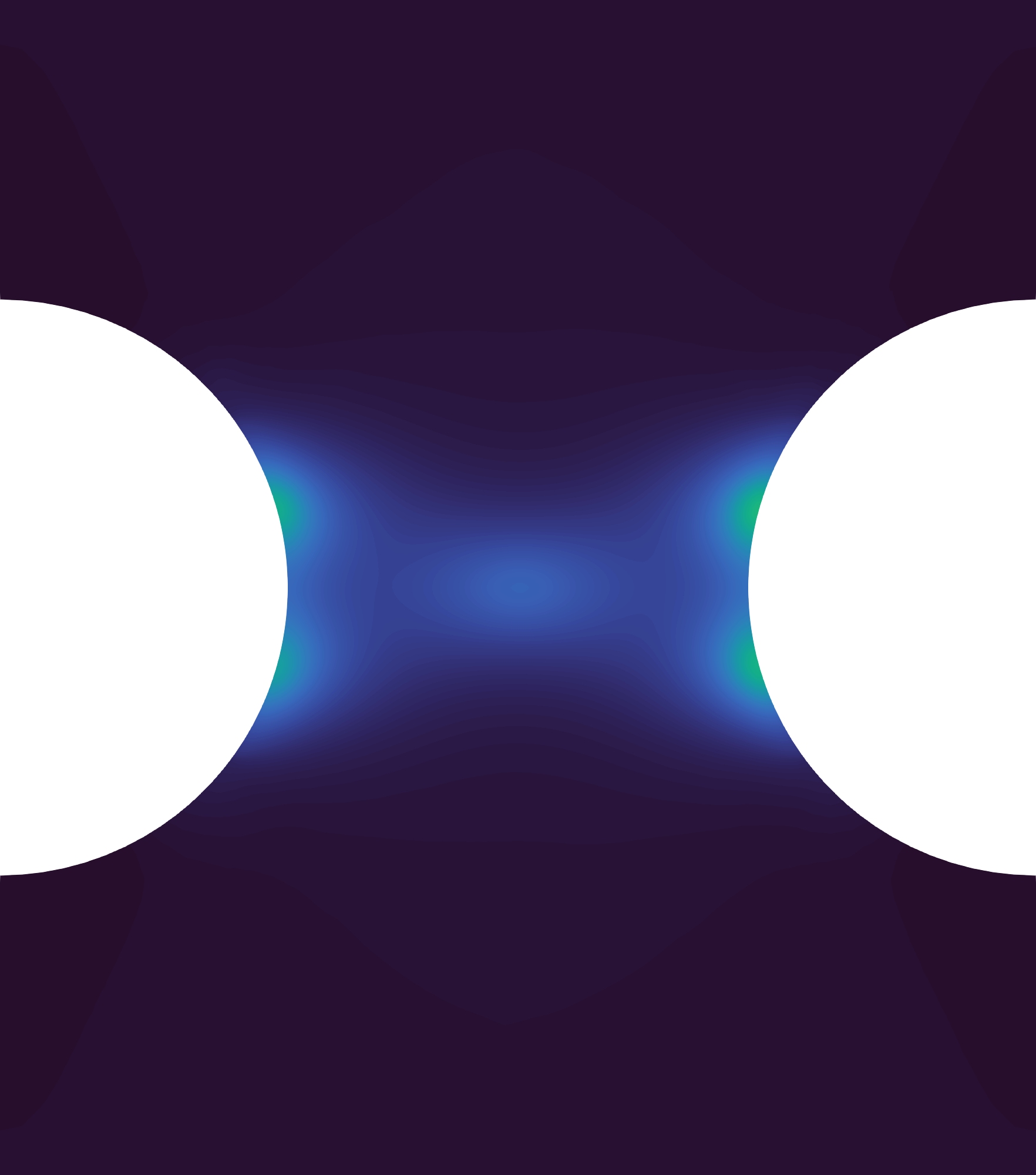}} &
        \subfloat[\label{fig:holenotch_5_fields:b}]{
            \includegraphics[width=0.23\textwidth]{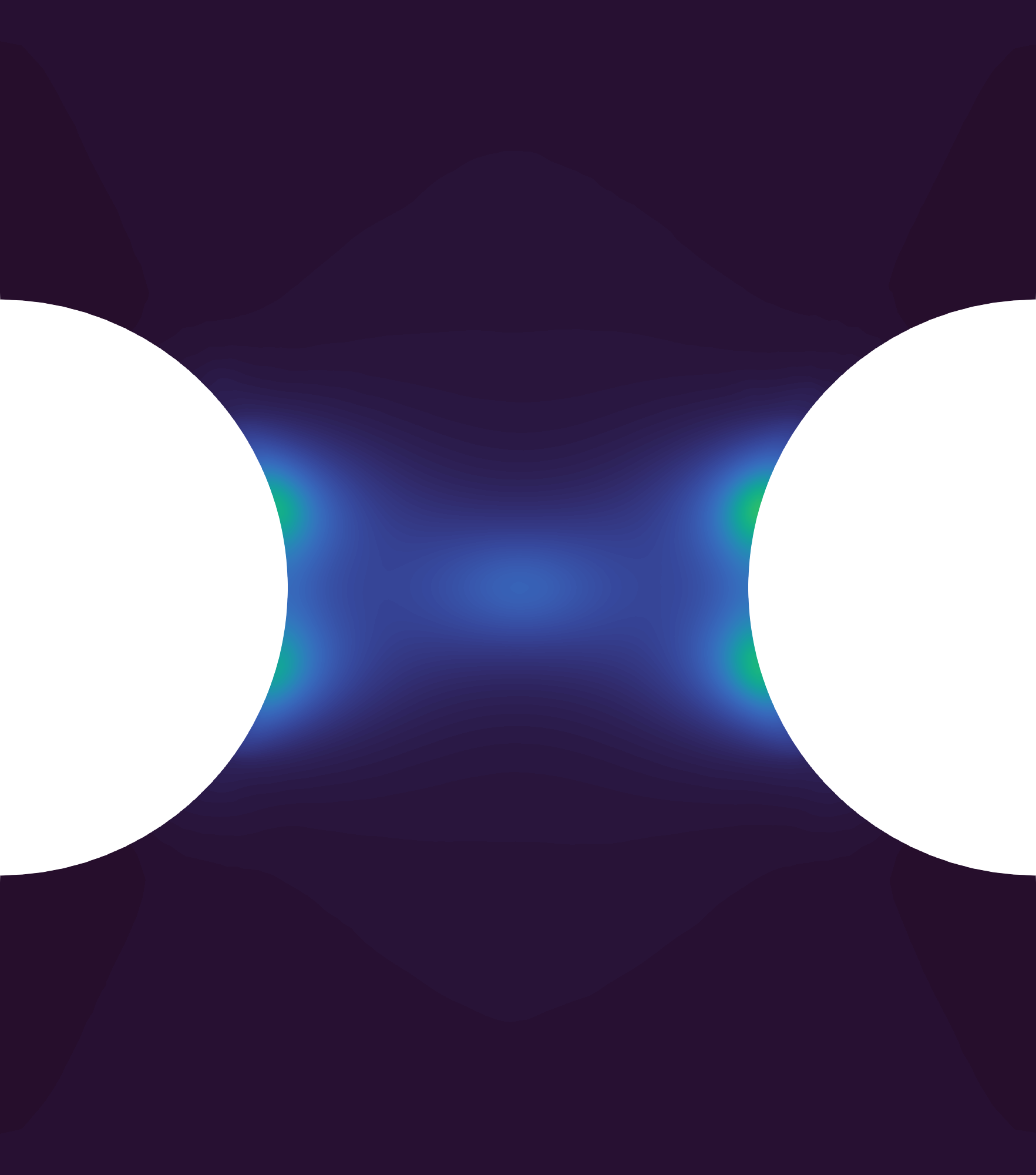}} &
        \subfloat[\label{fig:holenotch_5_fields:c}]{
            \includegraphics[width=0.23\textwidth]{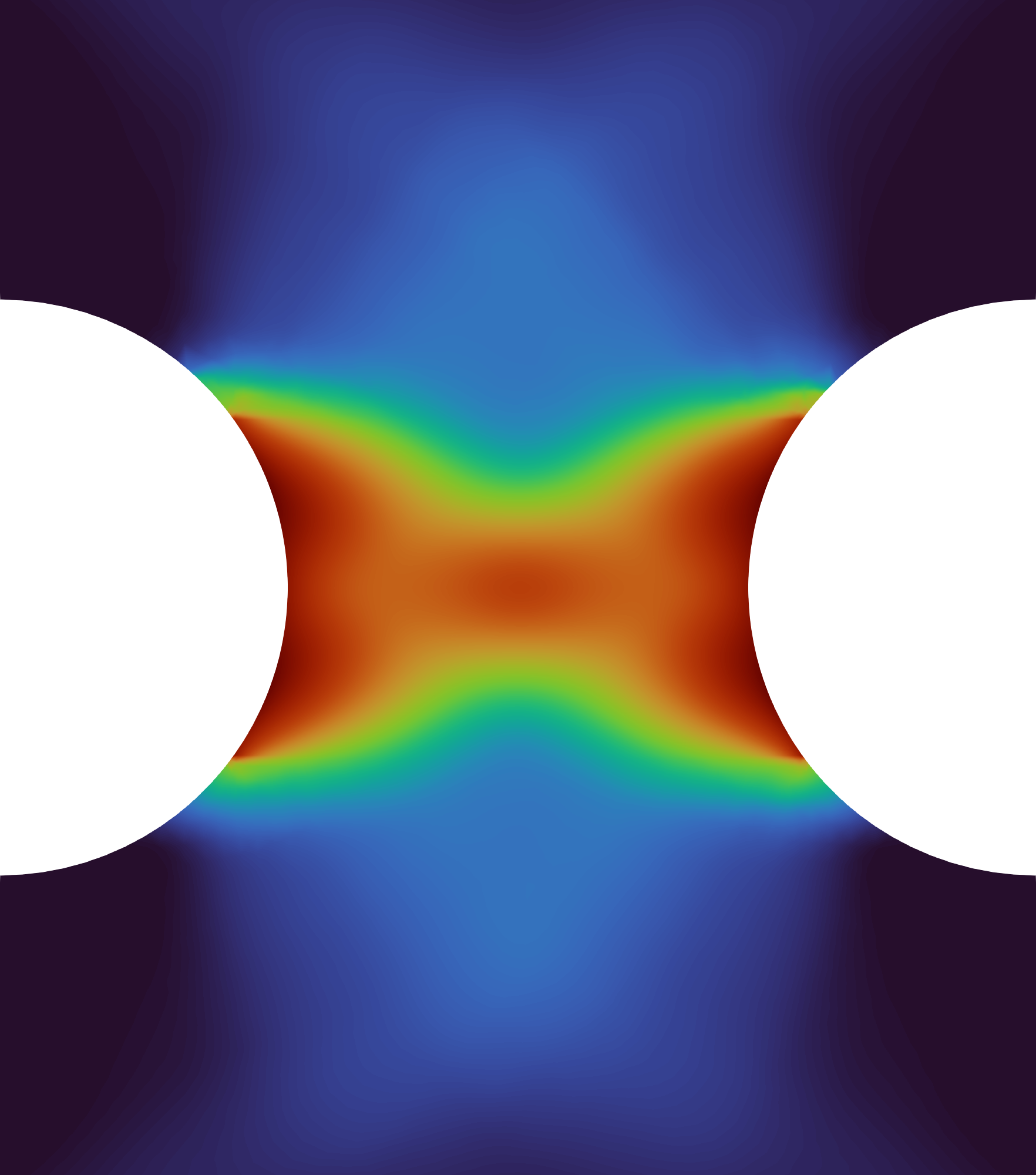}} &
        \subfloat[\label{fig:holenotch_5_fields:d}]{
            \includegraphics[width=0.23\textwidth]{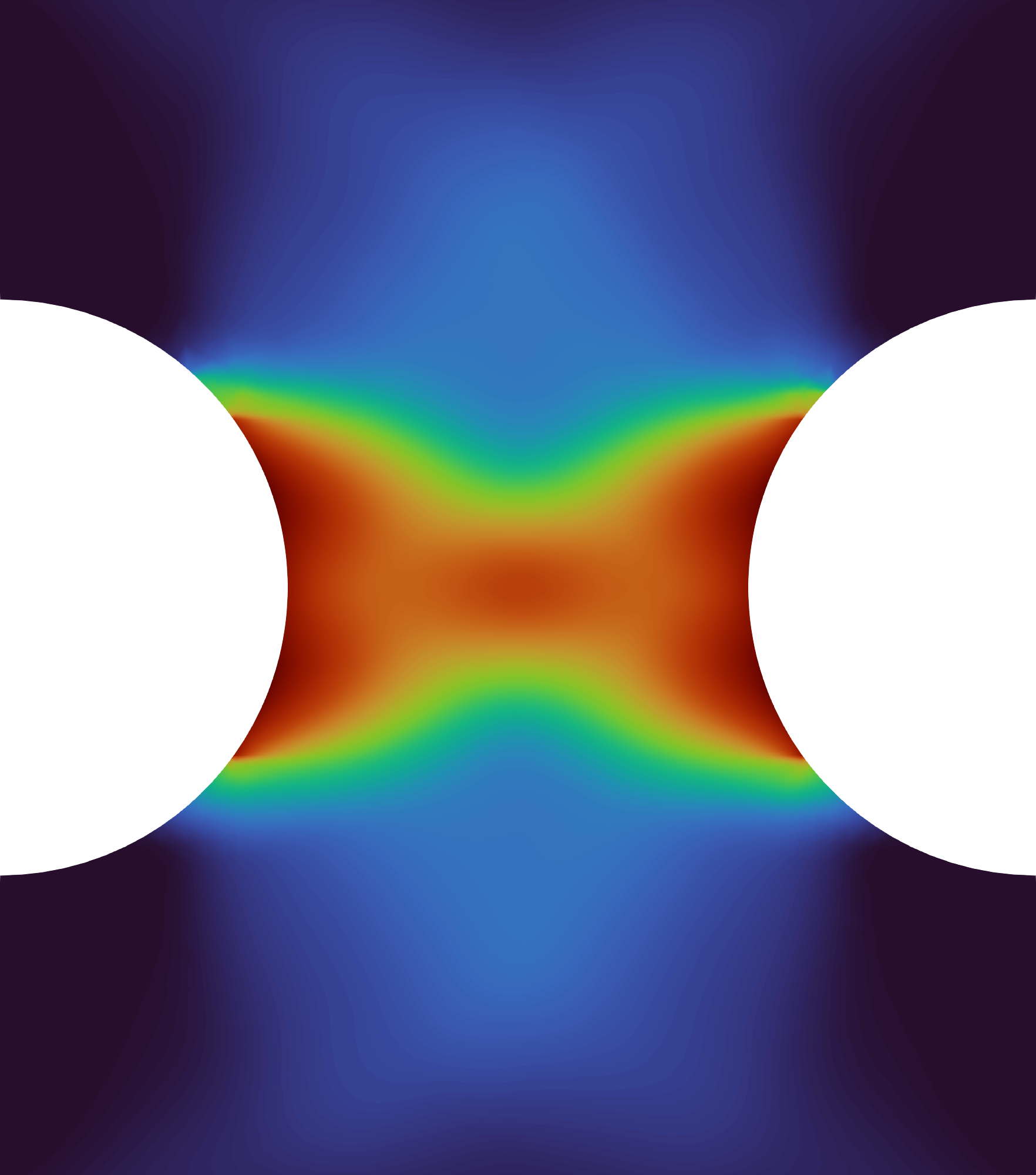}} \\
        \multicolumn{2}{c}{\includegraphics[width=0.25\textwidth]{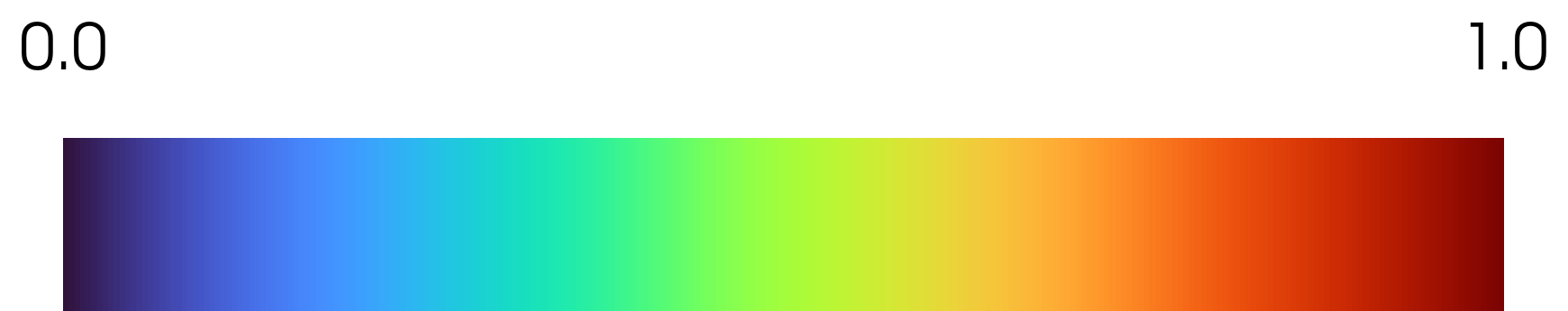}} &
        \multicolumn{2}{c}{\includegraphics[width=0.25\textwidth]{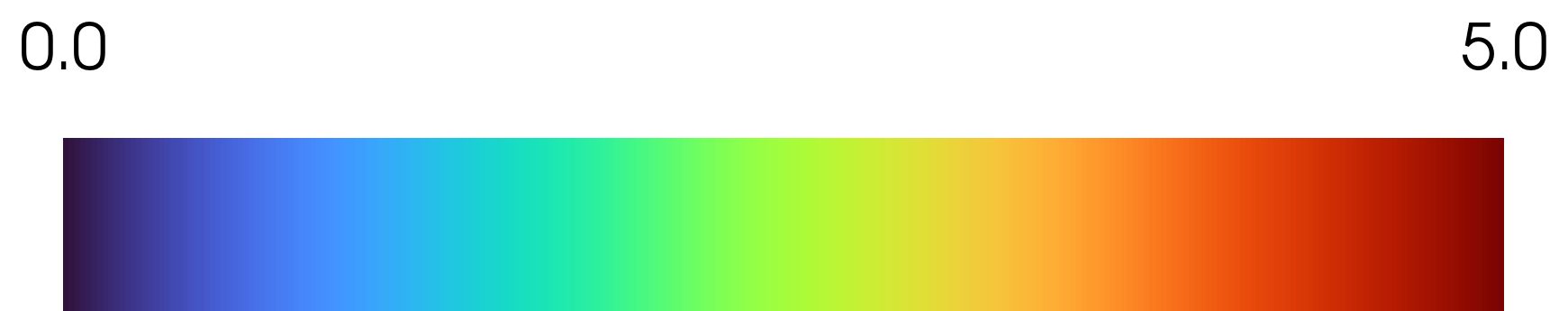}} \\
	\end{tabular}
    \caption[]{Symmetrically notched $r=\qty{5}{\milli\metre}$ specimen. Phase-field and accumulated equivalent plastic strain distributions in the ligament region immediately before rupture. The phase field $c$ is shown for \subref{fig:holenotch_5_fields:a} FEM and \subref{fig:holenotch_5_fields:b} DEM, while the accumulated equivalent plastic strain $\zeta$ is shown for \subref{fig:holenotch_5_fields:c} FEM and \subref{fig:holenotch_5_fields:d} DEM. A common colour scale is used for each field to facilitate direct comparison between the two methods.}
    \label{fig:holenotch_5_fields_a}
\end{figure}

\begin{figure}[htbp!]
	\centering
    \setlength{\tabcolsep}{3pt}
    \begin{tabular}{>{\centering\arraybackslash}m{0.23\textwidth}
                    >{\centering\arraybackslash}m{0.23\textwidth}
                    >{\centering\arraybackslash}m{0.23\textwidth}
                    >{\centering\arraybackslash}m{0.23\textwidth}}
        \subfloat[\label{fig:holenotch_5_fields:e}]{
            \includegraphics[width=0.23\textwidth]{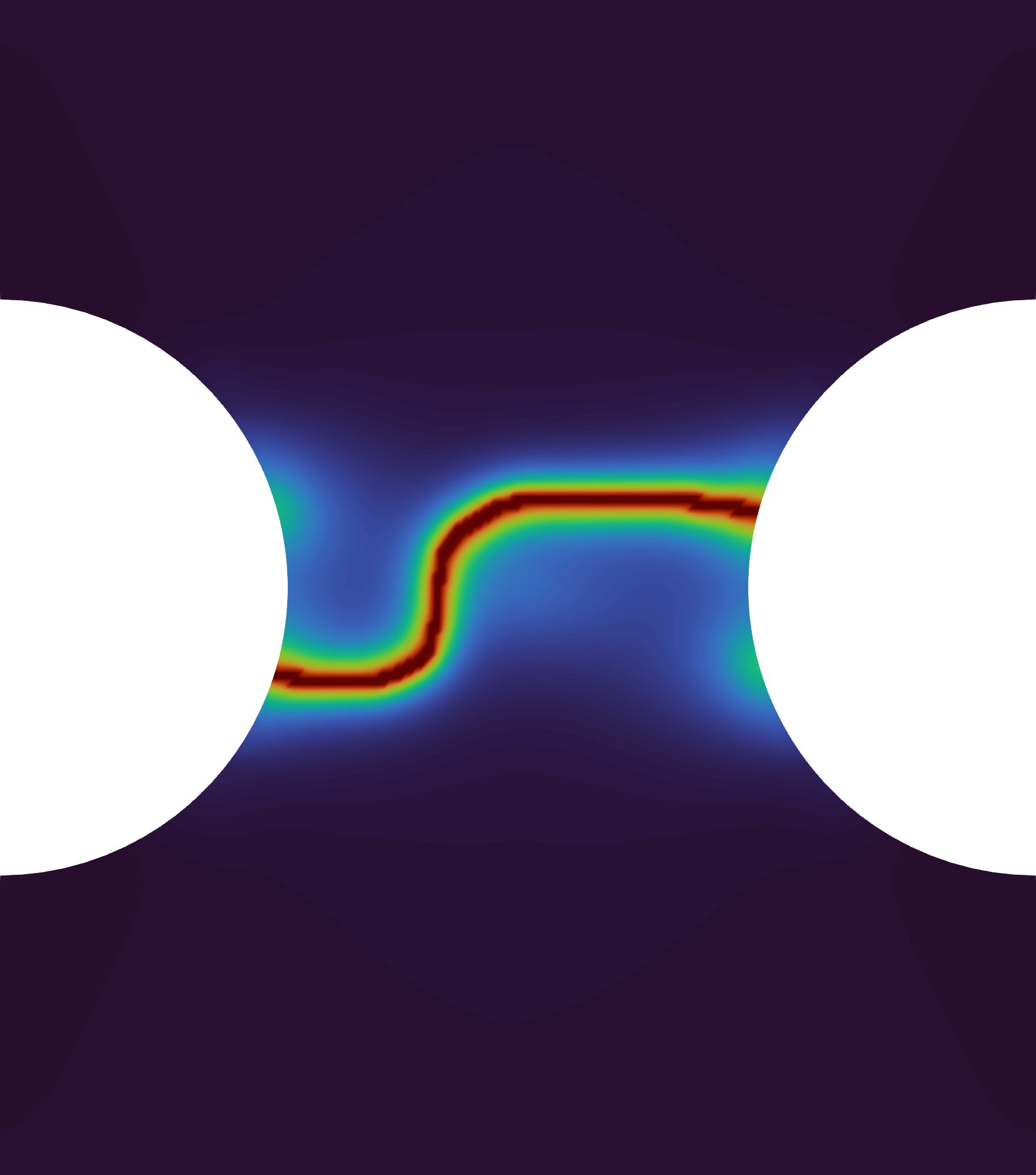}} &
        \subfloat[\label{fig:holenotch_5_fields:f}]{
            \includegraphics[width=0.23\textwidth]{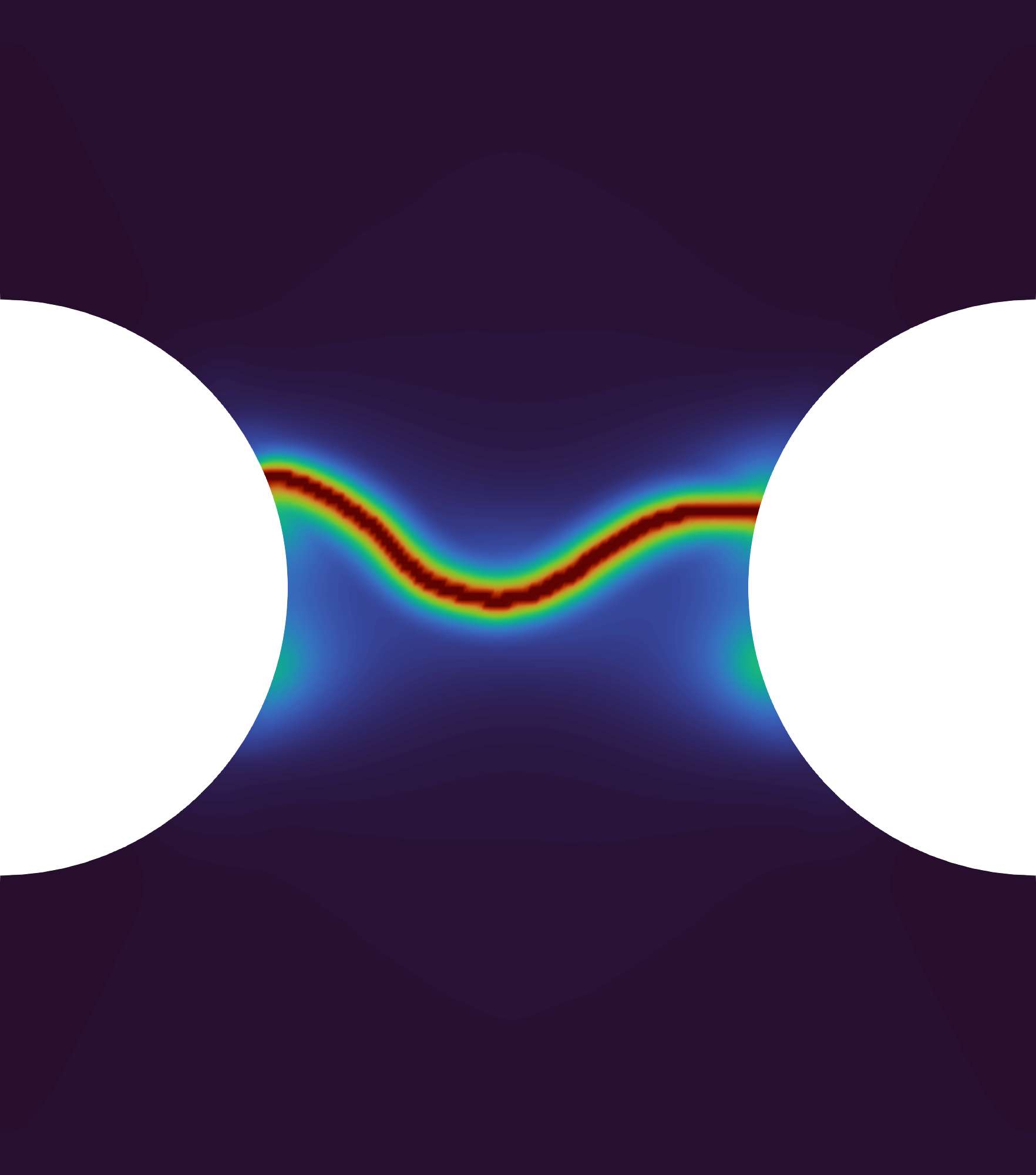}} &
        \subfloat[\label{fig:holenotch_5_fields:g}]{
            \includegraphics[width=0.23\textwidth]{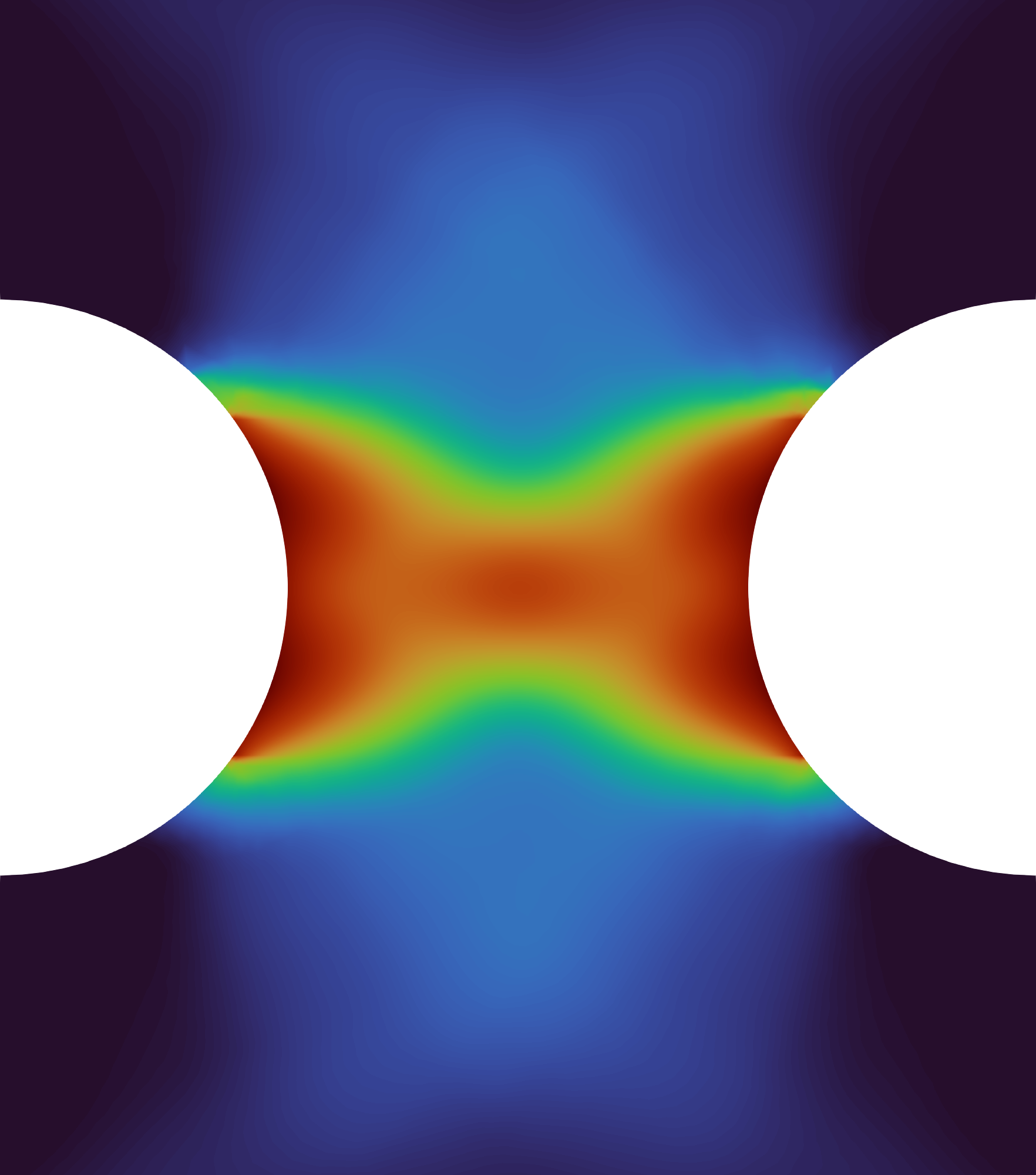}} &
        \subfloat[\label{fig:holenotch_5_fields:h}]{
            \includegraphics[width=0.23\textwidth]{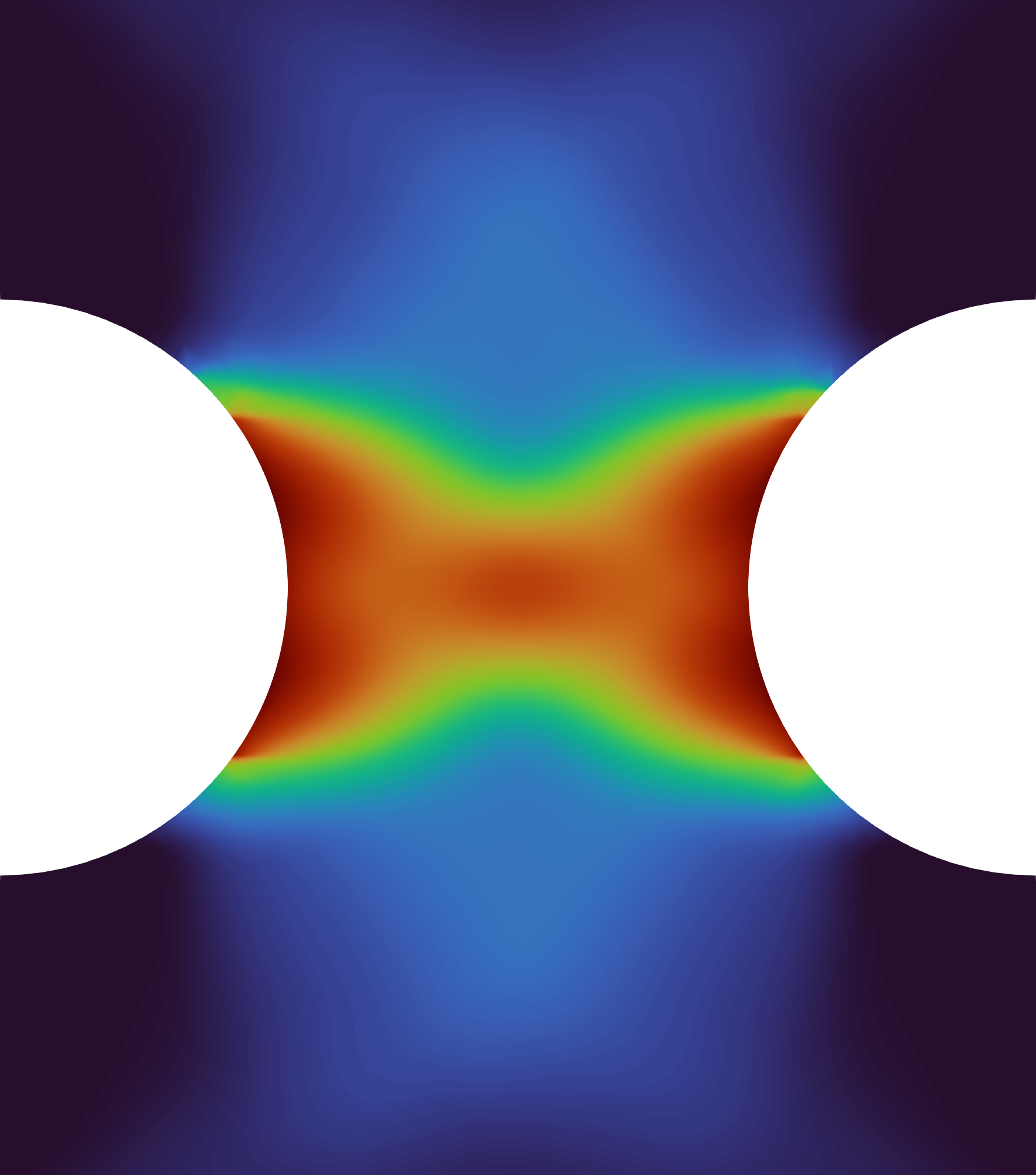}} \\
        \multicolumn{2}{c}{\includegraphics[width=0.25\textwidth]{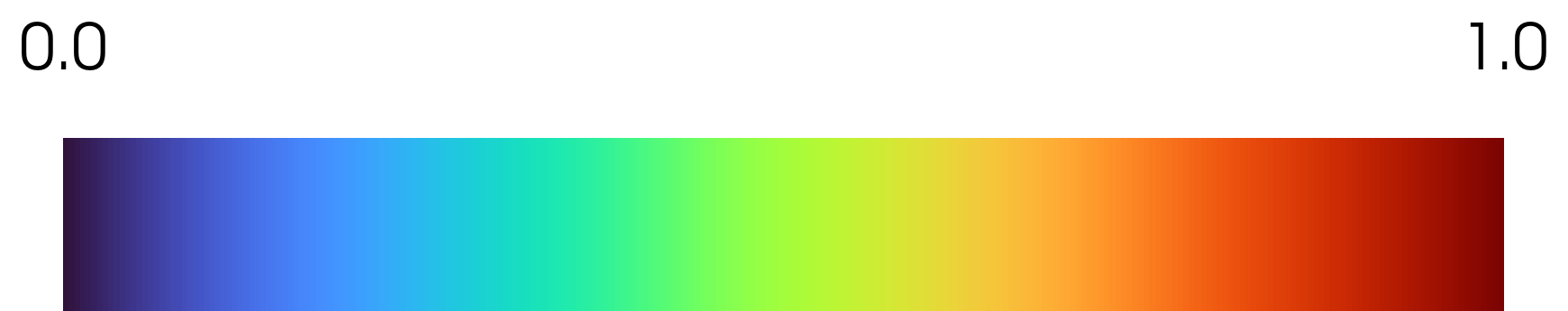}} &
        \multicolumn{2}{c}{\includegraphics[width=0.25\textwidth]{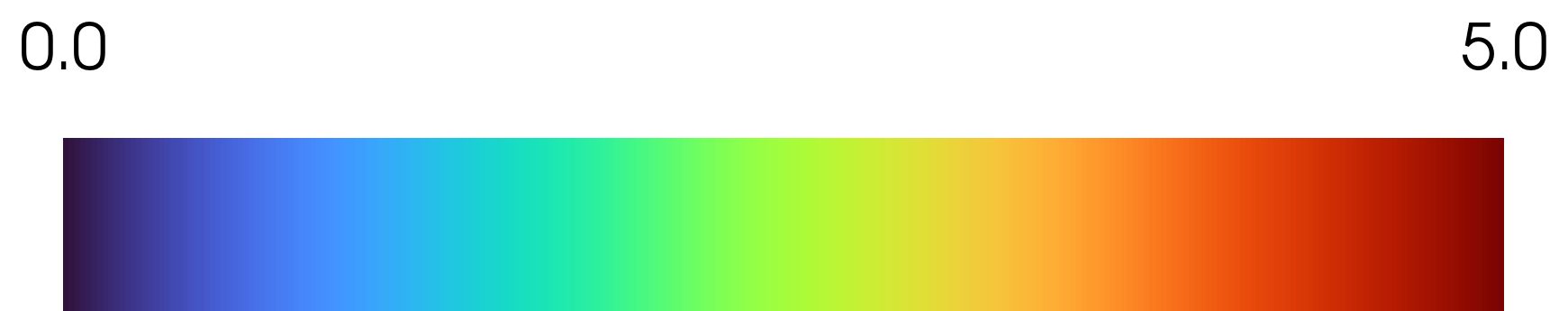}} \\
	\end{tabular}
    \caption[]{Symmetrically notched $r=\qty{5}{\milli\metre}$ specimen. Phase-field and accumulated equivalent plastic strain distributions in the ligament region immediately after rupture. The phase field $c$ is shown for \subref{fig:holenotch_5_fields:e} FEM and \subref{fig:holenotch_5_fields:f} DEM, while the accumulated equivalent plastic strain $\zeta$ is shown for \subref{fig:holenotch_5_fields:g} FEM and \subref{fig:holenotch_5_fields:h} DEM.}
    \label{fig:holenotch_5_fields_b}
\end{figure}

The phase-field and accumulated equivalent plastic strain distributions immediately before and after rupture are shown in \FIG{fig:holenotch_5_fields_a} and \FIG{fig:holenotch_5_fields_b}, respectively. Before rupture, both methods predict closely comparable localisation patterns. The accumulated equivalent plastic strain changes only slightly across the final transition, while the phase field undergoes the dominant change associated with complete crack linking. This suggests that the plastic localisation is largely established before complete separation. The subsequent failure is therefore dominated by rapid phase-field evolution within the plastically affected ligament. The abrupt failure and inclined crack development are also qualitatively consistent with the behaviour previously reported for the $r=\qty{5}{\milli\metre}$ notched benchmark.\cite{Ambati2015b}

\subsection{Ductile fracture of a V-notch specimen} \label{sec:num_exp_2}   

\begin{figure}[htbp!]
\centering
\setlength{\tabcolsep}{3pt}
\begin{tabular}{>{\centering\arraybackslash}m{0.33\textwidth} >{\centering\arraybackslash}m{0.67\textwidth}}
\subfloat[\label{fig:v_notched}]{
\includegraphics[width=0.29\textwidth]{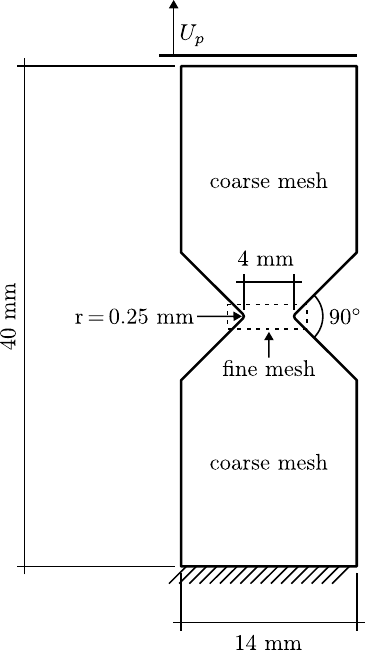}} &
\subfloat[\label{fig:diagonally_notched}]{
\includegraphics[width=0.67\textwidth]{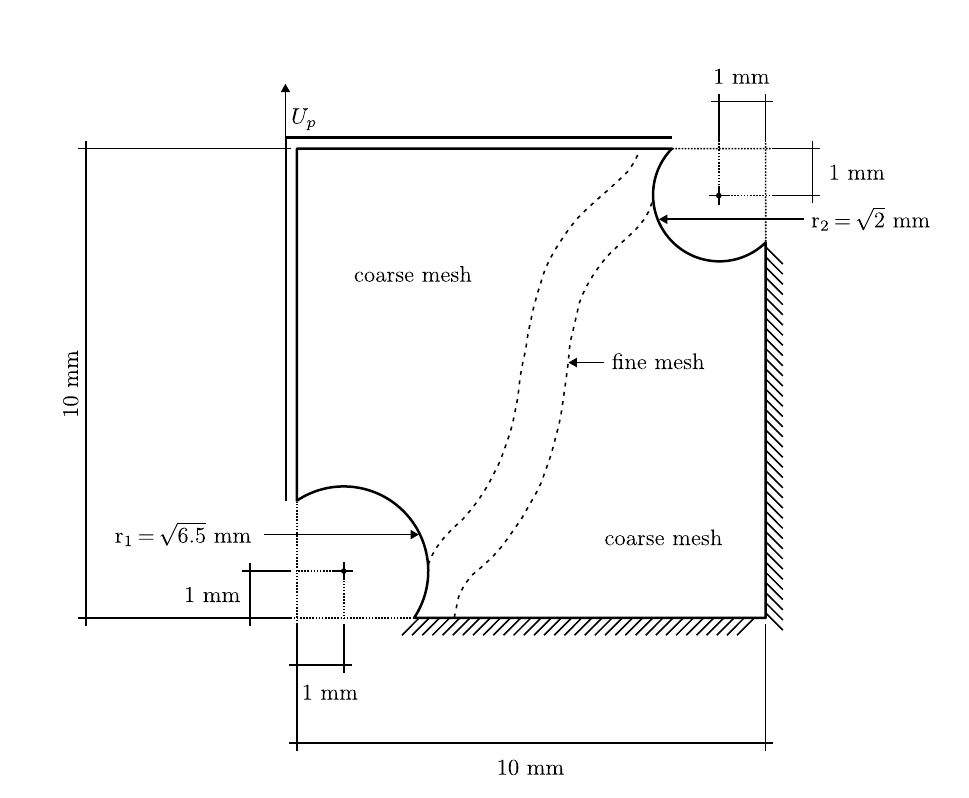}}
\end{tabular}
\caption[]{Geometry and boundary conditions for \subref{fig:v_notched} the V-notch specimen and \subref{fig:diagonally_notched} the diagonally notched specimen.}
\label{fig:notched_geometries}
\end{figure}

The V-notch specimen is used to assess whether the proposed DEM framework can reproduce symmetric crack propagation from opposing notch roots and subsequent linking across the central ligament. The predicted force-displacement response is also compared with the experimental measurements reported by Li et al.\cite{Li2011} The specimen geometry follows the V-notch benchmark considered by Miehe et al.\cite{Miehe2016} The geometry and boundary conditions are shown in \FIG{fig:v_notched}. The material parameters adopted in the present simulations are listed in \TAB{tab:vnotch_params}.

The FEM reference used \num{10000} load increments. The first $100$ increments covered the range up to $U_p=\qty{0.2}{\milli\metre}$, and the remaining \num{9900} increments covered the range up to $U_p=\qty{0.48}{\milli\metre}$. The computation required \qty{5.79}{\hour}. Each DEM run used \num{230} load increments. Of these, \num{30} increments covered the range up to $U_p=\qty{0.2}{\milli\metre}$ and \num{200} covered the range up to $U_p=\qty{0.48}{\milli\metre}$. Across the eight seeds, the mean DEM wall-clock time was \qty{19.0\pm2.5}{\hour}, with a range of \qtyrange{16.0}{22.8}{\hour}. No run reached the maximum of \num{500} coupled cycles at any increment.

\begin{figure}[htbp!]
\centering
\captionsetup[subfloat]{farskip=0pt,captionskip=1pt}
\setlength{\tabcolsep}{4pt}
\begin{tabular}{@{}c@{\hspace{4pt}}c@{}}
\subfloat[\label{fig:vnotch_plots:a}]{
\includegraphics[width=0.48\textwidth]{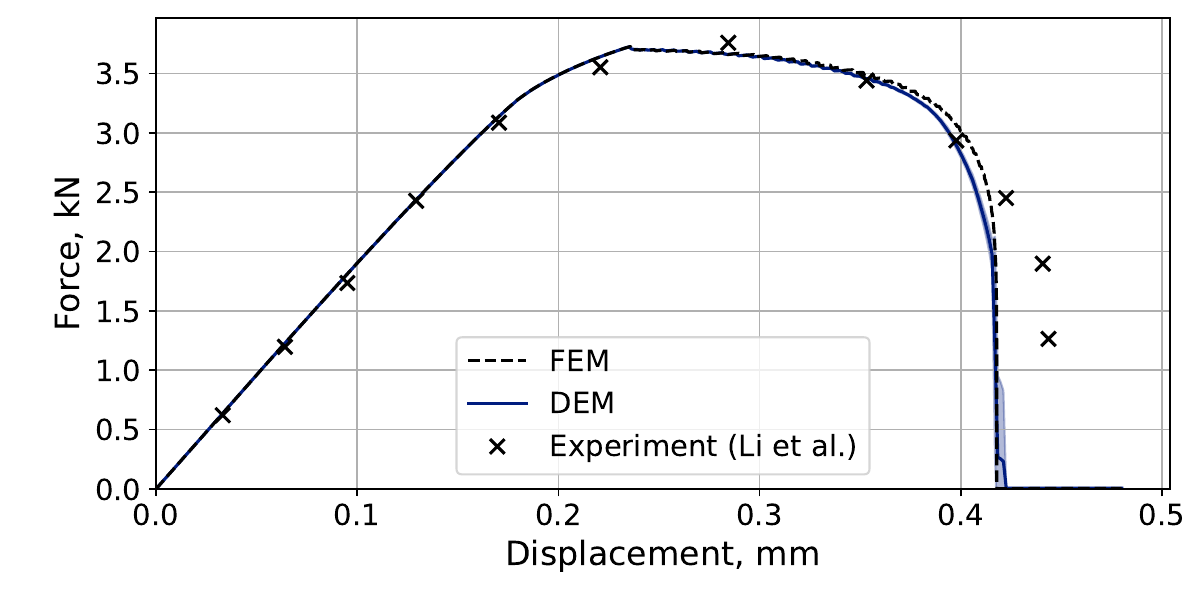}} &
\subfloat[\label{fig:vnotch_plots:b}]{
\includegraphics[width=0.48\textwidth]{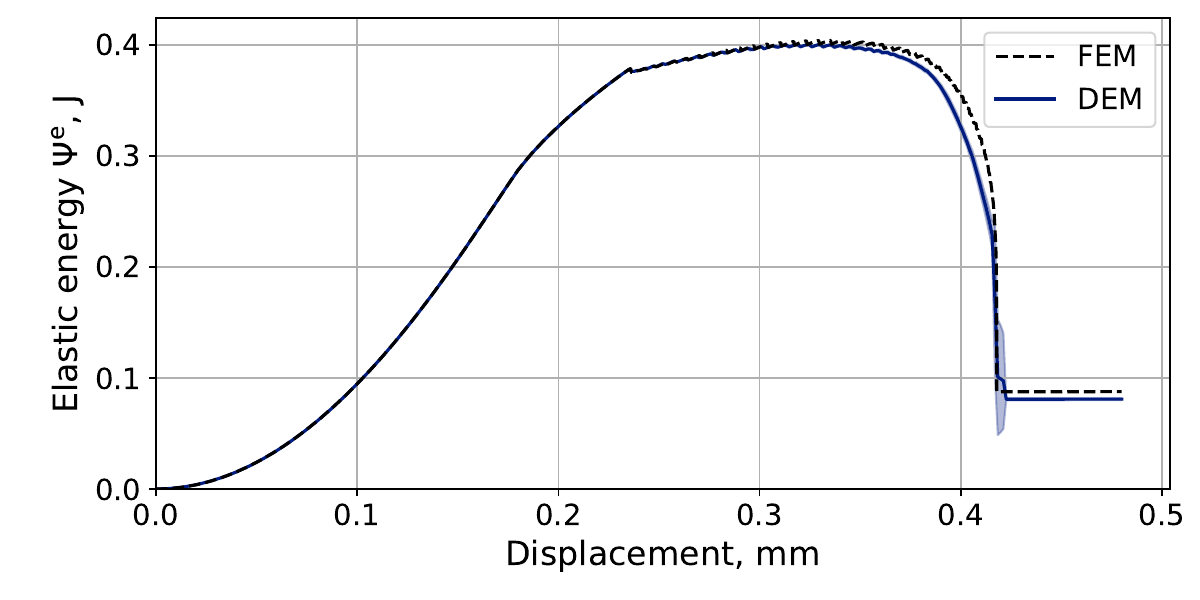}} \\[-2pt]
\subfloat[\label{fig:vnotch_plots:c}]{
\includegraphics[width=0.48\textwidth]{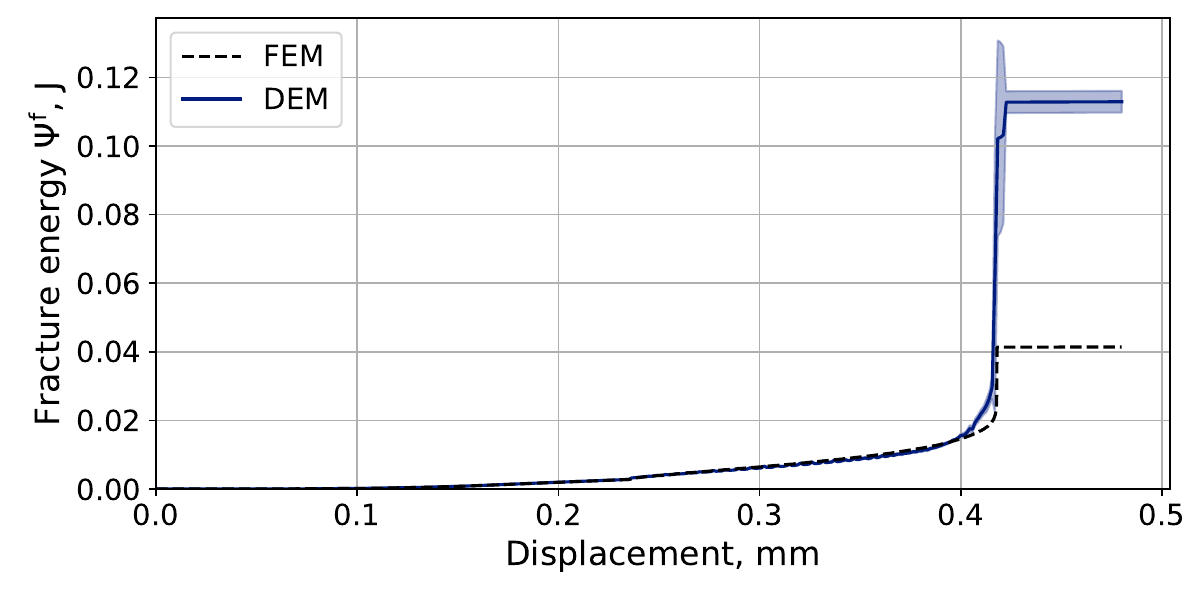}} &
\subfloat[\label{fig:vnotch_plots:d}]{
\includegraphics[width=0.48\textwidth]{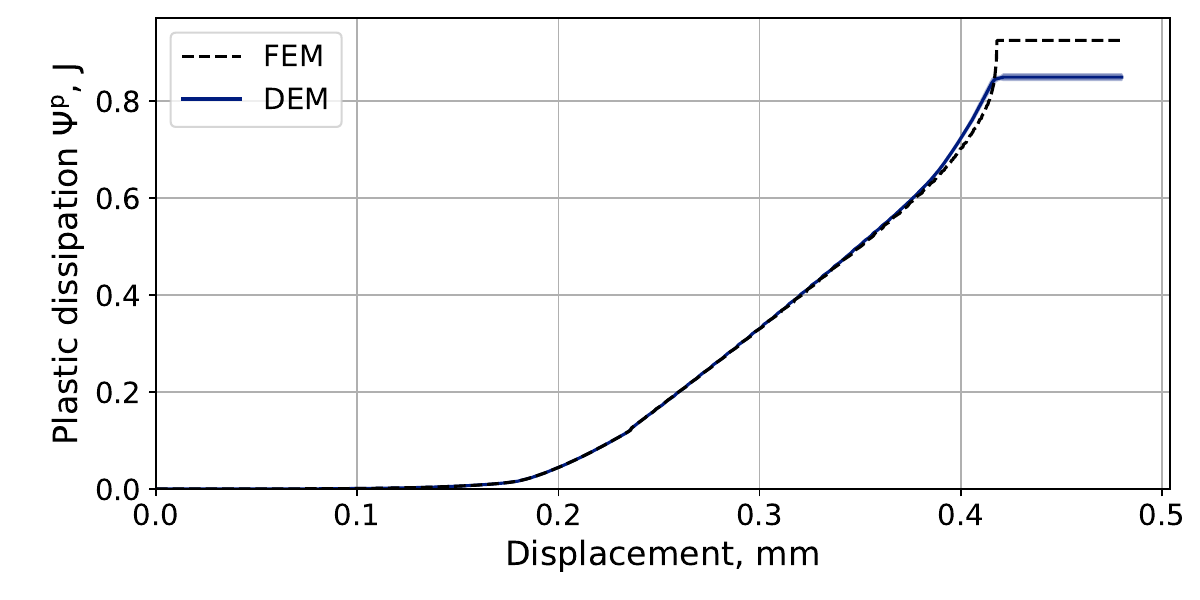}} \\[-2pt]
\multicolumn{2}{c}{%
\subfloat[\label{fig:vnotch_plots:e}]{
\includegraphics[width=0.96\textwidth]{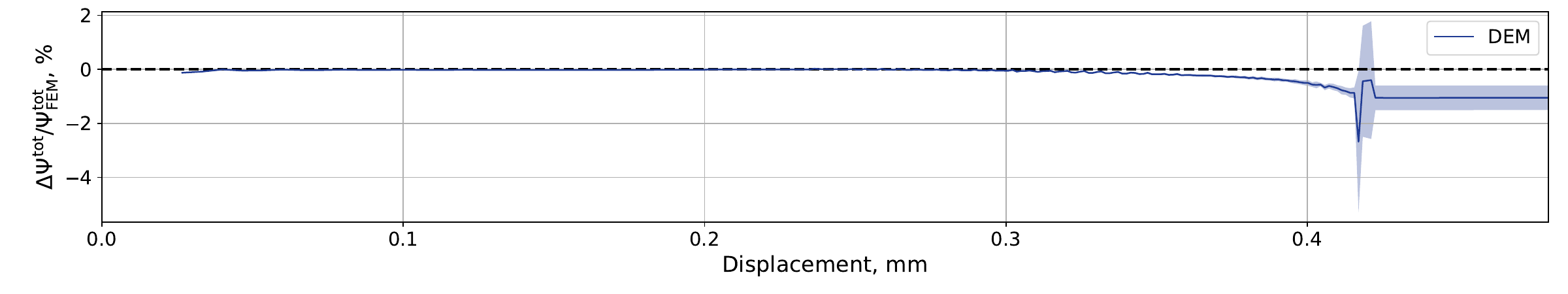}}} \\
\end{tabular}
\caption[]{V-notch specimen. Global response obtained with the proposed DEM and the FEM reference: \subref{fig:vnotch_plots:a} force-displacement response, including the experimental measurements of Li et al.\cite{Li2011}; \subref{fig:vnotch_plots:b} elastic energy $\Psi^{\mathrm{e}}$; \subref{fig:vnotch_plots:c} fracture energy $\Psi^{\mathrm{f}}$; and \subref{fig:vnotch_plots:d} plastic energy $\Psi^{\mathrm{p}}$; \subref{fig:vnotch_plots:e} relative difference in total energy, $100(\Psi^{\mathrm{tot}}_{\mathrm{DEM}}-\Psi^{\mathrm{tot}}_{\mathrm{FEM}})/\Psi^{\mathrm{tot}}_{\mathrm{FEM}}$, where $\Psi^{\mathrm{tot}}=\Psi^{\mathrm{e}}+\Psi^{\mathrm{f}}+\Psi^{\mathrm{p}}$. All quantities are plotted against the prescribed displacement $U_p$. The DEM curves represent the mean over eight seeds, and the shaded bands span $\pm1$ standard deviation.}
\label{fig:vnotch_plots}
\end{figure}

The global responses are compared in \FIG{fig:vnotch_plots}. The DEM and FEM force-displacement curves agree closely through the loading branch, peak load and most of the softening regime. Both numerical solutions also follow the experimental measurements well over this range. Near final rupture, the experiment shows a more gradual loss of load-carrying capacity, whereas the numerical responses drop more sharply. Overall, the DEM captures the main features of the measured response, with the largest difference occurring only during the final stage of failure.

The elastic-energy predictions remain in close agreement throughout the loading history. Near rupture, the DEM shows a somewhat larger fracture-energy contribution and a correspondingly smaller plastic contribution than the FEM reference. These differences partly offset one another, so the total-energy response remains close over most of the simulation. The largest deviation is confined to the short interval associated with final rupture. The narrow ensemble bands before this point also indicate that the stable response is only weakly affected by random initialisation. The increased spread is concentrated around the terminal instability, where the solution becomes more sensitive to the precise evolution of the fracture process.

\begin{figure}[htbp!]
\centering
\setlength{\tabcolsep}{3pt}
\begin{tabular}{>{\centering\arraybackslash}m{0.23\textwidth}>{\centering\arraybackslash}m{0.23\textwidth}>{\centering\arraybackslash}m{0.23\textwidth}>{\centering\arraybackslash}m{0.23\textwidth}}
\subfloat[\label{fig:vnotch_fields:a}]{
\includegraphics[width=0.23\textwidth]{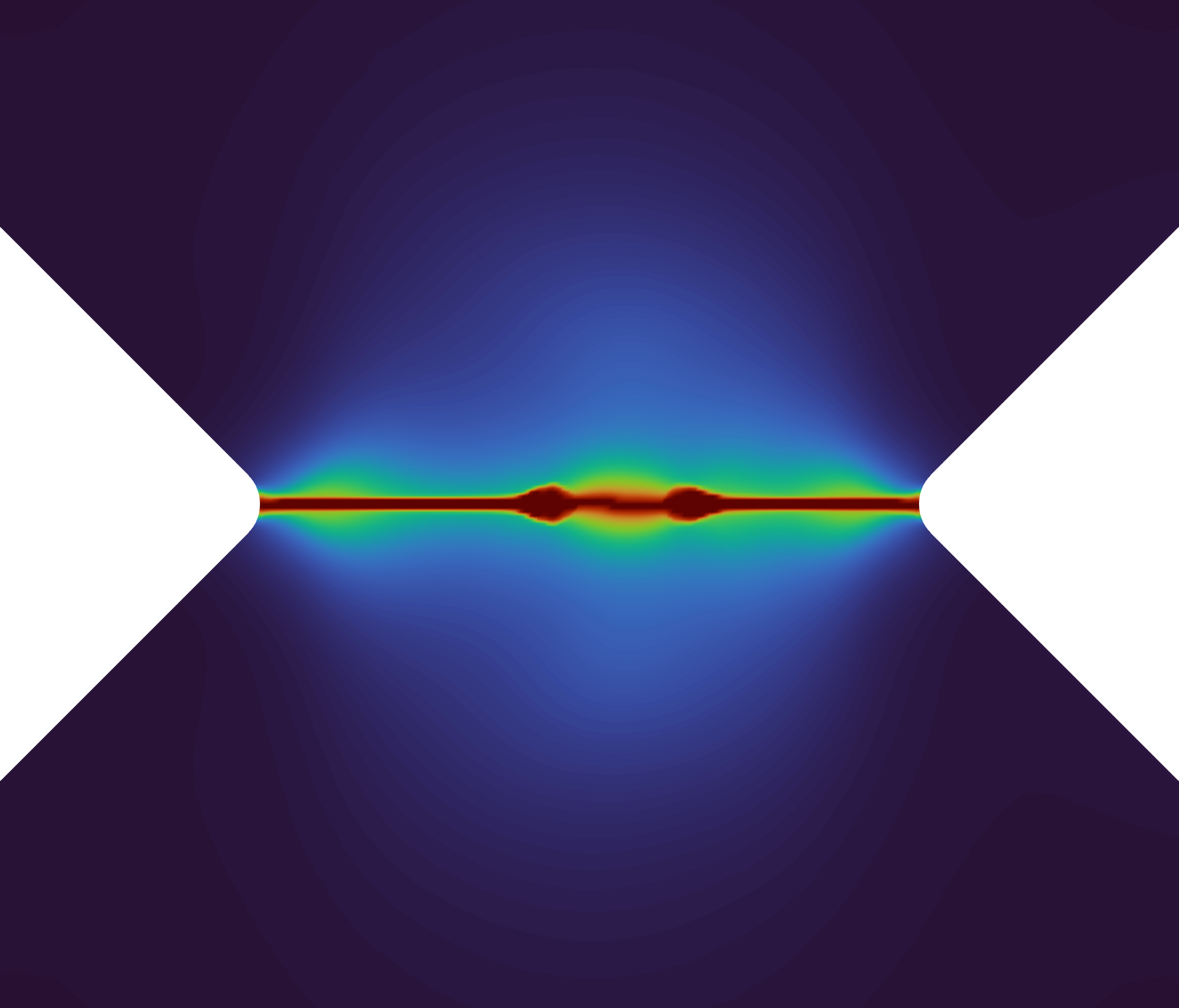}} &
\subfloat[\label{fig:vnotch_fields:b}]{
\includegraphics[width=0.23\textwidth]{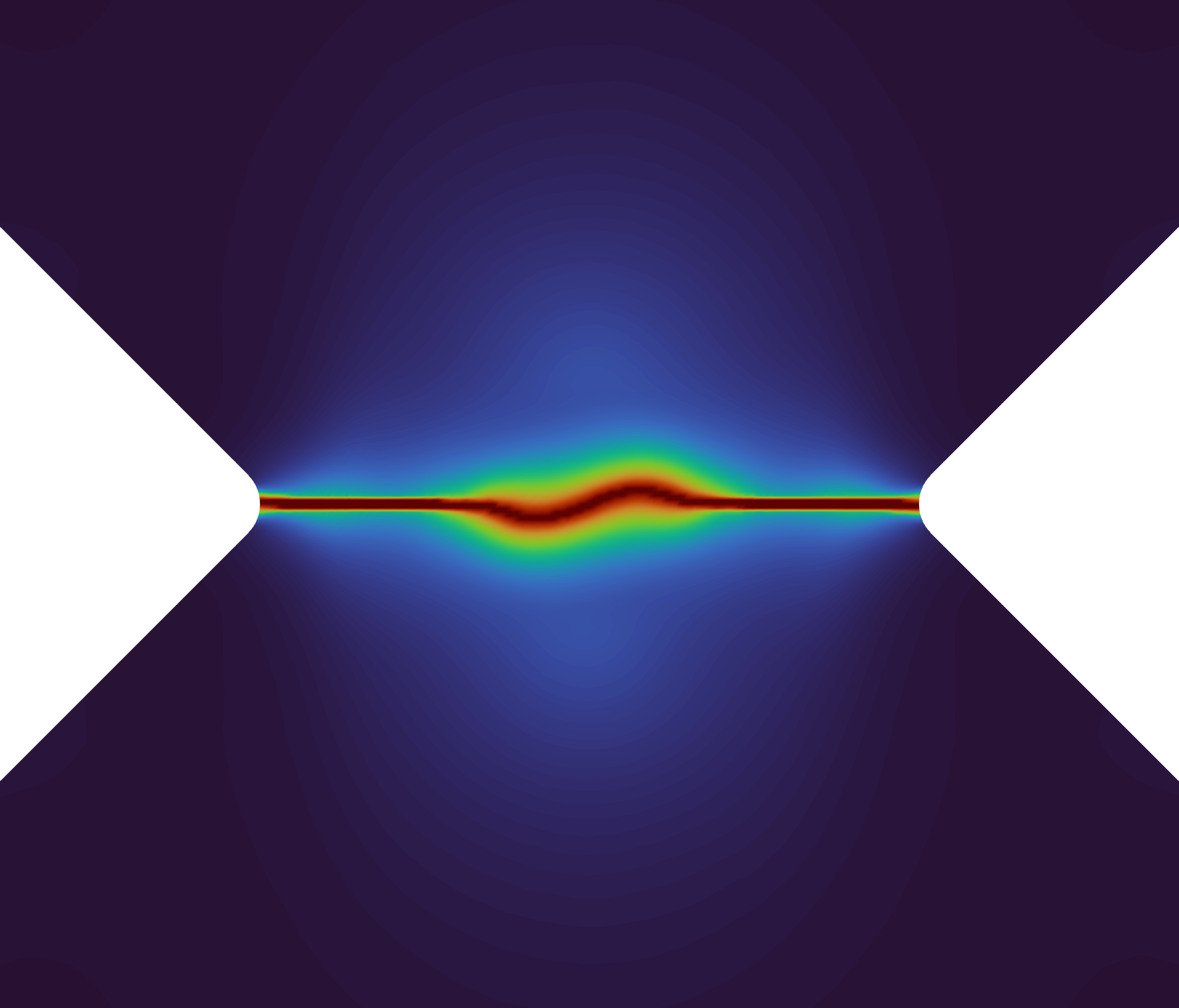}} &
\subfloat[\label{fig:vnotch_fields:c}]{
\includegraphics[width=0.23\textwidth]{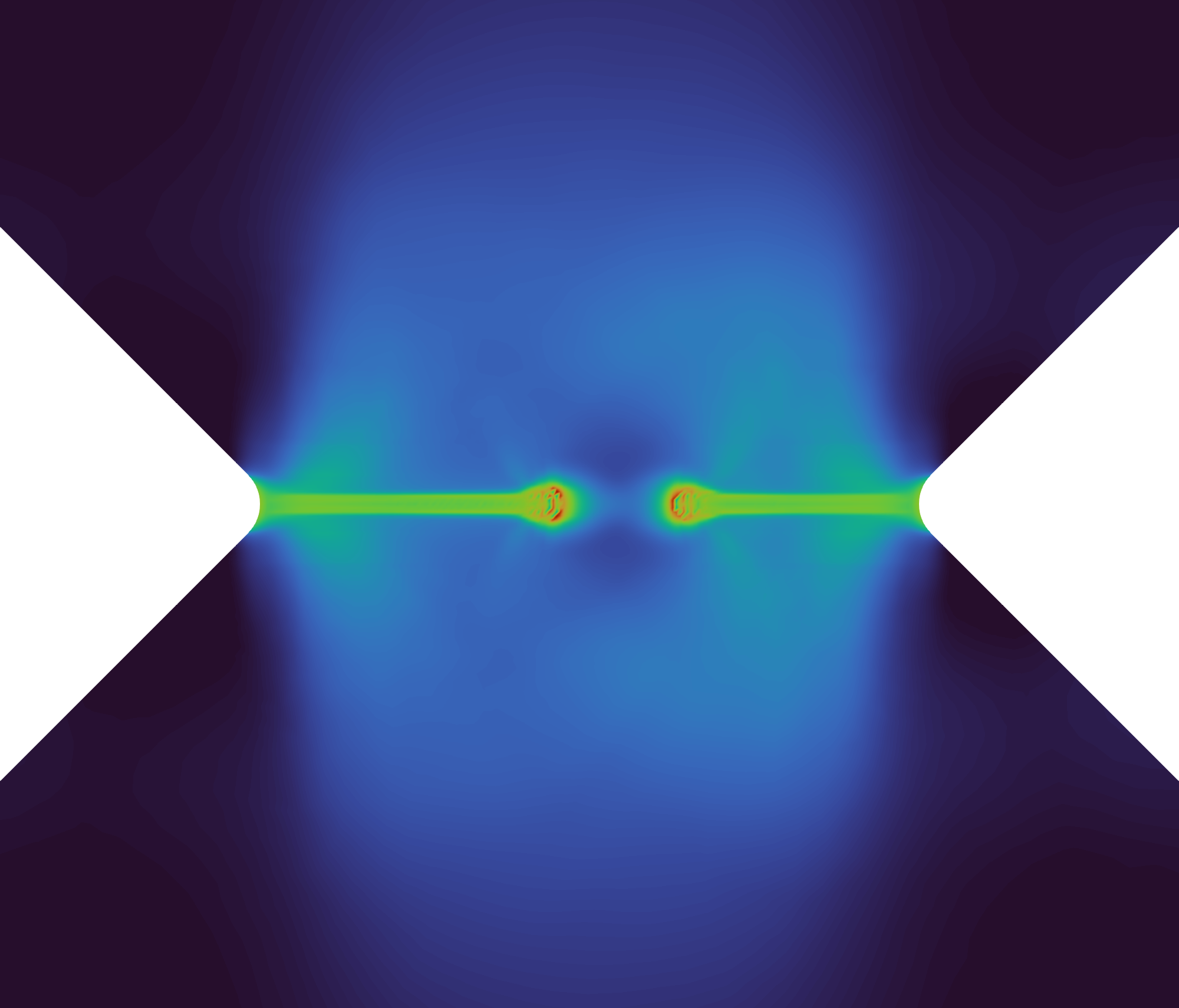}} &
\subfloat[\label{fig:vnotch_fields:d}]{
\includegraphics[width=0.23\textwidth]{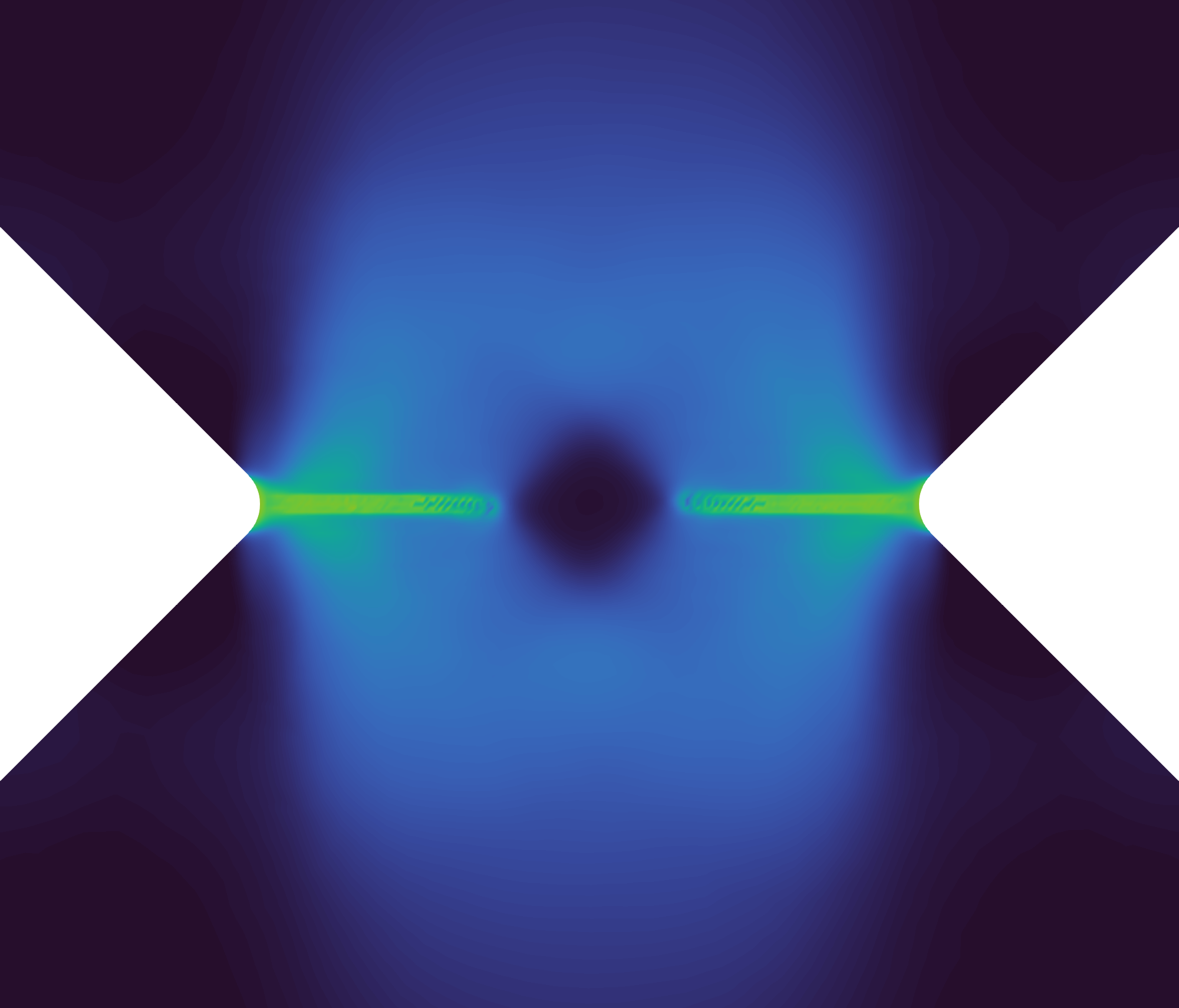}} \\
\multicolumn{2}{c}{\includegraphics[width=0.25\textwidth]{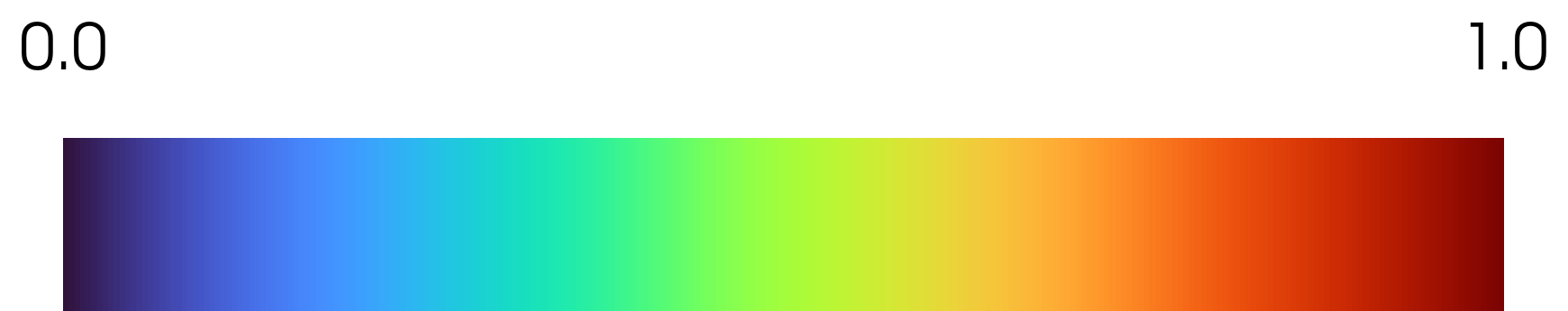}} &
\multicolumn{2}{c}{\includegraphics[width=0.25\textwidth]{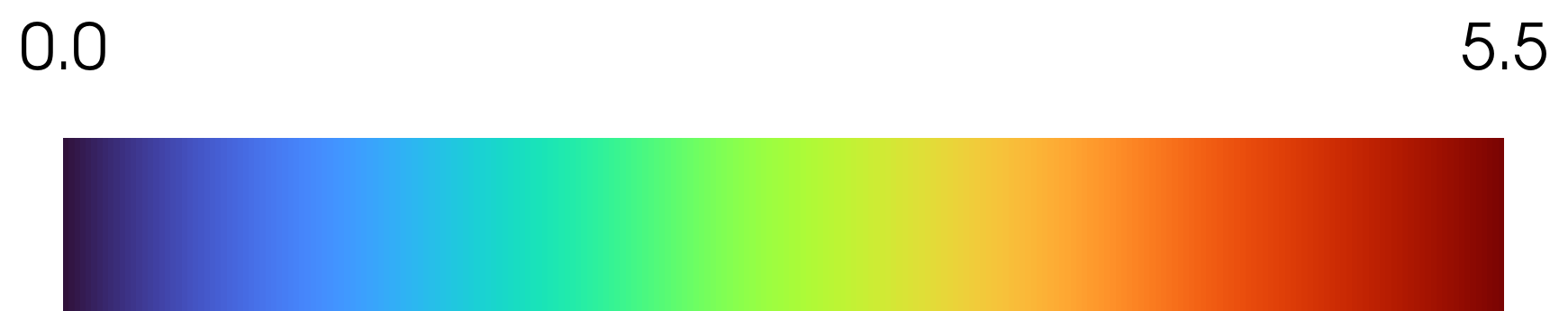}} \\
\end{tabular}
\caption[]{V-notch specimen. Phase-field and accumulated equivalent plastic strain distributions at rupture. The phase field $c$ is shown for \subref{fig:vnotch_fields:a} FEM and \subref{fig:vnotch_fields:b} DEM, while the accumulated equivalent plastic strain $\zeta$ is shown for \subref{fig:vnotch_fields:c} FEM and \subref{fig:vnotch_fields:d} DEM.}
\label{fig:vnotch_fields}
\end{figure}

The phase-field and accumulated equivalent plastic strain distributions at rupture are shown in \FIG{fig:vnotch_fields}. Both methods predict fracture localisation within the central ligament and a similar overall linking pattern. The FEM solution shows several small, localised degraded regions along the fracture zone, whereas the DEM produces a more continuous crack band. A similar difference is visible in the accumulated equivalent plastic strain field. The FEM solution exhibits stronger local concentrations in these regions, while the DEM distribution is smoother overall.

\begin{table}[htbp!]
\centering
\caption{Material parameters adopted for the V-notch specimen.}
\label{tab:vnotch_params}
\begin{tabular}{llll}
\toprule
\textbf{Parameter} & \textbf{Notation} & \textbf{Value} & \textbf{Unit} \\
\midrule
Young's modulus                & $E$                         & \num{68899.4}      & MPa  \\
Poisson's ratio                & $\nu$                       & $0.33$           & --   \\
Yield stress                   & $\sigma_{\mathrm{y}}$       & $530$            & MPa  \\
Hardening modulus              & $H$                         & $561$            & MPa  \\
Undegraded fracture toughness  & $\mathcal{G}_{c}^{0}$       & $90$             & N/mm \\
Phase-field length scale       & $l_0$                       & $0.2$            & mm   \\
Toughness-degradation interval & $\xi$                       & $0.2$            & --   \\
Residual toughness fraction    & $b$                         & \num{5e-3} & --   \\
Degradation threshold          & $\zeta^{\mathrm{cr}}$       & $0$              & --   \\
\bottomrule
\end{tabular}
\end{table}

\subsection{Ductile fracture of a diagonally-notched specimen} \label{sec:num_exp_3}

The diagonally notched specimen is used to assess the ability of the proposed DEM to reproduce a strongly curved crack path between two offset notches. The configuration requires progressive crack reorientation across the ligament under coupled plastic localisation and toughness degradation. Since the trajectory is not prescribed, this example provides a demanding test of the spatial field representations. The geometry and boundary conditions are shown in \FIG{fig:diagonally_notched}, and the material parameters are given in \TAB{tab:doublenotch_params}.

The FEM reference used \num{10000} load increments. The first \num{100} increments covered loading up to $U_p=\qty{0.23}{\milli\metre}$, while the remaining \num{9900} increments reached $U_p=\qty{0.53}{\milli\metre}$. The computation required \qty{6.79}{\hour}. Each DEM run used \num{230} load increments. Of these, \num{30} increments covered loading up to $U_p=\qty{0.23}{\milli\metre}$ and \num{200} reached $U_p=\qty{0.53}{\milli\metre}$. Across the eight seeds, the DEM wall-clock time was \qty{22.8\pm1.7}{\hour}, with a range of \qtyrange{19.6}{24.3}{\hour}. Seeds $3$, $4$ and $7$ each reached the \num{500}-cycle limit once, during their rupture increments. No seed reached the limit before rupture. The cycle limit was therefore confined to the terminal event and did not affect the stable pre-rupture response.

\begin{figure}[htbp!]
	\centering
    \captionsetup[subfloat]{farskip=0pt,captionskip=1pt}
    \setlength{\tabcolsep}{4pt}
    \begin{tabular}{@{}c@{\hspace{4pt}}c@{}}
        \subfloat[\label{fig:diagonal_plots:a}]{
            \includegraphics[width=0.48\textwidth]{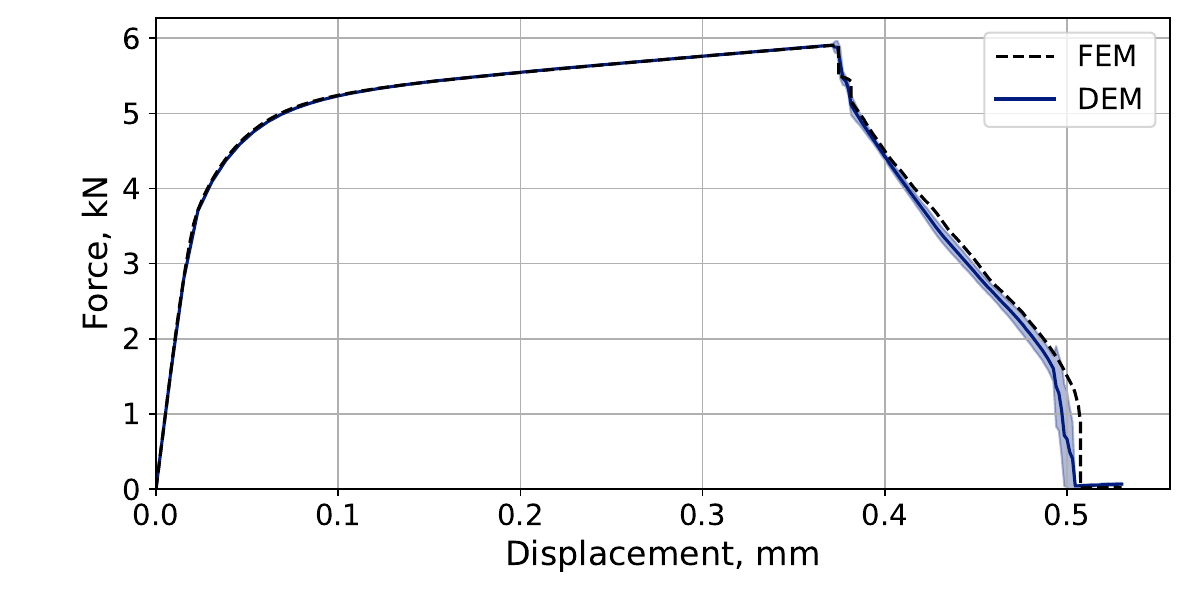}} &
        \subfloat[\label{fig:diagonal_plots:b}]{
            \includegraphics[width=0.48\textwidth]{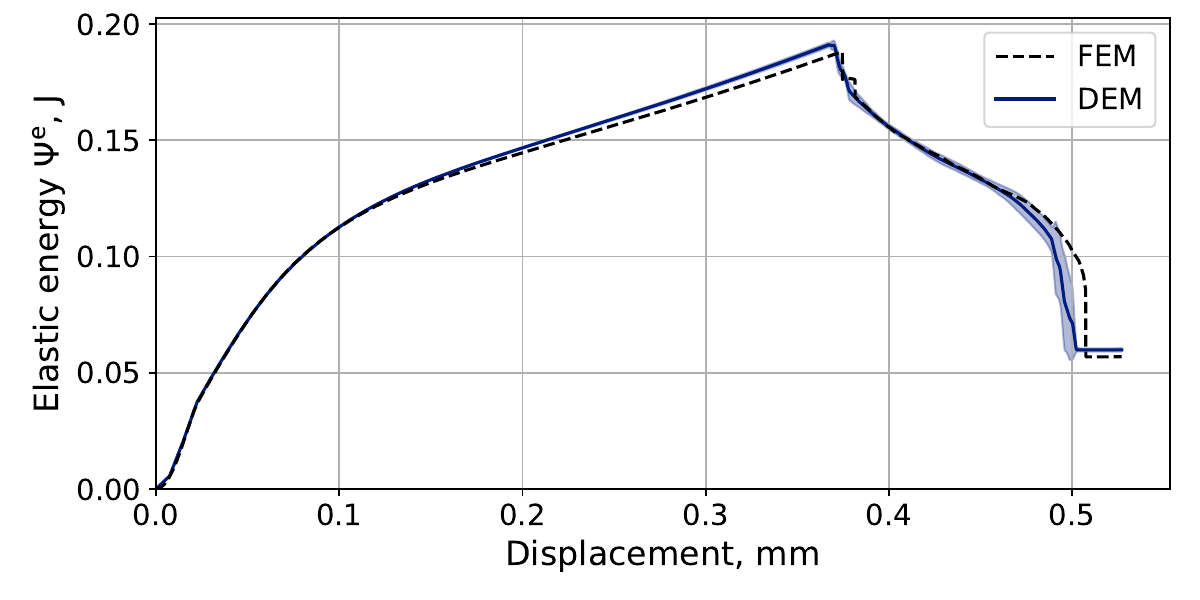}} \\[-2pt]
        \subfloat[\label{fig:diagonal_plots:c}]{
            \includegraphics[width=0.48\textwidth]{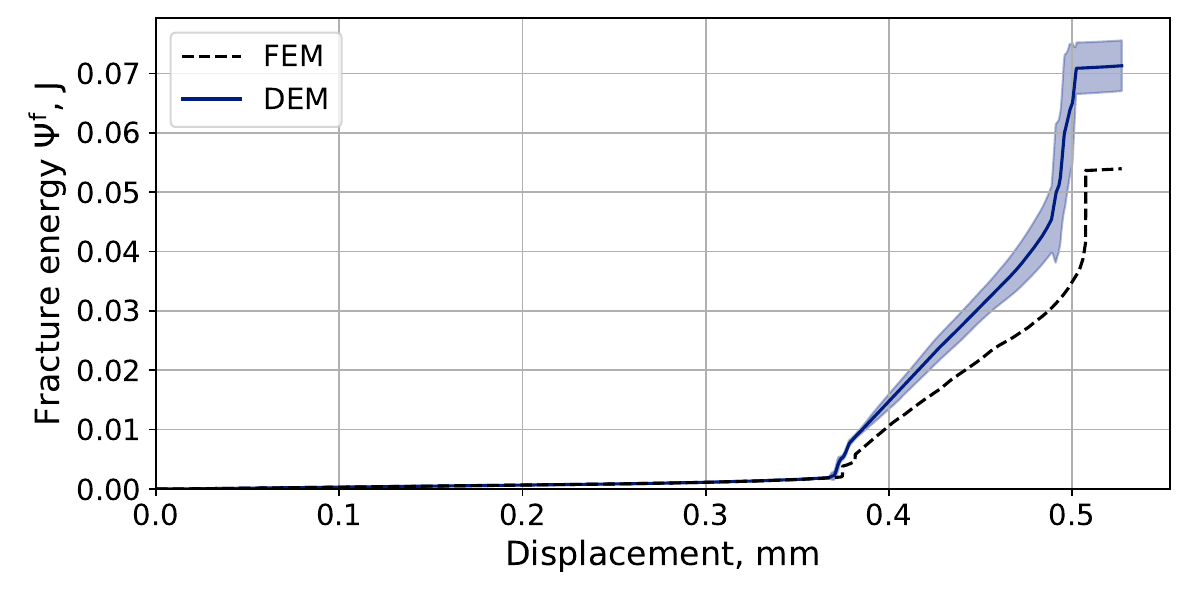}} &
        \subfloat[\label{fig:diagonal_plots:d}]{
            \includegraphics[width=0.48\textwidth]{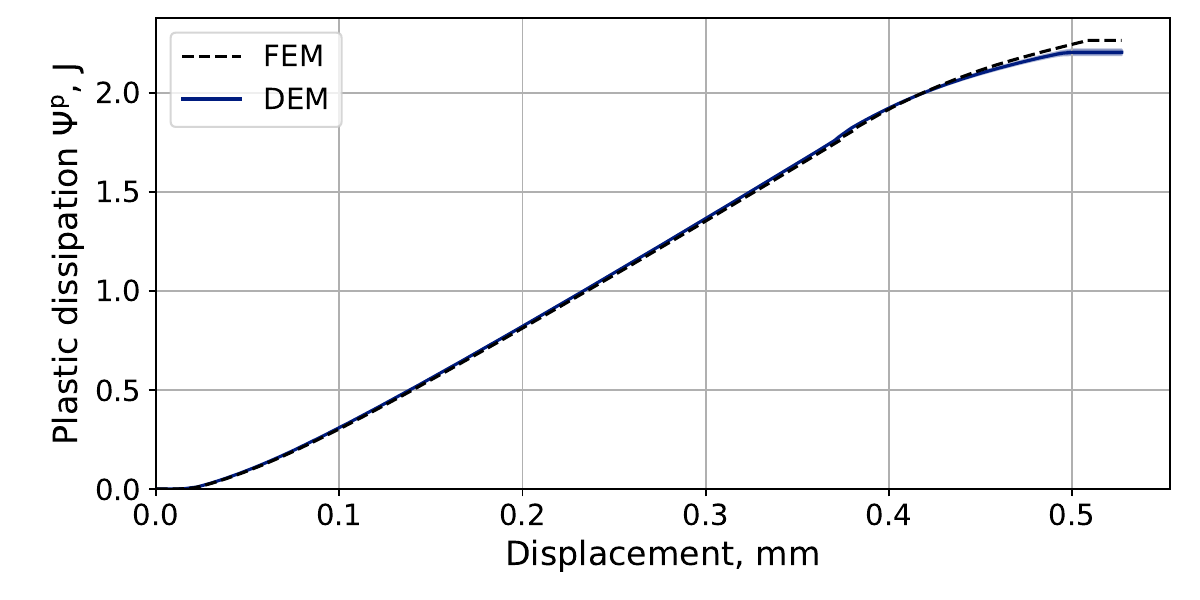}} \\[-2pt]
        \multicolumn{2}{c}{%
        \subfloat[\label{fig:diagonal_plots:e}]{
            \includegraphics[width=0.96\textwidth]{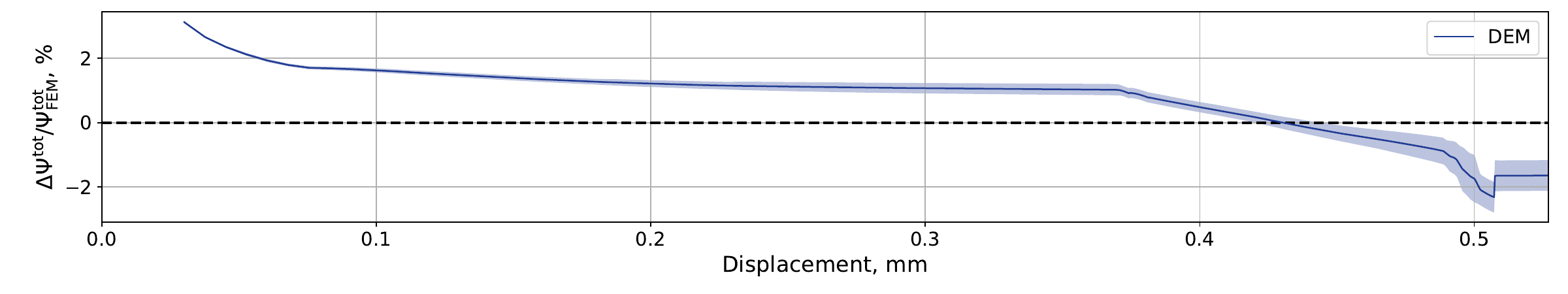}}} \\
	\end{tabular}
    \caption[]{Diagonally notched specimen. Global response obtained with the proposed DEM and the FEM reference: \subref{fig:diagonal_plots:a} force-displacement response; \subref{fig:diagonal_plots:b} elastic energy $\Psi^{\mathrm{e}}$; \subref{fig:diagonal_plots:c} fracture energy $\Psi^{\mathrm{f}}$; \subref{fig:diagonal_plots:d} plastic energy $\Psi^{\mathrm{p}}$; and \subref{fig:diagonal_plots:e} relative difference in total energy, $100(\Psi^{\mathrm{tot}}_{\mathrm{DEM}}-\Psi^{\mathrm{tot}}_{\mathrm{FEM}})/\Psi^{\mathrm{tot}}_{\mathrm{FEM}}$, where $\Psi^{\mathrm{tot}}=\Psi^{\mathrm{e}}+\Psi^{\mathrm{f}}+\Psi^{\mathrm{p}}$. All quantities are plotted against the prescribed displacement $U_p$. The DEM curves represent the mean over eight seeds, and the shaded bands span $\pm1$ standard deviation.}
    \label{fig:diagonal_plots}
\end{figure}

The global responses are compared in \FIG{fig:diagonal_plots}. The DEM ensemble closely reproduces the FEM force-displacement curve through loading, peak load and the subsequent softening regime. Rupture occurs at a slightly smaller prescribed displacement in DEM. The difference is limited to the terminal instability, while the peak load and preceding response remain closely matched. The narrow ensemble band over most of the loading history also indicates limited sensitivity to random initialisation before rupture.

The elastic and plastic contributions remain close to the FEM reference. The most noticeable difference occurs in the fracture energy. Once crack growth begins, the DEM predicts a somewhat larger fracture contribution and a slightly smaller terminal plastic contribution. The relative total-energy difference remains within a few per cent throughout the displayed loading range. It is slightly positive over most of the pre-rupture response and becomes modestly negative following the earlier DEM rupture. The differences between the individual energy components therefore partly compensate for one another.

The larger DEM fracture contribution accompanies a marginally more diffuse representation of the phase-field band. The substantially coarser DEM loading schedule may also contribute to the difference in energy partitioning. The load-step study in Section~\ref{sec:num_parametric} shows that the fracture and plastic energy contributions are sensitive to increment size, although it does not establish the cause of the discrepancy in the present specimen.

\begin{figure}[htbp!]
	\centering
    \setlength{\tabcolsep}{3pt}
    \begin{tabular}{>{\centering\arraybackslash}m{0.23\textwidth}
                    >{\centering\arraybackslash}m{0.23\textwidth}
                    >{\centering\arraybackslash}m{0.23\textwidth}
                    >{\centering\arraybackslash}m{0.23\textwidth}}
        \subfloat[\label{fig:diagonal_fields:a}]{
            \includegraphics[width=0.23\textwidth]{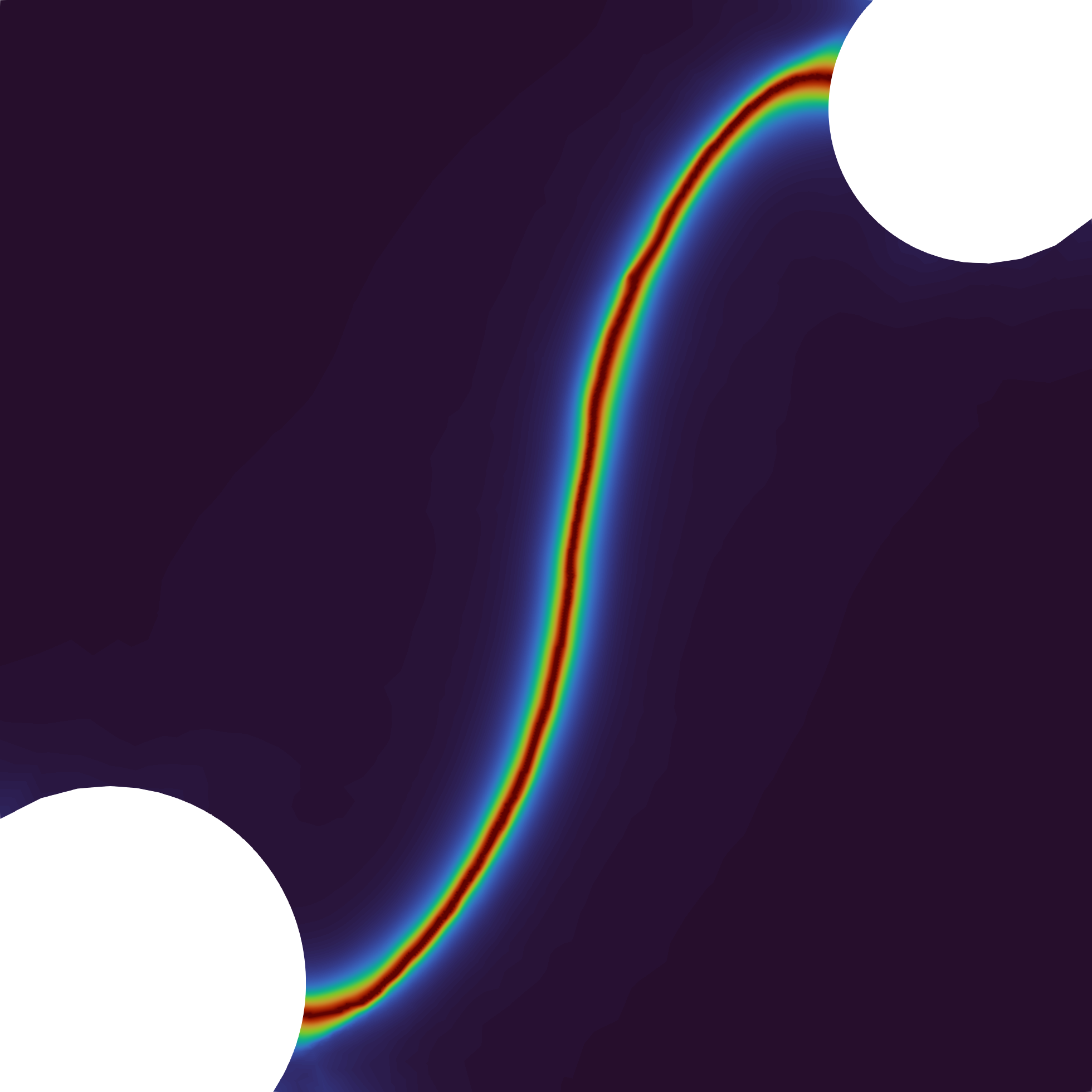}} &
        \subfloat[\label{fig:diagonal_fields:b}]{
            \includegraphics[width=0.23\textwidth]{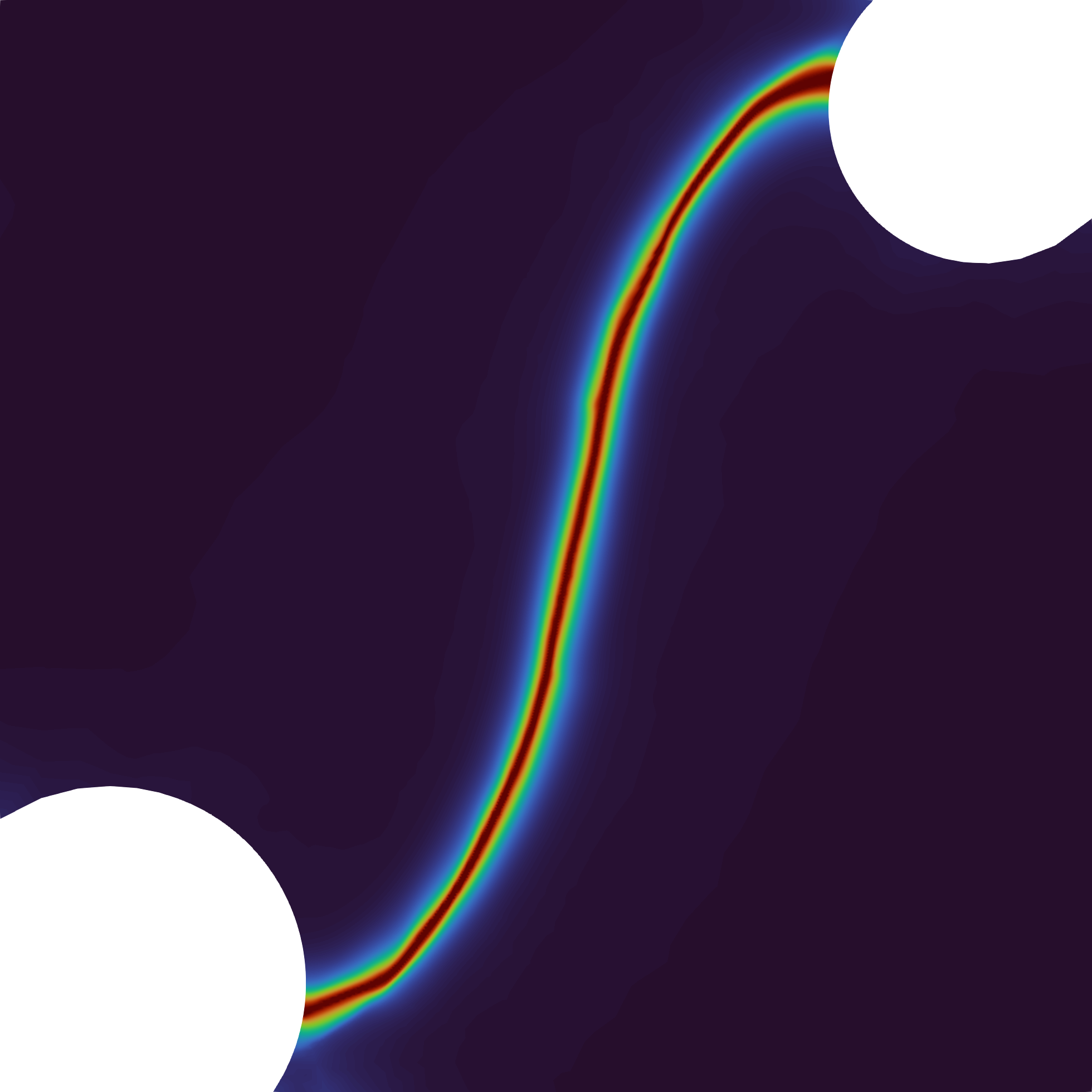}} &
        \subfloat[\label{fig:diagonal_fields:c}]{
            \includegraphics[width=0.23\textwidth]{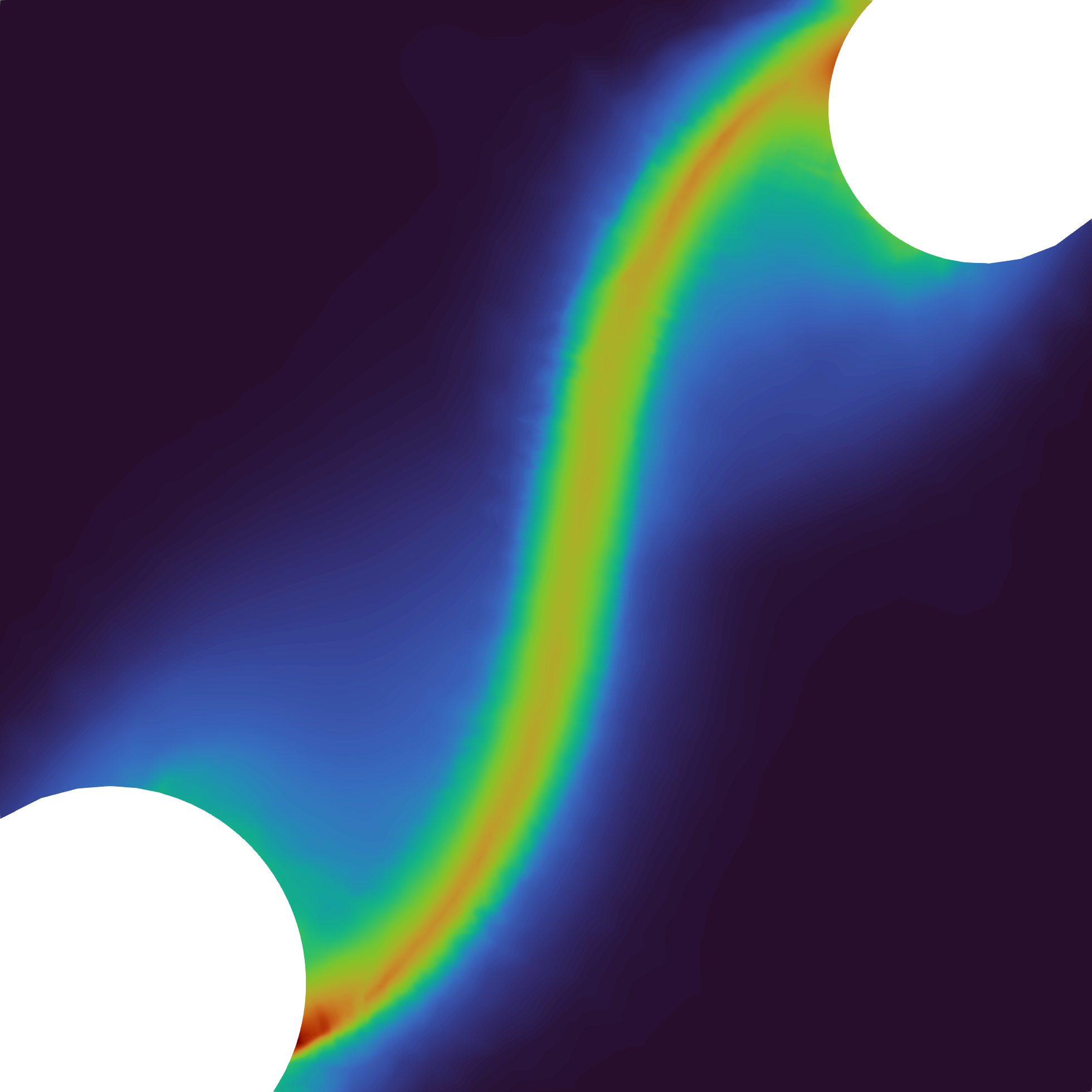}} &
        \subfloat[\label{fig:diagonal_fields:d}]{
            \includegraphics[width=0.23\textwidth]{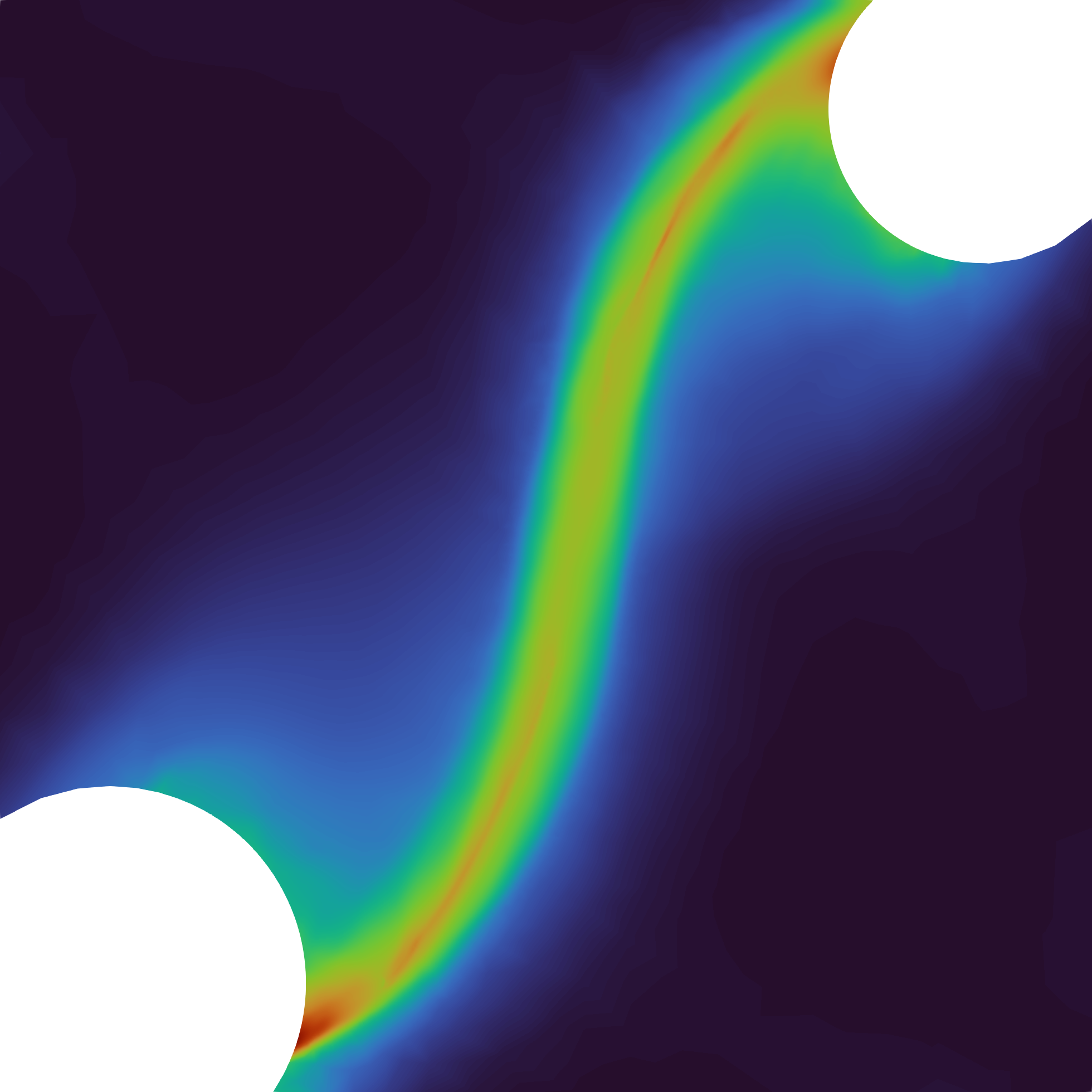}} \\
        \multicolumn{2}{c}{\includegraphics[width=0.25\textwidth]{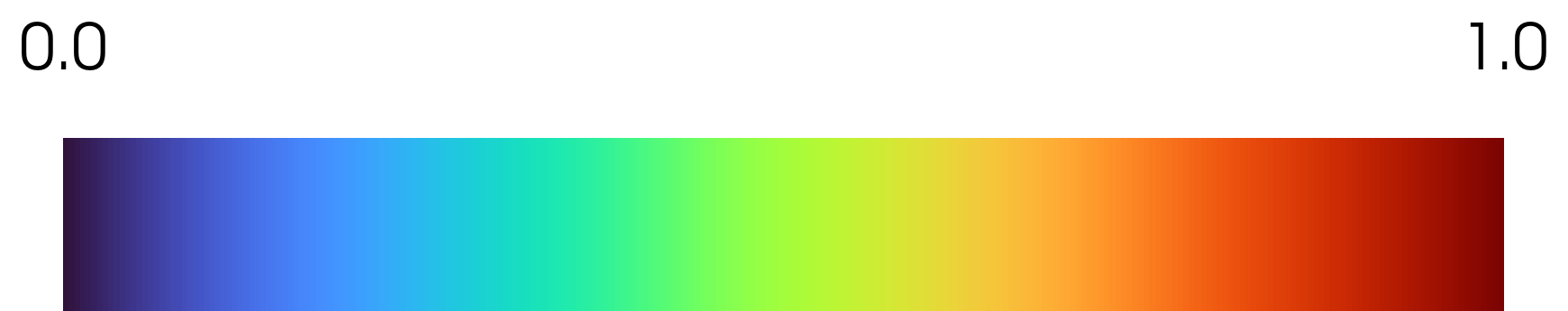}} &
        \multicolumn{2}{c}{\includegraphics[width=0.25\textwidth]{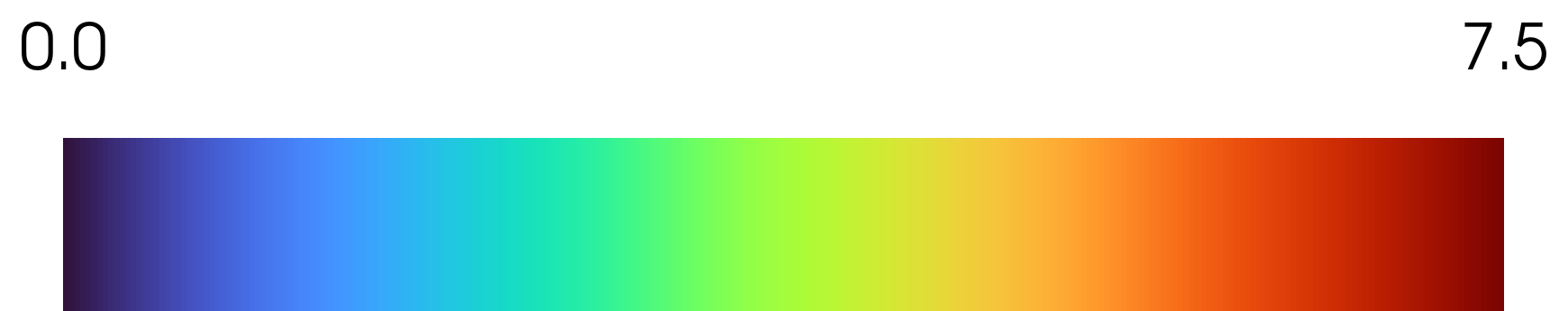}} \\
	\end{tabular}
    \caption[]{Diagonally notched specimen. Phase-field and accumulated equivalent plastic strain distributions at rupture. The phase field $c$ is shown for \subref{fig:diagonal_fields:a} FEM and \subref{fig:diagonal_fields:b} DEM, while the accumulated equivalent plastic strain $\zeta$ is shown for \subref{fig:diagonal_fields:c} FEM and \subref{fig:diagonal_fields:d} DEM. The DEM fields correspond to seed \num{8}.}
    \label{fig:diagonal_fields}
\end{figure}

The phase-field and accumulated equivalent plastic strain distributions at rupture are shown in \FIG{fig:diagonal_fields}. Both methods predict the same curved crack connecting the two offset notch roots. The linking locations and overall trajectory agree closely. The DEM phase-field band is marginally more diffuse, while its accumulated equivalent plastic strain is slightly lower in the central linking region. These differences are local and do not alter the predicted fracture mechanism. The close agreement is particularly relevant because the curved trajectory emerges from the coupled solution and is not imposed in advance.

\begin{table}[htbp!]
\centering
\caption{Material parameters adopted for the diagonally notched specimen.}
\label{tab:doublenotch_params}
\begin{tabular}{llll}
\toprule
\textbf{Parameter} & \textbf{Notation} & \textbf{Value} & \textbf{Unit} \\
\midrule
Young's modulus                & $E$                         & \num{195000}       & MPa  \\
Poisson's ratio                & $\nu$                       & $0.30$           & --   \\
Yield stress                   & $\sigma_{\mathrm{y}}$       & $443$            & MPa  \\
Hardening modulus              & $H$                         & $300$            & MPa  \\
Undegraded fracture toughness  & $\mathcal{G}_{c}^{0}$       & $209$            & N/mm \\
Phase-field length scale       & $l_0$                       & $0.2$            & mm   \\
Toughness-degradation interval & $\xi$                       & $0.35$           & --   \\
Residual toughness fraction    & $b$                         & \num{1e-2} & --   \\
Degradation threshold          & $\zeta^{\mathrm{cr}}$       & $0$              & --   \\
\bottomrule
\end{tabular}
\end{table}

\subsection{Displacement-step-size convergence study} \label{sec:num_parametric}

A displacement-step-size study is performed to assess the sensitivity of the DEM solution to the loading discretisation. The V-notch specimen from Section~\ref{sec:num_exp_2} is selected because its gradual softening is followed by a rapid terminal rupture. The first \num{30} increments, covering the initial loading response up to $U_p=\qty{0.2}{\milli\metre}$, are retained in every simulation. The subsequent range up to $U_p=\qty{0.48}{\milli\metre}$ is discretised using $N\in \{\numlist[list-separator = {, },list-final-separator = {, }]{50;100;150;200;250;300}\}$ increments. $6$ random seeds are considered for each value of $N$. This arrangement isolates the influence of the refined-stage discretisation while preserving the same initial loading history.

\begin{figure}[htbp!]
\centering
\captionsetup[subfloat]{farskip=0pt,captionskip=1pt}
\setlength{\tabcolsep}{4pt}
\begin{tabular}{@{}c@{\hspace{4pt}}c@{}}
\subfloat[\label{fig:convergence:a}]{\includegraphics[width=0.48\textwidth]{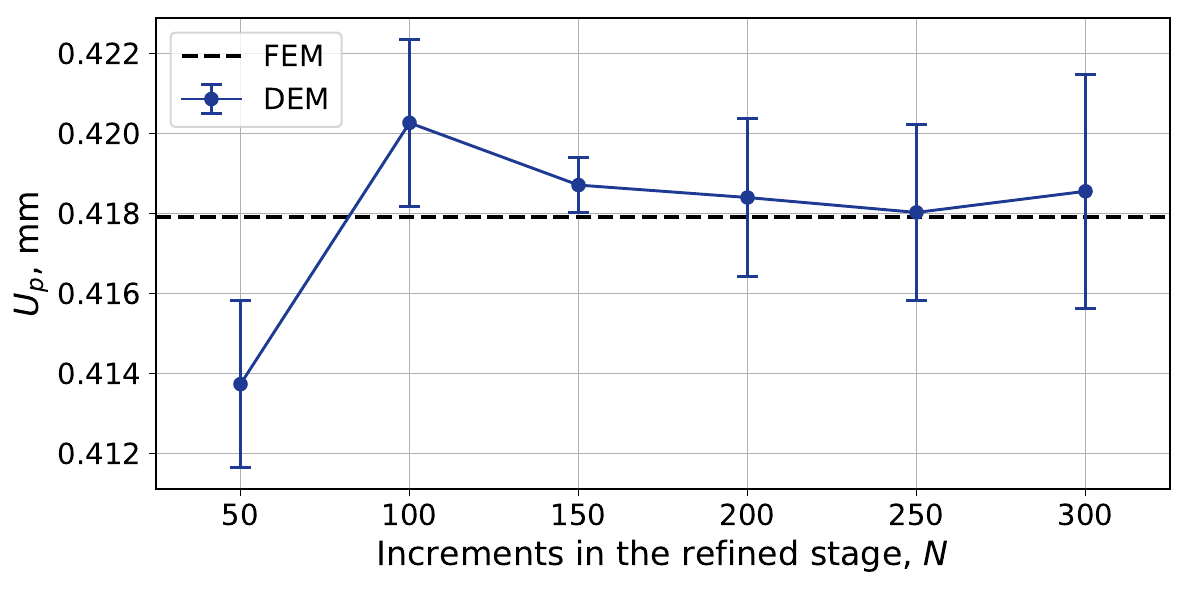}} &
\subfloat[\label{fig:convergence:b}]{\includegraphics[width=0.48\textwidth]{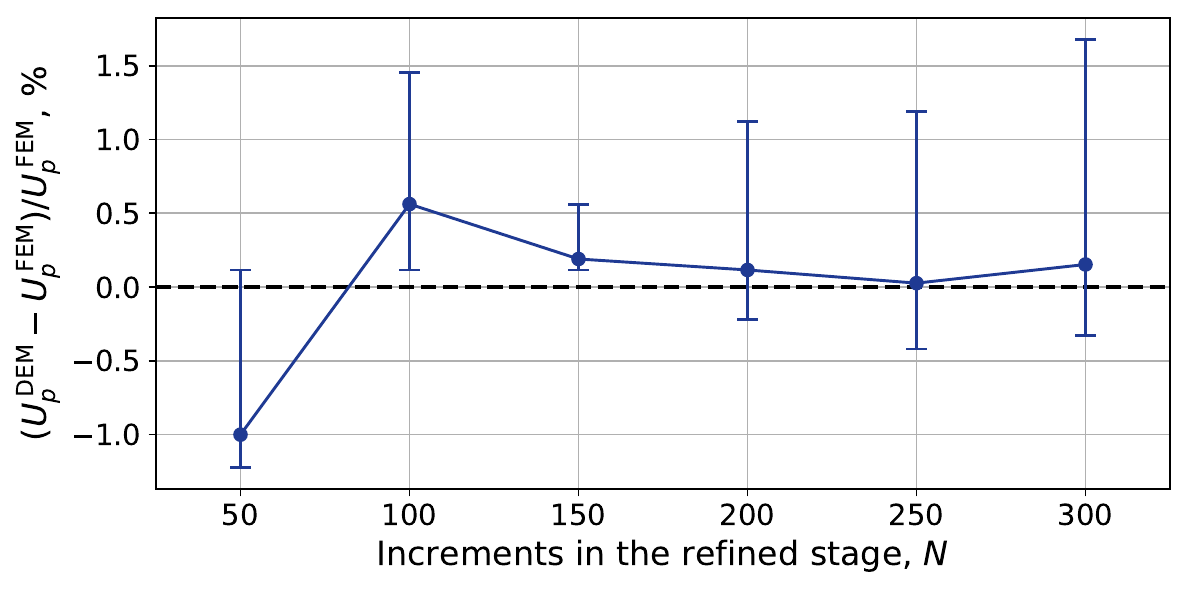}} \\[-2pt]
\subfloat[\label{fig:convergence:c}]{\includegraphics[width=0.48\textwidth]{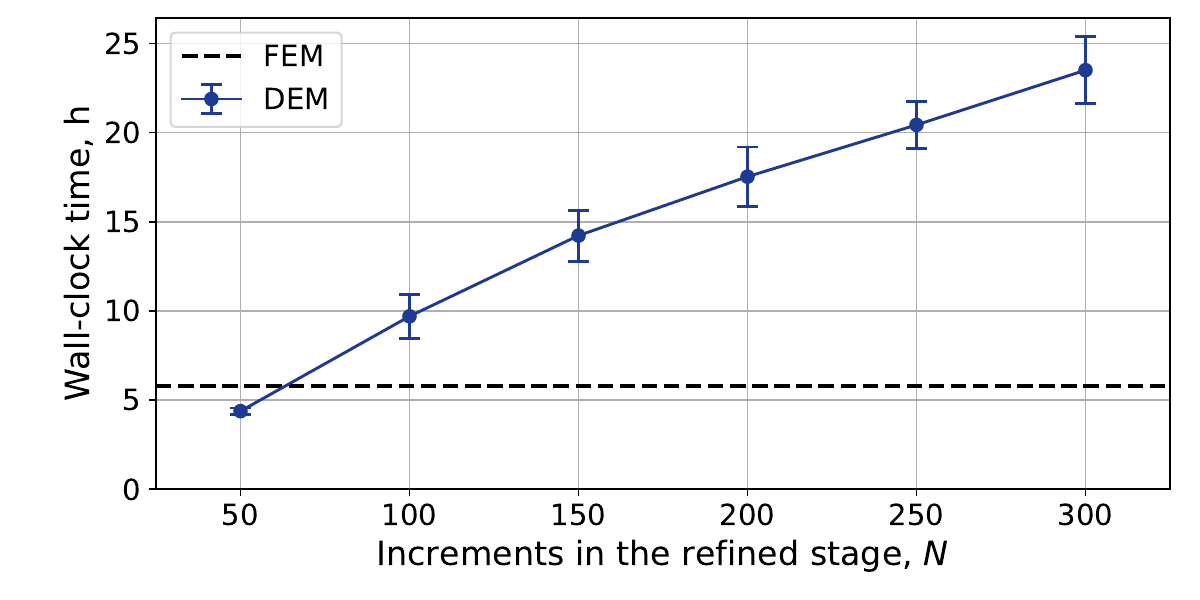}} &
\subfloat[\label{fig:convergence:d}]{\includegraphics[width=0.48\textwidth]{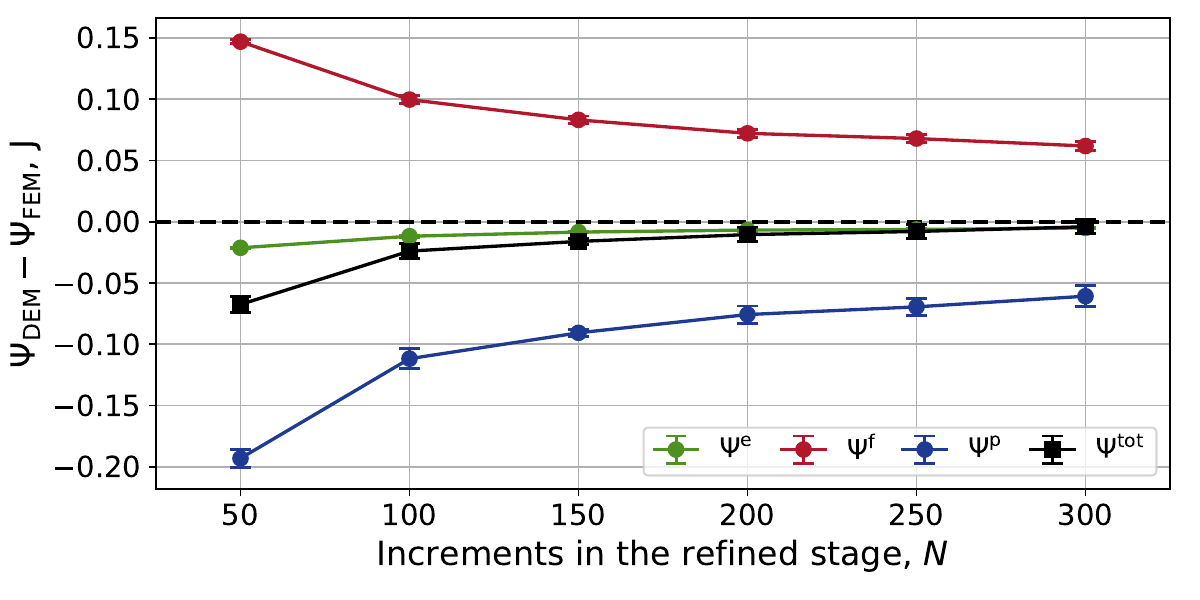}}
\end{tabular}
\caption[]{Displacement-step-size convergence of the DEM for the V-notch specimen. The refined loading stage from $U_p=\qty{0.2}{\milli\metre}$ to $U_p=\qty{0.48}{\milli\metre}$ is discretised using $N$ increments, with $6$ seeds considered for each value of $N$. \subref{fig:convergence:a} Rupture displacement, shown as the mean $\pm1$ standard deviation, together with the FEM reference; \subref{fig:convergence:b} signed relative error in rupture displacement, $100(U_{p}^{\mathrm{DEM}}-U_{p}^{\mathrm{FEM}})/U_{p}^{\mathrm{FEM}}$, with whiskers spanning the minimum and maximum values across the $6$ seeds; \subref{fig:convergence:c} wall-clock time per seed, shown as the mean $\pm1$ standard deviation, together with the FEM runtime; and \subref{fig:convergence:d} DEM-FEM differences in the final elastic, fracture, plastic and total energies.}
\label{fig:convergence_plots}
\end{figure}

The results are summarised in \FIG{fig:convergence_plots}. Since the plastic internal variables and the associated toughness degradation are updated incrementally, the terminal solution can depend on the sequence of prescribed displacement states. The effect is most apparent during rapid crack linking. Nevertheless, the mean rupture displacement becomes largely insensitive to further refinement once a sufficient number of increments is used.

The mean rupture displacement changes noticeably between $N=\num{50}$ and $N=\num{100}$, but stabilises from approximately $N=\num{150}$ onwards. For $N\geq\num{150}$, the mean rupture displacement remains within approximately \qty{0.2}{\percent} of the FEM reference. The largest error among all individual seeds and discretisations remains below approximately \qty{1.8}{\percent}. The seed-to-seed range does not decrease monotonically with $N$. The mean and the ensemble spread should therefore both be considered when assessing convergence of the abrupt terminal event.

The DEM wall-clock time increases approximately linearly with the number of refined-stage increments. The mean time rises from approximately \qty{4}{\hour} at $N=\num{50}$ to approximately \qty{24}{\hour} at $N=\num{300}$. Each additional increment requires a separate coupled optimisation and constitutive update, which is consistent with the observed scaling. The FEM wall-clock time is included in \FIG{fig:convergence:c} for context. However, the FEM and DEM calculations were performed on different hardware. The comparison should therefore not be interpreted as a hardware-normalised measure of relative computational efficiency.

The magnitudes of the final energy differences in \FIG{fig:convergence:d} generally decrease towards zero as the loading discretisation is refined. At the coarsest discretisation, the DEM predicts lower elastic and plastic contributions and a higher fracture contribution than FEM. This pattern is consistent with coarse increments resolving fewer intermediate coupled states during crack linking and with the lower plastic accumulation observed before complete fracture.

As $N$ increases, the elastic, fracture and plastic contributions all approach their FEM values. The magnitude of the final total-energy difference decreases from approximately \qty{0.07}{\joule} at $N=\num{50}$ to less than \qty{0.01}{\joule} at $N=\num{300}$. The fracture and plastic differences have opposite signs and therefore partly compensate. Consequently, agreement in total energy alone provides a less stringent measure than agreement in the individual components.

The $N=\num{200}$ discretisation used for the principal V-notch comparison lies within the regime in which the mean rupture displacement has stabilised. Its mean rupture-displacement error is approximately \qty{0.1}{\percent}, and its final total-energy difference is small. It therefore provides a practical balance between loading resolution and computational cost for this example. The study does not establish a universal optimal increment count, which will depend on the loading history, fracture instability and required accuracy.

\section{Conclusions} \label{sec:conclusions}

The numerical results show that the proposed DEM reproduces the principal features of the FEM solutions across the five ductile-fracture benchmarks considered. The force-displacement responses, energy evolution, crack trajectories and accumulated equivalent plastic strain fields generally remain in close agreement throughout stable loading and softening. The largest differences are concentrated around abrupt terminal fracture, where the incremental plastic history, toughness degradation and crack-linking process make the solution more sensitive to the loading discretisation. The V-notch step-size study confirms this behaviour: for $N\geq\num{150}$, the mean rupture displacement remains within approximately \qty{0.2}{\percent} of the FEM reference, while the largest error across the individual seeds and discretisations considered remains below approximately \qty{1.8}{\percent}. Across the random initialisations considered, seed-to-seed variability is generally small during the stable response and becomes more pronounced near abrupt fracture events.

Computational cost remains a significant limitation of the present implementation. The displacement-step-size study shows an approximately linear increase in wall-clock time with the number of refined-stage loading increments because every additional increment requires a new coupled optimisation and constitutive update. Extended softening may also require many alternating phase-field-displacement cycles, while the reaction-force stabilisation criterion can become ineffective after complete separation when the reaction approaches numerical noise. Future work should therefore investigate adaptive displacement stepping, more efficient coupled stopping criteria and alternative optimisation strategies that reduce the number of field evaluations required per increment. Improved use of GPU parallelism and more efficient training algorithms will also be important if the method is to become computationally competitive for larger problems.

Further development of the trainable field representations is also required. Preliminary architecture studies showed that simply increasing the size of the displacement network did not provide a clear improvement relative to the additional computational cost. Improving equilibrium-residual convergence may therefore require enriched or adaptive displacement representations rather than larger networks alone. For the phase field, the symmetrically notched $r=\qty{2.5}{\milli\metre}$ example produced a narrower localisation band than the FEM reference. Since the Gaussian width is defined through $\sigma_k=s_{\mathrm{RBF}}h_k$, future work should examine the selection of $s_{\mathrm{RBF}}$, adaptive or trainable RBF widths, and the interaction between RBF resolution and the physical phase-field length scale.

The present framework also provides a basis for incorporating more complex constitutive physics. Natural extensions include gradient-enhanced plasticity, finite-strain elastoplasticity and thermomechanically coupled ductile fracture formulations.\cite{Miehe2016,Svolos2025} Such developments would provide a more demanding assessment of the advantages and limitations of the proposed treatment of local constitutive history.

\section*{CRediT authorship contribution statement}

\textbf{N. Plungė}: Conceptualisation, Data curation, Formal analysis, Investigation, Methodology, Software, Validation, Visualization, Writing - original draft.  
\textbf{P. Brommer}: Conceptualisation, Funding acquisition, Methodology, Supervision, Writing - review \& editing. 
\textbf{R. S. Edwards}: Conceptualisation, Methodology, Supervision, Writing - review \& editing. 
\textbf{E. G. Kakouris}: Conceptualisation, Funding acquisition, Methodology, Supervision, Writing - review \& editing.

\section*{Declaration of competing interest}
The authors declare that they have no competing financial interests or personal relationships that could have appeared to influence the research presented in this article.

\section*{Data Availability Statement}

The input data supporting this study, together with the code used to produce the FEM and DEM results, are made available via the associated Zenodo repository.\cite{plunge2026zenodo} The main output data are reported in the figures, while additional data may be made available from the corresponding author upon reasonable request.

\section*{Acknowledgements}

The authors acknowledge the computing resources provided by the Scientific Computing Research Technology Platform at the University of Warwick. GPU computations were undertaken on the Sulis Tier 2 HPC platform, hosted by the University of Warwick Scientific Computing Research Technology Platform. Sulis is funded by Engineering and Physical Sciences Research Council (EPSRC) Grant EP/T022108/1 and the HPC Midlands+ consortium. The first author is supported by a PhD studentship funded by the EPSRC Centre for Doctoral Training in Modelling of Heterogeneous Systems, HetSys, Grant EP/S022848/1, project reference 2886009.

\bibliography{mybibfile.bib}

\end{document}